\documentclass[twocolumn,showpacs,preprintnumbers,amsmath,amssymb]{revtex4-1}

\usepackage{graphicx}
\usepackage{dcolumn}
\usepackage{bm}

\begin{document}

\preprint{APS/123-QED}

\title{Extension of the standard Heisenberg Hamiltonian to multispin exchange interactions}

\author{S.~Mankovsky, S.~Polesya  and H.~Ebert}
\affiliation{%
Department of Chemistry/Phys. Chemistry, LMU Munich,
Butenandtstrasse 11, D-81377 Munich, Germany \\
}%

\newcommand{\DONE}[1]{ \marginpar{\bf\large DONE} {\bf \em #1}  \marginpar{\bf\large DONE}}

\newcommand{\DISC}{ \marginpar{\bf\em\large DIS\-CUS\-SION} }  

\date{\today}

\begin{abstract}
  An extension of the Heisenberg Hamiltonian is discussed, that allows
  to go beyond the standard bilinear spin Hamiltonian taking into account various
  contributions due to multispin interactions
  having both chiral and non-chiral character. The parameters of the
  extended Hamiltonian are calculated from first principles within the
  framework of the multiple scattering Green function formalism giving
  access to an explicit representation of these parameters in real
  space. The discussions are focused on the chiral interactions,
  i.e.\ biquadratic and three-spin Dzyaloshinskii-Moriya like vector
  interactions $\vec{\cal{D}}_{ijij}$ (BDMI) and $\vec{\cal{D}}_{ijkj}$
  (TDMI), respectively, as well as  
  the three-spin chiral interaction (TCI)   $J_{ijk}$. 
  Although all parameters are driven by 
  spin-orbit coupling (SOC), some differences in their properties are 
  demonstrated by calculations for real materials.
In particular it is shown that the three-spin chiral interactions
  $J_{ijk}$ may be  topology as well as  SOC induced, 
  while the TDMI is associated  only with the SOC. 
  As the magnitude of the chiral interactions can
  be quite sizable, they can lead to 
  a stabilization of a noncollinear magnetic texture in some materials
  that is absent when these interactions are neglected. 
\end{abstract}

\maketitle

\section{ Introduction \label{IN}}

The Heisenberg spin Hamiltonian is nowadays a rather popular tool providing
a bridge between the electronic structure of magnetic materials and
their spin-dynamical and finite-temperature magnetic properties.
However, restriction of the classical model to only isotropic bilinear
exchange parameters is not always able to describe successfully the
experimental findings. In this case an extension of Heisenberg model is
used to take into account specific features of the system under
consideration. This concerns in particular the impact of spin-orbit
coupling (SOC) leading to magnetocrystalline anisotropy (MCA) and to the
spin-space anisotropy of the exchange coupling described by an 
exchange tensor $\underline{J}_{ij}$ instead of scalar
parameters. The latter can be reduced to the isotropic exchange
parameters $J_{ij}$ and chiral  Dzyaloshinskii-Moriya (DM) vector
$\vec{D}_{ij}$ representing the antisymmetric part of the exchange
tensor ${\underline{J}}_{ij}$.  

Still, this form of the Hamiltonian implies for example neglecting the dependence of
the exchange parameters on the relative orientation of the magnetic
moments in the system. To go beyond this bilinear approximation for the 
inter-atomic exchange interactions, one can take into account
higher-order contributions to the Heisenberg Hamiltonian, i.e.,
biquadratic, fourth-order three-spin,  
four-spin interactions, etc. terms \cite{HO63,HO64,AB67,IU74,IU76,Aks80,Bro84a,IUS14,ALU+08}.

The origin of higher-order interactions was discussed already many
years ago by various authors \cite{Kit60,TU77,MGY88}, focusing on those
being isotropic in spin space. 
Obviously, the dominating mechanism responsible for these terms can be
different for different materials. 
Kittel \cite{Kit60} discussing the transition from the antiferromagnetic
(AFM) to the ferromagnetic (FM) state
in metamagnetic materials (including metals) suggested an important role
of the biquadratic exchange interaction due to exchange
magnetostriction caused by a dependence of the exchange interaction on
the volume during an AFM/FM transition. 
MacDonald et al. \cite{MGY88} discussed the Hubbard model
Hamiltonian, which can be  transformed in the limit of large on-site
Coulomb interaction $U$  and assuming half-filling of the electron
energy bands implying electron localization around atomic sites
to a form equivalent to the Heisenberg spin Hamiltonian.
An expansion of the Hamiltonian in powers of the ratio $t/U$ gives
access to high-order terms of the spin Hamiltonian with bilinear and
four-spin exchange interactions $\sim t^2/U$ and $\sim t^4/U^3$,
respectively \cite{MGY88,BBMK08,BLHK16}.
In the absence of a magnetic field braking time reversal symmetry of
  the system the three-spin term should vanish as it is
antisymmetric with respect to time reversal transformation.
Tanaka and Uryu \cite{TU77} have derived the four-spin interactions
based on the Heitler-London theory by expanding the ground state energy
in terms of the overlap integrals between the orbitals of electrons
located at different lattice sites.
Detailed calculations of bilinear and biquadratic exchange interactions
within a real-space tight-binding framework have been performed for FM Fe by
Spisak and Hafner \cite{SH97} who demonstrate a significant contribution
of the biquadratic exchange interactions to the Curie temperature.

During the last decade the interest in skyrmions grew rapidly because  
their specific magnetic texture stabilized by chiral spin
interactions makes them attractive for various spintronic applications
(see e.g. \cite{MMR+16a,KWR+16, DBB+16}).
Most investigations in the field were restricted to the bilinear
Dzyaloshinskii-Moriya interaction (DMI) and focused on 
materials for which a strong DMI can be expected \cite{SPR+14,PMB+14,DBB+16}.

The DMI is caused by spin-orbit coupling (SOC)
and is non-zero in non-centrosymmetric systems only. Competing with
isotropic FM or AFM interactions it leads to a deviation from the
collinear magnetic state by creating a helimagnetic structure in the
absence of an external magnetic field, characterized by  
a non-zero vector spin chirality $\vec{\chi}_{ij} = \hat{s}_i \times  \hat{s}_j$ .

Recently, first-principles investigations have been performed going
beyond the bilinear approximation,    
taking into account higher-order chiral interactions
\cite{BSL19,LRP+19a} in the extended Heisenberg model.
The calculation of the chiral biquadratic DMI-like interaction (BDMI) for deposited dimers
\cite{BSL19} has demonstrated that its magnitude can be comparable to that
of the conventional bilinear Dzyaloshinskii-Moriya interaction, implying the
non-negligible role of biquadratic contributions.
In addition, the first-principles investigations on the magnetic properties of Fe
monatomic chains on a Re(0001) substrate have 
shown  \cite{LRP+19a} that chiral four-spin interactions can be responsible for the
opposite chirality of the spin spirals when compared to that determined
by DMI.

Another type of chiral interaction, the three-spin chiral interaction
(TCI) term, was discussed as a possible source for the formation
of chiral magnetic phases \cite{PP04,BCK+14}.
This three-spin interaction term in the
Heisenberg Hamiltonian, $\sum_{i,j,k} J_{ijk} \hat{s}_i\cdot (\hat{s}_j
\times \hat{s}_k)$, gives a non-zero contribution only for a
non-coplanar magnetic structure, 
i.e. in the case of non-zero scalar chirality, defined as a
counter-clock-wise triple scalar product $\chi_{ijk} =
\hat{s}_k\cdot (\hat{s}_i \times \hat{s}_j)$.
This can lead to the transition to a chiral spin liquid state, for which 
the time-reversal symmetry is broken spontaneously by the appearance of
long-range order of scalar chirality even in the absence of long-range
magnetic order or an external magnetic field \cite{WWZ89,Rok90,FFR91,SC95}. 
 Describing a transition in a frustrated quantum
  spin system from a spin liquid to a chiral spin liquid state
  within the framework of the Hubbard model, it was shown that expanding
 the Hubbard Hamiltonian in powers of $t/U$ leads to a third-order term
  which is proportional to the 
flux $\Phi_{ijk}$  enclosed by the three-spin loop
\cite{KB11,LT17},
giving rise to the three-spin interaction  \cite{KB11,LT17}  
entering the spin Hamiltonian
represented by 
 $J_{ijk} = (24/U^2)|t_{ij}||t_{jk}||t_{ki}|\sin(\Phi_{ijk}/ \Phi_0)$ 
 (with $\Phi_0 = \hbar c/e$) 
\cite{Rok90,SC95,SPD04,BCK+14}.
Note that the phase $\Phi_{ijk}/\Phi_0$ is generated by the external
magnetic field breaking time-reversal symmetry in the system.
In the presence of an inhomogeneous magnetic texture in the system, the
finite geometric quantum phase of the electron wave function can appear
due to the $s-d$-interaction, which can be described in terms of the
emergent effective electromagnetic potential leading to an effective
magnetic field $B^{eff} \sim \chi_{ijk}$\cite{FJTM11,TF14} giving rise to the three-spin energy contribution $\sim \chi_{ijk}^2$, that can be
associated with the topology-induced three-spin chiral-chiral exchange
interaction \cite{GHH+20}. These interactions have been introduced and
evaluated on the basis of first-principles electronic structure
calculations for the B20-type compounds MnGe and FeGe. 
The authors report also
about another type of three-spin interaction having topological origin, so
called spin-chiral interactions, which are however non-zero only if
spin-orbit interaction is taken into account.

Discussing skyrmion-hosting materials, the formation of skyrmion magnetic
texture is usually ascribed to the DMI, implying the lack of the
inversion symmetry in these systems.
However, recently it was suggested that the magnetic frustration could
stabilize skyrmions even in materials  with centrosymmetric lattices.
This idea was proposed and discussed by
various authors within theoretical investigations \cite{OCK12,BLHK16,HOM17}. 
In these works complex superstructures or the
skyrmion-lattice state are characterized by multiple ordering wave
vectors (multiple-Q), allowing to characterize a non-coplanar
magnetic structure via a double-Q description. This approach
applied to metallic systems allowed to demonstrate that the
non-coplanar vortex state can be stabilized having lower energy than the
helimagnetic structure expected due to RKKY interactions
\cite{SMM12}. Solenov et al. \cite{SMM12} showed that such a non-coplanar
state can be stabilized even in the absence of SOC, i.e. without
the DMI. The authors attribute this feature of a double-Q state to
the chirality-induced emergent magnetic field associated with a persistent
electric current in such systems (see e.g.\ \cite{BD05,LFBM18a,Tat18}), which is
proportional to the scalar chirality in the system.
In terms of the extended spin Hamiltonian, the above mentioned property
can be attributed to the three-spin interaction term also proportional
to the scalar chirality in the system.

In this contribution we present a coherent computational scheme that allows to
calculate the parameters of the extended Heisenberg Hamiltonian to any
order. The impact of higher order terms going beyond the bilinear level
and their anisotropy is discussed on the basis of corresponding
numerical results for various systems.

\section{Electronic structure}

Following our previous work \cite{MPE19}, we consider the change of the
grand canonical potential caused by the formation of a modulated spin
structure seen as a perturbation.  
This quantity is represented in terms of the Green function $G_0(E)$ for 
the FM reference state and its modification due to the perturbation. Neglecting all
temperature effects, and denoting the corresponding 
change in the Green function $\Delta G(E)$ one can write for the change in energy:  
\begin{eqnarray}
\Delta {\cal E} & \approx &  -\frac{1}{\pi} 
\mbox{Im}\, \mbox{Tr}  \int^{E_F} dE \,
(E - E_F) \, \Delta G(E) \;,
\label{Eq_DeltaF_1}
\end{eqnarray}
with the expansion 
\begin{eqnarray}
\Delta G(E) & = & G_0 \Delta V G_0  \nonumber \\
&&+  G_0 \Delta V G_0 \Delta V G_0 \nonumber \\
&&+ G_0 \Delta V G_0 \Delta V G_0 \Delta V G_0 \nonumber \\
&& +  G_0\Delta V G_0\Delta V G_0 \Delta V G_0 \Delta V G_0  + ... \;,
\label{Eq_GF_expansion}
\end{eqnarray}
for $\Delta G(E)$, where $\Delta V$ is the perturbation operator
associated with the modulated spin
structure. For the sake of readability we dropped the energy argument
for the unperturbed Green function $G_0(E)$.

Using the FM state as a reference state,   
the perturbation connected with the tilting of rigid magnetic moments
on lattice sites $i$ has the real space representation \cite{ME17,MPE19}
\begin{eqnarray}
 \Delta V(\vec{r}) &=&  \sum_i \beta \big( \vec{\sigma}\cdot\hat{s}_i
  -  \sigma_z\big) B_{xc}(\vec{r}) \;,
\label{Eq_perturb}
\end{eqnarray}
where ${\vec B}_{xc}(\vec r)$ is the spin-dependent part of the 
exchange-correlation potential,
 $\vec{\sigma}$  is the vector of $4 \times 4$ Pauli matrices and 
$\beta$ is one of the  standard Dirac matrices \cite{Ros61,EBKM16}. 
It is assumed here 
that  ${\vec B}_{xc}(\vec r)$ on site $i$ is aligned 
along the orientation of the spin moment $\hat{s}_i$, 
i.e.\  $\vec{B}_{xc}(\vec{r}) = {B}_{xc}(\vec{r}) \hat{s}_i$.

A very convenient and flexible way to represent the electronic
Green function $G_0(E)$ in Eqs.\
(\ref{Eq_DeltaF_1})-(\ref{Eq_GF_expansion}) is provided by the so-called
KKR (Korringa-Kohn-Rostoker) or multiple-scattering formalism.
Adopting this approach a real space expression for
$G_0(\vec{r},\vec{r}\,',E)$ can be written in a fully relativistic way as  \cite{EBKM16}:
\begin{eqnarray}
G_0(\vec{r},\vec{r}\,',E) & = &
\sum_{\Lambda_1\Lambda_2} 
Z^{n}_{\Lambda_1}(\vec{r},E)
                              {\tau}^{n n'}_{\Lambda_1\Lambda_2}(E)
Z^{n' \times}_{\Lambda_2}(\vec{r}\,',E)
 \nonumber \\
 & & 
-  \sum_{\Lambda_1} \Big[ 
Z^{n}_{\Lambda_1}(\vec{r},E) J^{n \times}_{\Lambda_1}(\vec{r}\,',E)
\Theta(r'-r)  \nonumber 
\\
 & & \qquad\quad 
+ J^{n}_{\Lambda_1}(\vec{r},E) Z^{n \times}_{\Lambda_1}(\vec{r}\,',E) \Theta(r-r')
\Big] \delta_{nn'} \; .
\label{Eq_KKR-GF}
\end{eqnarray}
Here 
$Z^{n}_{\Lambda_1}(\vec{r},E)$ and 
$J^{n}_{\Lambda_1}(\vec{r},E)$ 
are
the regular and irregular  solutions of the
single site Dirac equation and
  ${\underline{\tau}}^{n n'}$ is the 
  so-called scattering path operator matrix \cite{EBKM16}.
  Substituting the expression in  Eq.\ (\ref{Eq_KKR-GF}) into Eq.\
  (\ref{Eq_GF_expansion}) and using  Eq.\ (\ref{Eq_DeltaF_1}) one
  obtains in a straight and natural way a real space expression for the 
 energy change $\Delta \cal E$, which will
 be used below to derive expressions for the exchange coupling
 parameters entering the extended Heisenberg Hamiltonian.

The dependence of the energy on the 
magnetic configuration calculated from first principles
via Eqs.\ (\ref{Eq_DeltaF_1})  to (\ref{Eq_perturb})
will be mapped  onto the
extended Heisenberg Hamiltonian 
\begin{eqnarray}
  H &=&  - \sum_{i,j}  J^{s}_{ij} (\hat{s}_i \cdot \hat{s}_j)
 - \sum_{i,j}  \vec{{D}}_{ij} \cdot (\hat{s}_i \times
      \hat{s}_j) \nonumber \\ 
 &&    
 - \frac{1}{3!}\sum_{i,j,k}^N
                   J_{ijk} \hat{s}_i\cdot (\hat{s}_j \times \hat{s}_k) \; ,  \nonumber \\          
 &&      -  \frac{2}{p!}\sum_{i,j,k,l}  J^{s}_{ijkl} (\hat{s}_i \cdot
        \hat{s}_j)(\hat{s}_k \cdot \hat{s}_l)  \nonumber \\  
 &&  
 - \frac{2}{p!} \sum_{i,j,k,l}  \vec{{\cal D}}_{ijkl} \cdot (\hat{s}_i \times
      \hat{s}_j) (\hat{s}_k \cdot \hat{s}_l)  \;.
\label{Eq_Heisenberg_general}
\end{eqnarray}
Here
 the prefactors  of the various sums account 
 for multiple counting contributions
occurring upon summation over all lattice sites, where $p$ specifies 
the number of interacting atoms; 
i.e.\ $p = 2$, $p = 3$ and $p = 4$ correspond
to the biquadratic, three-spin and four-spin interactions,
respectively. The prefactor $\frac{2}{p!}$ occurs due 
to the chosen normalization of
the exchange parameters, that leads to the same prefactor for the
biquadratic term in the Hamiltonian as in the case of the bilinear term.
Note that we follow the more common convention \cite{LKG84} for the
bilinear exchange interaction parameters, while also other conventions
are used in the literature \cite{USPW03}, as was pointed out
previously \cite{PMB+14}. 
However, it should be stressed 
that for the sake of simplicity working out 
the  expressions for the exchange parameters
in the following, the prefactors are not
taken into account, as they appear coherently for the model as well as 
for the  first-principles representations of the energy
 and cancel each other in the final
expression for the exchange parameters.

\section{Four-spin exchange interactions}

Extending the spin Hamiltonian to go beyond the classical Heisenberg model,
we discuss first the four-spin exchange interaction terms  $J_{ijkl}$
and  $\vec{\cal D}_{ijkl}$.
The isotropic exchange as well as the z-component of the DMI-like four-spin
interactions can be given in terms of the fourth-rank tensor $J^{\alpha
  \beta  \gamma \delta}_{ijkl}$ which accounts also for pair ($k=i,\, l=j$, so-called 
biquadratic) and three-spin ($l=j$) interactions.  The tensor elements
$J^{\alpha \beta  \gamma \delta}_{ijkl}$ can be calculated using the
fourth-order term of the Green function expansion in Eq.\ (\ref{Eq_GF_expansion}).
Substituting this expression into Eq.\
(\ref{Eq_DeltaF_1}) and using the sum rule $\frac{dG}{dE} = - GG$ for the
Green function, one obtains after integration by parts the forth-order
term of the total energy change $\Delta {\cal E}^{(4)}$ given by:
\begin{eqnarray}
\Delta {\cal E}^{(4)} &=& - \frac{1}{\pi} \mbox{Im}\,\mbox{Tr} \int^{E_F}
                    dE\, \nonumber \\
  && \times \Delta V
   G_0  \Delta V G_0 \Delta V G_0 \Delta V G_0 \;.
\label{Eq_Free_Energy-4}
\end{eqnarray}
Using the ferromagnetic state with $\vec{M}||\hat{z}$ as a reference
state, and considering the spin-spiral $\vec{s}_i = (\sin \theta \cos(\vec{q} \cdot
\vec{R}_i), \sin\theta \sin(\vec{q} \cdot \vec{R}_i), \cos \theta)$ as
the source for the perturbation $\Delta V$ at small $\vec{q}$ values,
only the $x$ and $y$ components of the exchange tensor get involved (see
also Ref. \cite{MPE19}). Following the 
scheme used to derive an expression for the bilinear exchange
interactions \cite{MPE19}, the fourth-order derivative 
with respect to the $\vec{q}$-vector gives the
elements of the exchange tensor represented via the expression (see
Appendix B) 
\begin{eqnarray}
  J^{\alpha\beta\gamma\delta}_{ijkl}  &=&  \frac {1}{2\pi} \mbox{Im}\,\mbox{Tr}  \int^{E_F} dE \, 
               \nonumber \\       
&& \Big[
\underline{T}^{i, \alpha}(E)\, \underline{\tau}^{ij}(E)
   \underline{T}^{j, \beta}(E)\, \underline{\tau}^{jk}(E)    \nonumber    \\
       && \times  \underline{T}^{k, \gamma}(E)\, \underline{\tau}^{kl}(E)
\underline{T}^{l, \delta}(E)\, \underline{\tau}^{li}(E)
 \Big] \; ,
\label{Eq:J_XYZL} 
\end{eqnarray}
where the matrix elements of the torque operator $T^{i,\alpha}_{\Lambda\Lambda'}$
are defined as follows:\cite{EM09a}
%
\begin{eqnarray}
 T^{i,\alpha}_{\Lambda\Lambda'} & = & \int_{\Omega_i} d^3r  \, Z^{i \times}_{\Lambda}(\vec{r},E)\, \Big[\beta \sigma_{\alpha} B_{xc}^i(\vec{r})\Big] \, Z^{i}_{\Lambda'}(\vec{r},E)\;.  \label{Eq:ME}
\end{eqnarray}

\subsection{Non-chiral exchange interactions}

The four-spin scalar interaction, and as 
special cases, also the fourth-order three-spin term with $l = j$,
and the biquadratic exchange interaction term with $k=i$ and $l=j$, 
can also be written in a form often used in the literature,
i.e. they can be represented in terms of scalar products of spin directions: 
\begin{eqnarray}
  H^{(4)}_s &=& - \sum_{i,j,k,l}  J^{s}_{ijkl} (\hat{s}_i \cdot
                         \hat{s}_j)(\hat{s}_k \cdot \hat{s}_l) \;.
\label{Eq_Heisenberg_four-spin_2}
\end{eqnarray}
The parameters $J^{s}_{ijkl}$ (where $s$ means 'symmetric') are
represented by the symmetric part of the exchange tensor of 4-th rank in
Eq.\ (\ref{Eq:J_XYZL}) and are given by the expression (see Appendix B)
\begin{eqnarray}
 J^{s}_{ijkl}  &=& \frac{1}{4}( J_{ijkl}^{xxxx} +
                     J_{ijkl}^{yyyy} +  J_{ijkl}^{xxyy} +
                     J_{ijkl}^{yyxx} )\;.
\label{Eq_Heisenberg_4spin-scalar}
\end{eqnarray}
 The expression for a three-spin or a biquadratic configuration should
  have a form which can be seen as being composed of two closed loops, e.g.
\begin{eqnarray}
  J^{\alpha\beta\gamma\delta}_{ijil}  &=&  \frac {1}{2\pi} \mbox{Im}\,\mbox{Tr}  \int^{E_F} dE \, 
               \nonumber \\       
&& \Big(
\underline{T}^{i, \alpha}(E)\, \underline{\tau}^{ij}(E)
   \underline{T}^{j, \beta}(E)\, \underline{\tau}^{ji}(E)  \Big)    \nonumber    \\
       && \times  \Big( \underline{T}^{i, \gamma}(E)\, \underline{\tau}^{il}(E)
\underline{T}^{l, \delta}(E)\, \underline{\tau}^{li}(E)
 \Big) \; ,
\label{Eq:J_XYZL} 
\end{eqnarray}
  such that each can be associated with the pair interaction of atoms
  $i-j$ and $i-l$ and has a form similar to the one appearing in the case of
  bilinear interactions. 
  This form has a momentum 
  representation determined by the expression $
  (\underline{\tau}_{\vec{k}}\,\underline{\tau}_{\vec{k}\pm\vec{q}})\,
  (\underline{\tau}_{\vec{k}'}\,\underline{\tau}_{\vec{k}'\pm\vec{q}})$,
  with $\underline{\tau}_{\vec{k}}$  standing for the $\vec{k}$-dependent
  scattering path matrices,
  which should ensure the $q^2$ dependence of each expression associated
  with the scalar product of spin moments. This form of the expression
  implies also that only the site-off-diagonal 
  terms of Green function are involved in its calculation.
  
  Considering in addition the case $l = j$ one arrives at an
  expression for the biquadratic interactions 
\begin{eqnarray}
  J^{\alpha\beta\gamma\delta}_{ijij}  &=&  \frac {1}{2\pi} \mbox{Im}\,\mbox{Tr}  \int^{E_F} dE \, 
               \nonumber \\       
&& \Big(
\underline{T}^{i, \alpha}(E)\, \underline{\tau}^{ij}(E)
   \underline{T}^{j, \beta}(E)\, \underline{\tau}^{ji}(E)  \Big)    \nonumber    \\
       && \times  \Big( \underline{T}^{i, \gamma}(E)\, \underline{\tau}^{ij}(E)
\underline{T}^{j, \delta}(E)\, \underline{\tau}^{ji}(E)
 \Big) \; ,
\label{Eq:J_XYZL} 
\end{eqnarray}
for which the symmetry with respect to permutation of two spin moments
is just a consequence of the invariance of the trace of a product
of matrices with respect to cyclic 
permutation of the matrices.
The biquadratic exchange interaction terms (with $k=i$ and $l=j$) can be
seen as a linear term of an expansion of the bilinear  
exchange parameters in  powers of $(\hat{s}_i \cdot \hat{s}_j)$, in
order to take into account the dependence of these exchange parameters on
the relative orientation of the interacting spin magnetic moments on sites
$i$ and $j$. Focusing first on the scalar-interaction terms, this leads
to the expression 
\begin{eqnarray}
  {H}^s &=&  - \sum_{i,j}
                    \tilde{J}_{ij}(\theta_{ij}) (\hat{s}_i
                    \cdot \hat{s}_j) \nonumber \\
 &=&  - \sum_{i,j}
                    J_{ij} (\hat{s}_i
                    \cdot \hat{s}_j) - \sum_{i,j}  J^{s}_{ijij} (\hat{s}_i \cdot \hat{s}_j)(\hat{s}_i \cdot \hat{s}_j).
\label{Eq:J_Biguad} 
\end{eqnarray}
Note that both, bilinear and biquadratic terms  in
 Eq.\ (\ref{Eq:J_Biguad}) have two contributions from the same
 pair of atoms, i.e. $i = a, \,j = b$ and  $i = b, \,j = a$, with the
 interactions $J_{ab}= J_{ba}$ and $J_{abab}= J_{baba}$, 
 as it has been  mentioned above.

\subsection{Chiral multispin DMI-like exchange interactions}

Discussing chiral interactions, we start with the exchange
interactions represented by the vector characterizing 
the DMI-like interaction between two spin moments, $i$ and $j$, but
taking into account the magnetic configuration of surrounding atoms,
leading to the extension of the Heisenberg Hamiltonian written in the
following form 
\begin{eqnarray}
  {H}^{a(4)} &=&   \sum_{i,j,k,l}  \vec{{\cal D}}_{ijkl} \cdot (\hat{s}_i \times
      \hat{s}_j) (\hat{s}_k \cdot \hat{s}_l) \;.  
\label{Eq_Heisenberg_H4_DMIXY}
\end{eqnarray}
%
Assuming the magnetization direction of the reference system along the
$z$ axis, we distinguish between 
the $x$ and $y$ components of this chiral interaction on the one hand side, and its $z$
component on the other hand as they require different approaches for
their calculation. This is in full analogy to the DMI discussed recently \cite{ME17,
  MPE19}.

\subsubsection{DMI-like interactions: z-component} 

The z-component of the four-spin chiral interaction $\vec{{\cal
    D}}_{ijkl}$, when all site indices $i, j, k, l$ may be different, 
is represented by the antisymmetric part of the exchange tensor
characterizing the interaction between sites $i, j, k$ and  $j$. In full
analogy to the DMI, ${\cal  D}_{ijkj}^z$ can be written as follows
\begin{eqnarray}
{\cal D}^{z}_{ijkl}= \frac{1}{4}(J^{xyxx}_{ijkl} + J^{xyyy}_{ijkl} -
  J^{yxxx}_{ijkl} -  J^{yxyy}_{ijkl}) 
\label{Eq_D4_antisym}
\end{eqnarray}
with the tensor elements $J^{\alpha \beta \gamma \delta}_{ijkj}$
determined via Eq.\ (\ref{Eq:J_XYZL}).

In the following, we will focus on the three-spin DMI-like interactions TDMI (implying
$l = j$) and biquadratic vector interactions (with $l = j, k = i$),
which were calculated and discussed recently for some systems with
special geometry \cite{BSL19,LRP+19a} in comparison with the DMI.
Using Eq.\ (\ref{Eq_D4_antisym}) for the special case $l = j, k = i$ one
has for the $z$ component of the biquadratic interaction the
expression (see Appendix B): 
\begin{eqnarray}
{\cal D}^{z}_{ijij}= \frac{1}{4}(J^{xyxx}_{ijij} + J^{xyyy}_{ijij} -
  J^{yxxx}_{ijij} -  J^{yxyy}_{ijij})  \;.
\label{Eq_Heisenberg_D4}
\end{eqnarray}

\subsubsection{DMI-like interactions: x- and y-compoment}

To calculate the $x$ and $y$ components of the four-spin and as a
special case the TDMI and BDMI terms in a system magnetized along the $z$ direction,
we follow the scheme suggested by the authors for the calculation of the DMI parameters
\cite{ME17,  MPE19}, which exploited the DMI-governed behaviour of the
spin-wave dispersion having a finite slope at the $\Gamma$ point of the
Brillouin zone. However, in the present case a more general form of perturbation
is required, 
that allows for of the spin moments 
entering the scalar product in the four-spin energy terms
for  a  tilting towards the $x$ and $y$ axes. 
  For this purpose we 
  assume a 2D spin modulation according to the expression
\begin{eqnarray}
  \hat{s}_i &=&
 (\mbox{sin}(\vec{q}_1 \cdot \vec{R}_i) \;\mbox{cos}(\vec{q}_2 \cdot \vec{R}_i)\;,
                \mbox{sin}(\vec{q}_2 \cdot \vec{R}_i) \;, \nonumber \\
&&  \mbox{cos}(\vec{q}_1 \cdot
 \vec{R}_i)\mbox{cos}(\vec{q}_2 \cdot \vec{R}_i) ) \;,
\label{spiral2}
\end {eqnarray}
which is characterized by two wave vectors, $\vec{q}_1$ and $\vec{q}_2$,
orthogonal to each other, as for example $\vec{q}_1 = q_1\hat{y}$ and
$\vec{q}_2 = q_2\hat{x}$. 
The microscopic expression for the $x$ and $y$ components of
$\vec{\cal D}_{ijkl}$ describing the most general, four-spin interaction, can
be obtained on the basis of the third-order term of the Green function  expansion in
Eq.\ (\ref{Eq_GF_expansion}) leading to a corresponding third-order energy
contribution
\begin{eqnarray}
\Delta {\cal E}^{(3)} &=& -\frac{1}{\pi} \mbox{Im}\,\mbox{Tr} \int^{E_F}
                    dE (E - E_F)\, \nonumber \\
  && \times G_0  \Delta V G_0 \Delta V G_0 \Delta V G_0 \; .
\label{Eq_Free_Energy-3}
\end{eqnarray}
%
This is achieved by taking the first  derivative with respect to the
  wave-vectors $\vec{q}_{1(2)}$,
   the second-order  derivative
   with respect to the wave-vector  $\vec{q}_{2}$ 
   and considering finally the limit $q_{1(2)} \to 0$.
The components ${\cal D}^{y}_{ijkj}$ and  ${\cal D}^{x}_{ijkj}$
are determined this way by the first-order derivative with respect to the
  wave-vector $\vec{q}_1$ and  $\vec{q}_2$, respectively.
The non-zero elements of the 
first-order derivative in the limit $q_{1(2)} \to 0$ imply an
antisymmetric character of the interactions between the magnetic moments
on sites $i$ and $j$ in Eq.\ 
(\ref{Eq_Heisenberg_H4_DMIXY}), similar to the case of the conventional DMI.
At the same time, the non-zero second-order derivative with respect to 
$\vec{q}_{2}$ correspond to a scalar interaction between the magnetic
moments on sites $k$ and $l$, which 
is symmetric with respect to a sign change of the wave vector. 
The same properties should apply to the corresponding 
contribution to the model spin Hamiltonian in Eq.\
(\ref{Eq_Heisenberg_H4_DMIXY}). Equating for the ab-initio and model
approaches the corresponding terms 
proportional to $({\vec R}_i - {\vec R}_j)_y ({\vec R}_k - {\vec
  R}_l)^2_x$ and $({\vec R}_i - {\vec R}_j)_x ({\vec R}_k - {\vec
  R}_l)^2_x$ (we keep a similar form in both cases for the sake of
convenience) gives access to the elements ${\cal D}^{y,x}_{ijkl}$ and
${\cal D}^{y,y}_{ijkl}$, as well as ${\cal D}^{x,x}_{ijkl}$ and
${\cal D}^{x,y}_{ijkl}$, respectively, of the four-spin chiral interaction. 
As we focus here on TDMI and BDMI, they can be obtained as the special
cases $l = j$ and $l = j, k = i$, respectively.
With this, the elements of the TDMI vector can be written as follows
\begin{eqnarray}
  {\cal D}^{\alpha,\beta}_{ijkj}  &=&   \epsilon_{\alpha\gamma} \frac {1}{8\pi}\mbox{Im}\, \mbox{Tr} \int^{E_F} dE (E - E_F)\, 
               \nonumber \\       
                     &&\Big[ \underline{O}^{i}\, \underline{\tau}^{ij}
                        \underline{T}^{j, \gamma}\,  \underline{\tau}^{jk}
                        \underline{T}^{k, \beta}\, \underline{\tau}^{kj}
                  \underline{T}^{j, \beta}\, \underline{\tau}^{ji}
                        \;\nonumber \\
                     && - \underline{T}^{i, \gamma}\, \underline{\tau}^{ij}
                        \underline{O}^{j}  \underline{\tau}^{jk}
                        \underline{T}^{k, \beta}\, \underline{\tau}^{kj}
                  \underline{T}^{j, \beta}\,
                        \underline{\tau}^{ji}\Big] \;          \,\nonumber\\
                     && + \Big[ \underline{O}^{i}\, \underline{\tau}^{ij}
                        \underline{T}^{j, \beta}\,  \underline{\tau}^{jk}
                        \underline{T}^{k, \beta}\, \underline{\tau}^{kj}
                  \underline{T}^{j, \gamma}\, \underline{\tau}^{ji}
                        \;\nonumber \\
                     && -   \underline{T}^{i, \gamma}\, \underline{\tau}^{ij}
                       \underline{T}^{j, \beta}  \underline{\tau}^{jk}
                        \underline{T}^{k, \beta}\, \underline{\tau}^{kj}
                \underline{O}^{j} \,  \underline{\tau}^{ji}\Big] \;
\label{Eq:D-4-4_XYZ} 
\end{eqnarray}
with $\alpha, \beta = x,y$, and $\epsilon_{\alpha\gamma}$ the elements
of the transverse Levi-Civita tensor $ \underline{\epsilon}
= \begin{bmatrix}  0 & 1 \\  -1 & 0   \end{bmatrix} $.
The matrix elements of the torque operator $T^{i,\alpha}_{\Lambda\Lambda'}$
occurring in Eq.\ (\ref{Eq:D-4-4_XYZ}) are given by Eq.\ (\ref{Eq:ME}),
and the overlap integrals $O^{j}_{\Lambda\Lambda'}$  are defined in an analogous way \cite{EM09a}:
%
\begin{eqnarray}
 O^{j}_{\Lambda\Lambda'} & = & \int_{\Omega_j} d^3r  \,
 Z^{j \times}_{\Lambda}(\vec{r},E) \, Z^{j}_{\Lambda'}(\vec{r},E).  \label{Eq:ME1}
\end{eqnarray}

The expression in  Eq.\ (\ref{Eq:D-4-4_XYZ}) gives access to the $x$ and $y$
components of the DMI-like three-spin interactions in Eq.\
(\ref{Eq_Heisenberg_H4_DMIXY}) 
\begin{eqnarray}
 {\cal D}^{\alpha}_{ijkj}= {\cal D}^{\alpha,x}_{ijkj} +
 {\cal D}^{\alpha,y}_{ijkj} \;.
\label{Eq_Heisenberg_D4}
\end{eqnarray}

An expression for the BDMI also follows
directly from Eq.\ (\ref{Eq:D-4-4_XYZ}) using the restriction $k = i$.
This leads to the elements ${\cal D}^{\alpha,\beta}_{ijij}$  determining
chiral biquadratic exchange interactions (similar to the case of
four-spin interactions), which can be written in the following form  
\begin{eqnarray}
  {\cal D}^{\alpha,\beta}_{ijij}  &=&   \epsilon_{\alpha\gamma} \frac {1}{8\pi} \mbox{Im}\, \mbox{Tr} \int^{E_F} dE (E - E_F)\, 
               \nonumber \\       
                     &&\Big[ \Big(\underline{O}^{i}\, \underline{\tau}^{ij}
                        \underline{T}^{j, \gamma}\,  \underline{\tau}^{ji}
                        \underline{T}^{i, \beta}\, \underline{\tau}^{ij}
                  \underline{T}^{j, \beta}\, \underline{\tau}^{ji}
                        \;\nonumber \\
                     && - \underline{T}^{i, \gamma}\, \underline{\tau}^{ij}
                        \underline{O}^{j}  \underline{\tau}^{ji}
                        \underline{T}^{i, \beta}\, \underline{\tau}^{ij}
                  \underline{T}^{j, \beta}\,
                        \underline{\tau}^{ji} \Big)
               \nonumber \\       
                     &&+ \Big( \underline{O}^{i}\, \underline{\tau}^{ij}
                        \underline{T}^{j, \beta}\,  \underline{\tau}^{ji}
                        \underline{T}^{i, \beta}\, \underline{\tau}^{ij}
                  \underline{T}^{j, \gamma}\, \underline{\tau}^{ji}
                        \;\nonumber \\
                     && - \underline{T}^{i, \gamma}\, \underline{\tau}^{ij}
                        \underline{T}^{j, \beta}  \underline{\tau}^{ji}
                        \underline{T}^{i, \beta}\, \underline{\tau}^{ij}
                  \underline{O}^{j}\,
                        \underline{\tau}^{ji}  \Big) \Big] \;.
\label{Eq:D-4-2_XYZ} 
\end{eqnarray}

\subsection{Chiral exchange: three-spin exchange interactions}

Here we discuss the three-spin chiral exchange interaction entering
a corresponding extension term to the Heisenberg Hamiltonian 
\begin{eqnarray}
  H^{(3)} &=&  - \sum_{i \neq j\neq k}^N
                   J_{ijk} \hat{s}_i\cdot (\hat{s}_j \times \hat{s}_k) \; .
\label{Eq_Heisenberg_3-spin}
\end{eqnarray}
As it follows from this expression, the contribution due to the
three-spin interaction is non-zero only in case of a non-co-planar and non-collinear
magnetic structure characterized by the triple  product
$\hat{s}_i\cdot (\hat{s}_j \times \hat{s}_k)$
involving the spin moments on three different lattice sites.

Considering the torque acting in a FM system
on the magnetic moment of any atom $i$, which is
associated with the three-spin interactions, one can evaluate its
projection onto an arbitrary direction  $\hat{u}$, $T^{\rm
  (3)}_{ijk,\hat{u}}= -(\partial H^{(3)}/\partial
\hat{s}_i) \cdot (\hat{u} \times \hat{s}_i)$ , which is equal to  $J_{ijk}
(\hat{e}_{k}\cdot \hat u)$. This value is non-zero only in the case of
a non-zero scalar product $(\hat{s}_{k}\cdot \hat u)$, implying that
a non-vanishing torque on spin $\hat{s}_i$ created by the spin $\hat{s}_j$
coupled via the three-spin interaction, requires a tilting
of the third spin moment $\hat{e}_k$ to have a non-zero projection on
the torque direction. In contrast to that, the torque $T^{\rm
  DM}_{ij,\hat{u}} = \vec{D}_{ij}\cdot \hat u$ \cite{MBM+09} acting
due to the spin of atom $j$ on the spin moment of atom $i$ via the DMI,
is non-vanishing 
even in the system with all spin moments being collinear.
This makes clear that in order to work out the expression for the $
J_{ijk}$ interaction term, a more complicated multi-Q modulation
\cite{SMM12,OCK12,BLHK16} of the magnetic 
structure is required when compared to  a helimagnetic structure
characterized by a wave vector $\vec{q}$, which was used to
derive expressions for the $x-$ and $y-$components of the DMI
\cite{ME17,MPE19}.
 In addition, 
similarly to the DMI that gives a non-zero contribution to
  the energy due to its anti-symmetry  with respect
  to permutation, the energy due to the TCI, for a fixed spin
  configuration of all three atoms involved, 
  is non-zero only if   $J_{ijk}\neq  J_{ikj}$, etc. 
  Otherwise, the terms $ijk$ and $ikj$
  cancel each other due to the relation 
  $\hat{s}_i\cdot (\hat{s}_j
 \times \hat{s}_k) = - \hat{s}_i\cdot (\hat{s}_k \times \hat{s}_j)$.

Thus, to derive an expression for the TCI,
we use the 2D non-collinear spin texture described by Eq.\ (\ref{spiral2}),
which is characterized by two wave 
vectors oriented along two mutually perpendicular directions, 
as for example  $\vec{q}_1 = (0,q_y,0)$
and $\vec{q}_2 = (q_x,0,0)$
 (for more details see Appendix B).

In this case the spin chirality driven by the
three-spin interaction should lead to the asymmetry of the energy
$E(\vec{q}_1, \vec{q}_2)$ with respect to a sign change of any of the vectors
$\vec{q}_1$ and $\vec{q}_2$, as a consequence of full antisymmetry
of the scalar spin chirality.  
As a result, the three-spin interactions can be derived assuming a
non-zero slope of the energy dispersion $E(\vec{q}_1, \vec{q}_2)$ as
function of the two wave vectors, in the limit $\vec{q}_{1(2)} = 0 $.

Substituting the spin modulation in Eq.\ (\ref{spiral2}) into the spin
Hamiltonian in Eq.\ (\ref{Eq_Heisenberg_3-spin}) 
associated with the three-spin interaction, the second-order
derivative of the energy $E^{(3)}(\vec{q}_1, \vec{q}_2)$ with respect to
$q_1$ and $q_2$ wave vectors in the limit $q_1 \to 0$,  $q_2 \to 0$ is
given by the expression
\begin{eqnarray}
  && \frac{\partial^2}{\partial\vec{q}_1\partial\vec{q}_2} E_{H}^{(3)} 
                                                                       \nonumber \\
  && = - \sum_{i \neq j\neq k}
J_{ijk} \big( \hat{z} \cdot [(\vec{R}_i - \vec{R}_j) \times(\vec{R}_k -
  \vec{R}_j) ] \big) \; .
\label{Eq_Heisenberg_3-spin_deriv}
\end{eqnarray}

The microscopic energy term of the electron system, giving access to the
chiral three-spin interaction in the spin Hamiltonian is determined by the
second-order term of the free energy expansion given by the expression
\begin{eqnarray}
\Delta {\cal E}^{(2)} &=& -\frac{1}{\pi} \mbox{Im}\,\mbox{Tr} \int^{E_F}
                    dE (E - E_F)\, \nonumber \\
  && G_0 \Delta V G_0 \Delta V G_0 \;.
\label{Eq_Free_Energy-3}
\end{eqnarray}
To make a connection between the two
approaches associated with the ab-initio and model spin Hamiltonians, we
consider a second-order term with respect to the perturbation 
$\Delta V$ induced by the spin modulation in Eq.\ (\ref{spiral2}).
Taking the first-order derivative with respect to $q_1$ and $q_2$ in
the limit $q_1 \to 0$,  $q_2 \to 0$, and equating the terms proportional
to $\big( \hat{z} \cdot [(\vec{R}_i - \vec{R}_j) \times(\vec{R}_k -
  \vec{R}_j) ] \big)$ with the corresponding terms in the spin Hamiltonian, one
 obtains the following expression for the three-spin interaction
\begin{eqnarray}
  J_{ijk}  &=&   \frac {1}{8\pi} \mbox{Im}\, \mbox{Tr} \int^{E_F} dE (E - E_F)\,  \nonumber \\
 &&                    
\Big[ \underline{T}^{i, x}\, \underline{\tau}^{ij}
\underline{T}^{j, y}\, \underline{\tau}^{jk}
               \underline{O}^{k}\,  \underline{\tau}^{ki} 
- \underline{T}^{i, y}\, \underline{\tau}^{ij}
 \underline{T}^{j, x}\, \underline{\tau}^{jk}
     \underline{O}^{k}\,  \underline{\tau}^{ki}
               \nonumber \\       
&&     - \underline{T}^{i, x}\, \underline{\tau}^{ij}
\underline{O}^{j}\, \underline{\tau}^{jk}
               \underline{T}^{k, y}\, \underline{\tau}^{ki}\;
+ \underline{T}^{i, y}\, \underline{\tau}^{ij}
\underline{O}^{j}\, \underline{\tau}^{jk}
               \underline{T}^{k, x}\, \underline{\tau}^{ki} \;
               \nonumber \\       
&& +
   \underline{O}^{i} \, \underline{\tau}^{ij} 
\underline{T}^{i, x}\, \underline{\tau}^{jk}
               \underline{T}^{k, y}\, \underline{\tau}^{ki}\;
 - \underline{O}^{j}\,  \underline{\tau}^{ij}
\underline{T}^{i, y}\,\underline{\tau}^{jk}
               \underline{T}^{k, x}\, \underline{\tau}^{ki} \Big] \;
\label{Eq:J_XYZ} 
\end{eqnarray}

\section{Numerical results}

In order to illustrate the expressions developed above by their application to
realistic systems, corresponding  calculations on various representative systems have
been performed. 
 Some numerical details of these calculations are
described in Appendix A.

\subsection{Four-spin and biquadratic exchange interactions}

\begin{figure}
\includegraphics[width=0.35\textwidth,angle=0,clip]{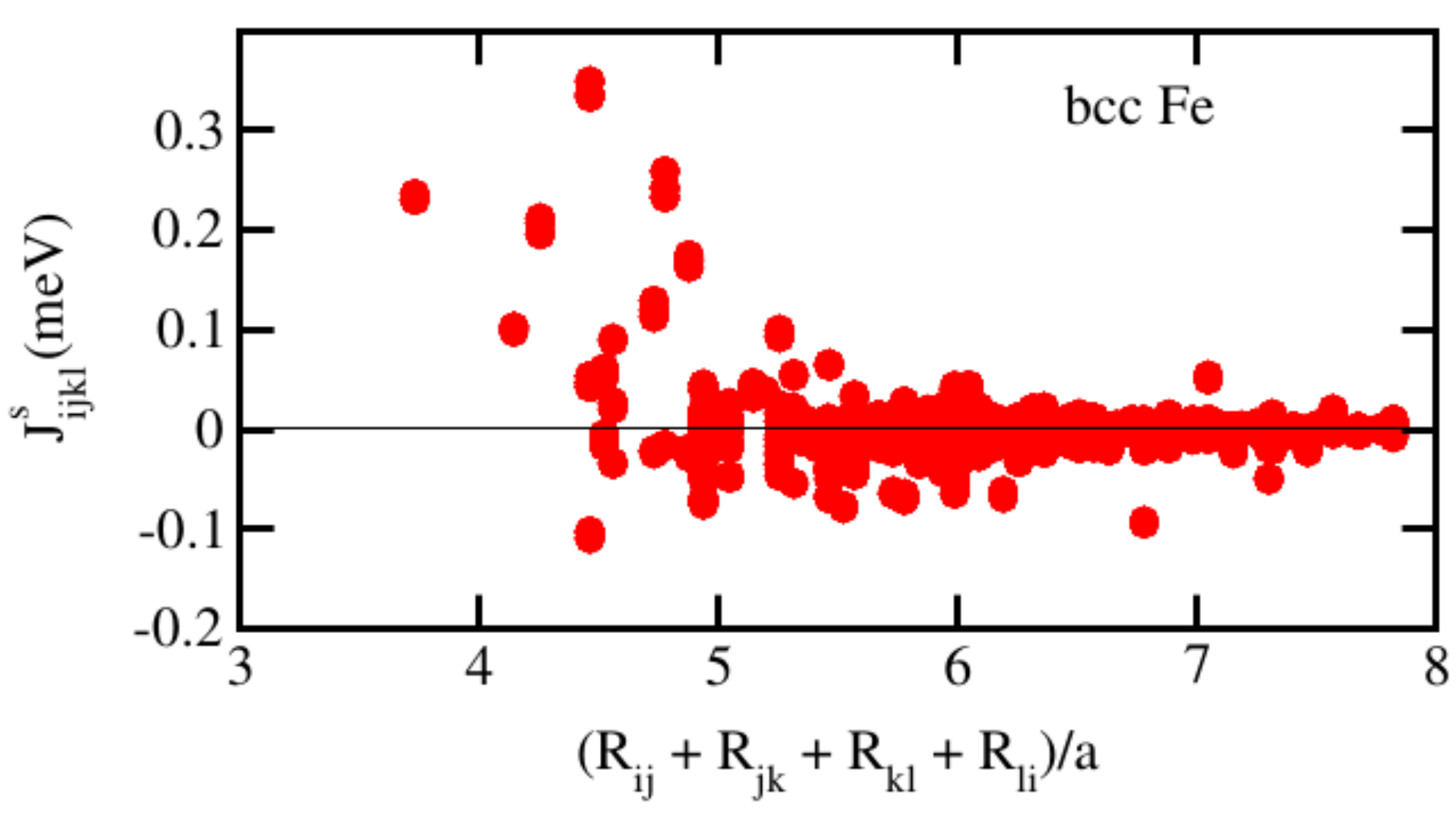}\;(a)
\includegraphics[width=0.35\textwidth,angle=0,clip]{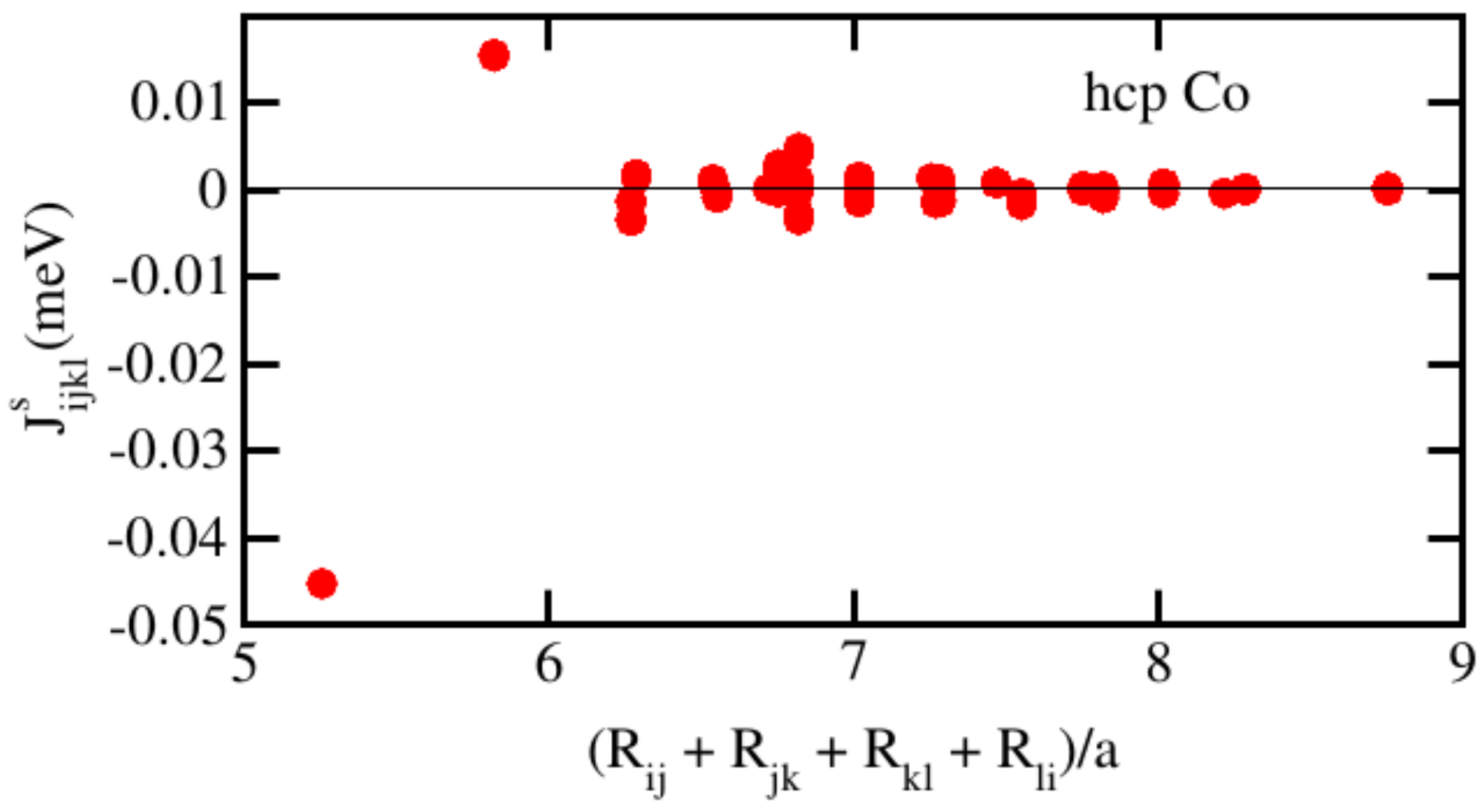}\;(b)
\includegraphics[width=0.35\textwidth,angle=0,clip]{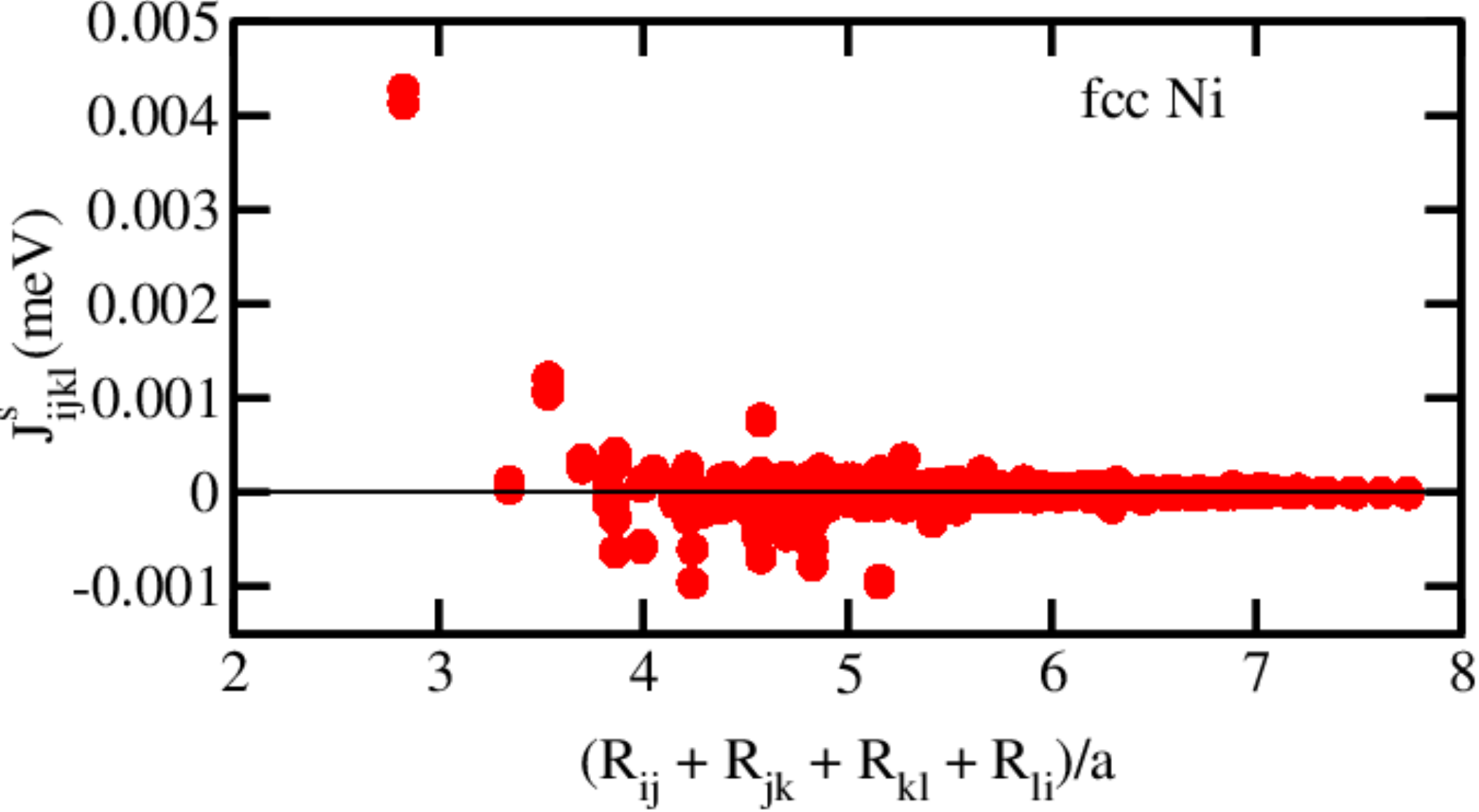}\;(c)
\caption{\label{fig:FOURSPIN-pure} The four-spin interaction exchange
  parameters  $J^{s}_{ijkl}$ according to Eq.\
  (\ref{Eq_Heisenberg_4spin-scalar}) calculated for the FM hcp Co (a),
  bcc Fe (b) 
  and fcc Ni (c) with the magnetization along the $\hat{z}$-axis.
    }  
\end{figure}

Figure \ref{fig:FOURSPIN-pure} represents an example for the four-spin exchange 
parameters $J^s_{ijkl}$ calculated on the basis of Eq.\
(\ref{Eq_Heisenberg_4spin-scalar}) for the three $3d$ bulk ferromagnetic
systems bcc Fe, hcp Co and fcc Ni. The results are plotted as a function of the distance $R_{ij}
  + R_{jk} + R_{kl} + R_{li}$, including only the
  interactions corresponding to $i \neq j\neq k \neq l$, i.e., all sites are
  different. For these systems the
  exchange parameters are about two orders of magnitude smaller than the
  first-neighbor bilinear exchange interactions. However, in general their
  contribution can be non-negligible due to the large number of such four-spin
  loops. Therefore, in some particular cases they should be taken into account.

\begin{figure})
\includegraphics[width=0.35\textwidth,angle=0,clip]{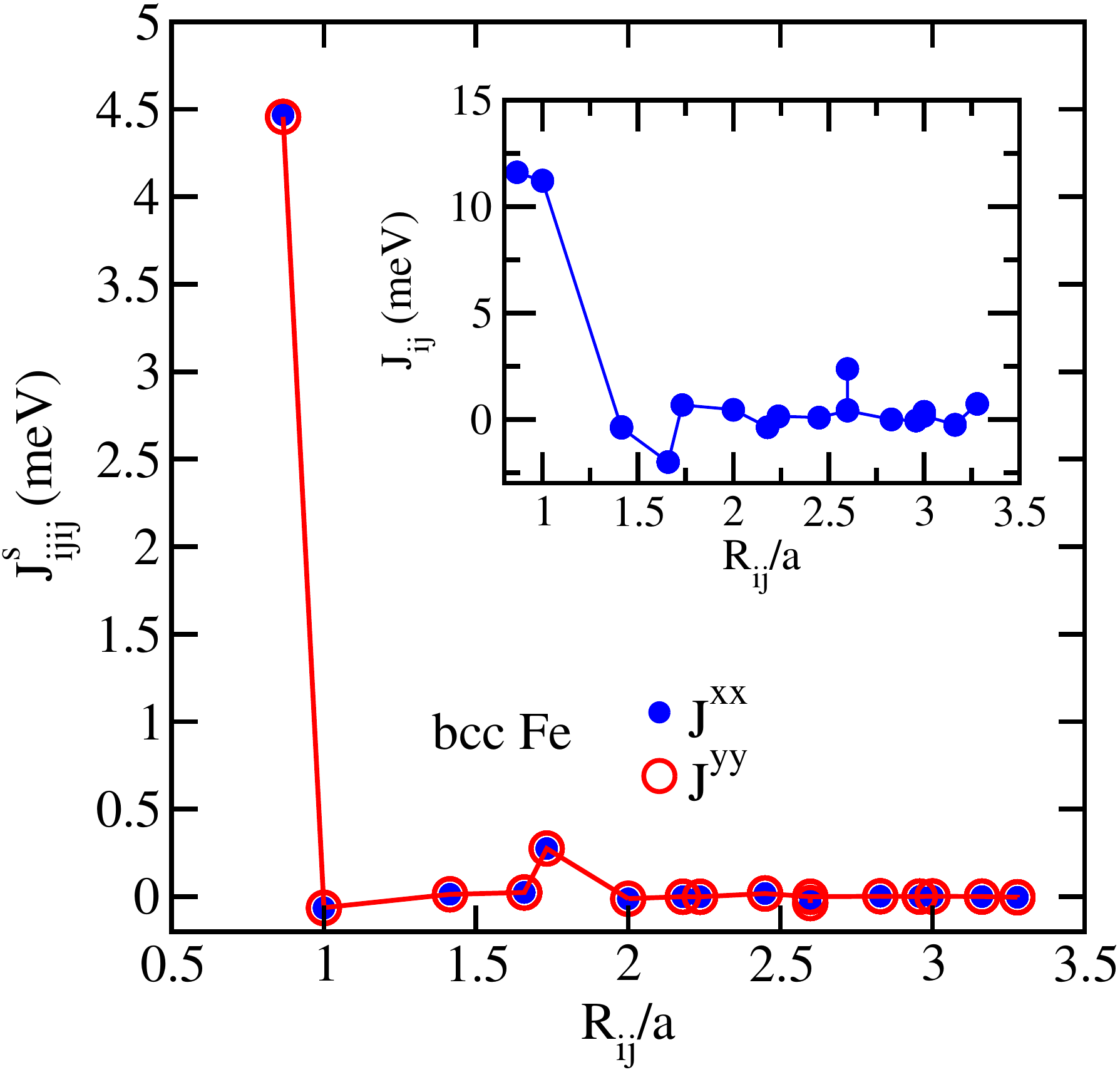}\;(a)
\includegraphics[width=0.35\textwidth,angle=0,clip]{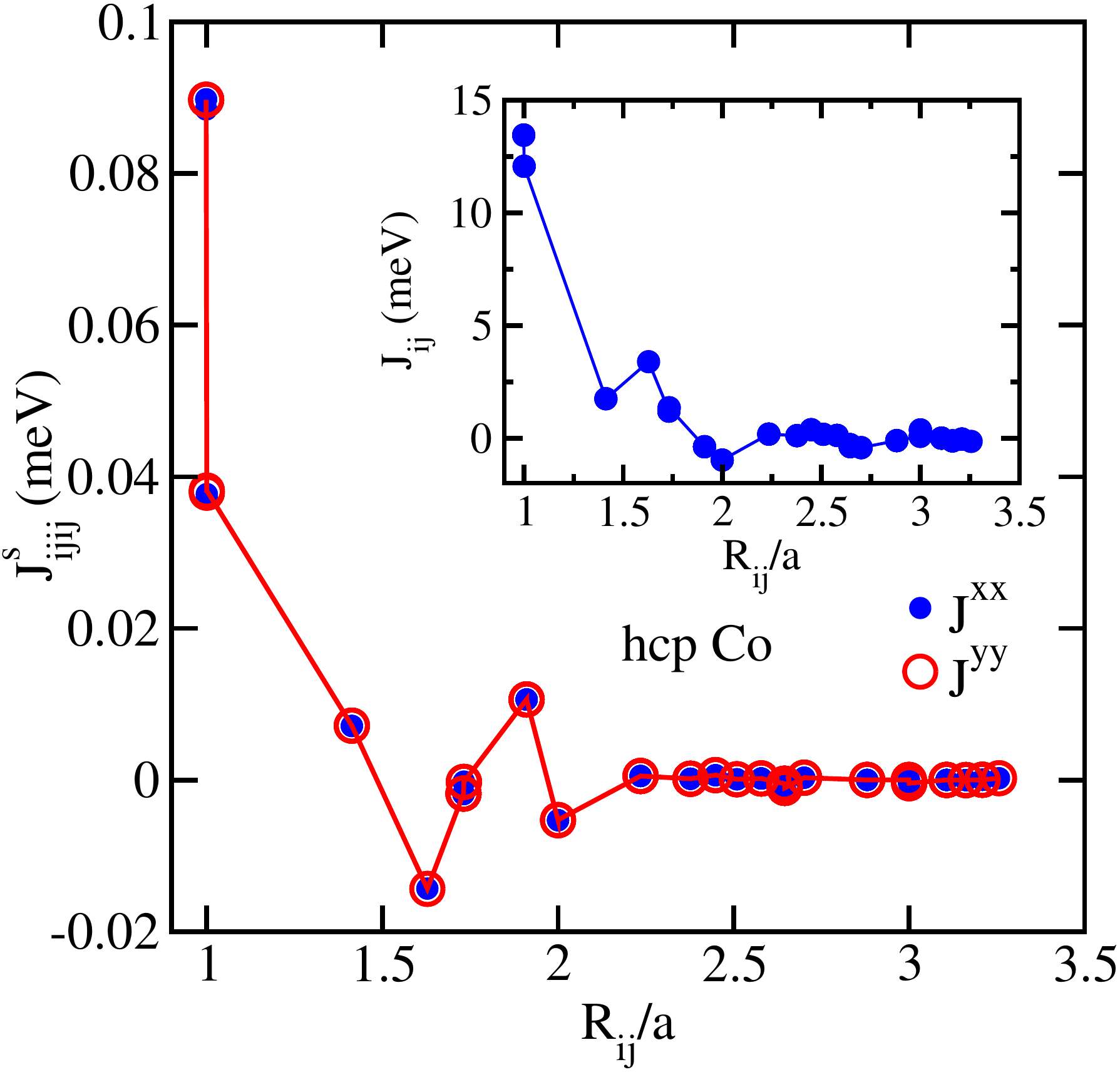}\;(b)
\includegraphics[width=0.35\textwidth,angle=0,clip]{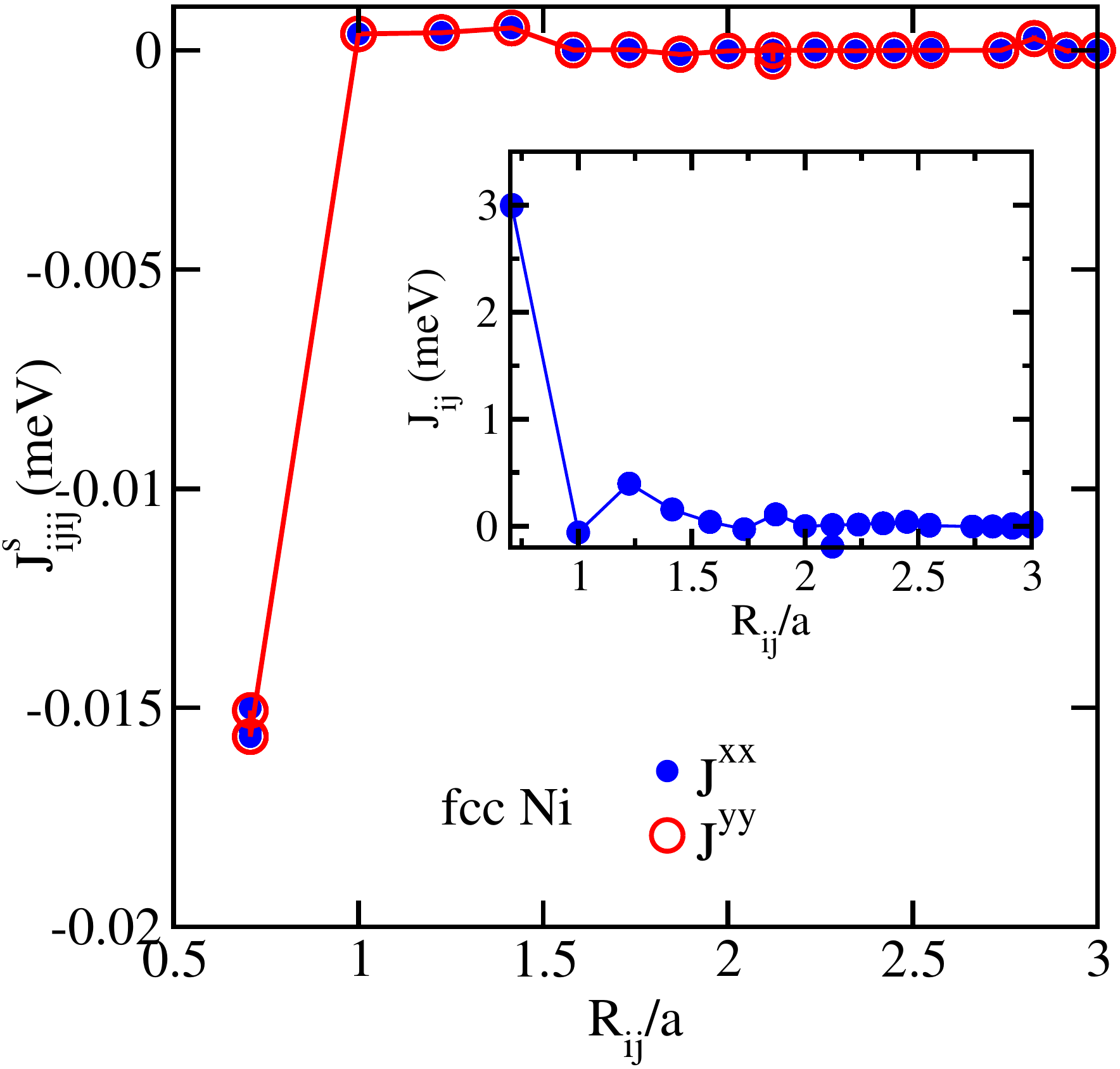}\;(c)
\caption{\label{fig:FeCoNi_biquad}  Scalar biquadratic exchange
  interactions in bcc  Fe (a), hcp Co (b) and fcc Ni (Ni). The insets 
  show the bilinear exchange interaction parameters calculated for the
  FM state with the magnetization along the $\hat{z}$-axis.
    }   
\end{figure}

\begin{figure}
\includegraphics[width=0.35\textwidth,angle=0,clip]{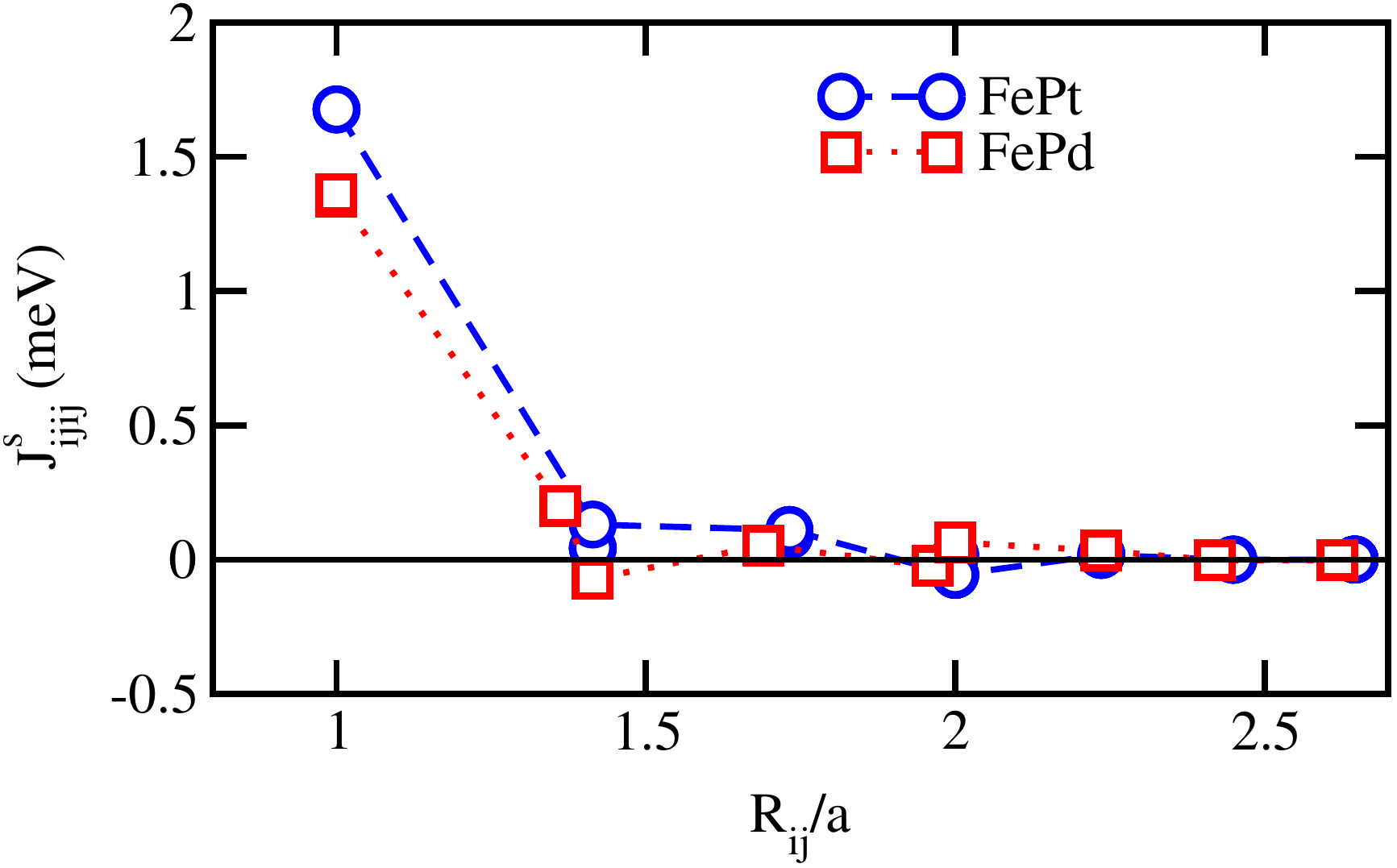}\;
\caption{\label{fig:FePt_biquad} Scalar biquadratic Fe-Fe exchange
  interactions in the FM-ordered FePt and FePd with the magnetization
  along the $\hat{z}$-axis. 
    }  
\end{figure}

Examples for the scalar biquadratic exchange interaction parameters $J^s_{ijij}$ are shown in
Fig.\ \ref{fig:FeCoNi_biquad} for bcc Fe, hcp Co and fcc Ni, and in
Fig.\ \ref{fig:FePt_biquad} for the compounds FePt and FePd having CuAu
crystal structure. For comparison, the insets give the corresponding bilinear
isotropic exchange interactions. One can see rather
strong first-neighbor interactions in bcc Fe and in the compounds FePt
and FePd. This confirms the previous theoretical results for bcc Fe 
\cite{SH97}, and demonstrates the non-negligible character of biquadratic
interactions. This is of course a material-specific property.  

\subsection{DMI-like multispin exchange interactions}

The properties of the chiral multispin exchange interaction parameters in Eq.\
(\ref{Eq_Heisenberg_H4_DMIXY}) can be compared with the DM interactions
as both are vector quantities.
Similarly to the DMI, these parameters are caused by SOC,
i.e. they vanish in the case of SOC = 0. This feature is indeed
demonstrated by our test calculations. 
The calculations have been performed for bulk bcc Fe, for (Pt/$X$/Cu)$_n$
multilayers with $X =$ Mn, Fe and Co, and for an Fe overlayer deposited
on TMDC (transition metal dichalcogenide) monolayers, e.g. 1H-TaTe$_2$
and 1H-WTe$_2$.
The model multilayer system is composed of Pt, $X$ and Cu
on subsequent (111) layers of the fcc lattice, without
structural relaxation. In the case of the Fe/TMDC systems the structural
relaxation has been performed both within the layers as well as in the
$z$ direction perpendicular to the layer plane.

\begin{figure}
\includegraphics[width=0.37\textwidth,angle=0,clip]{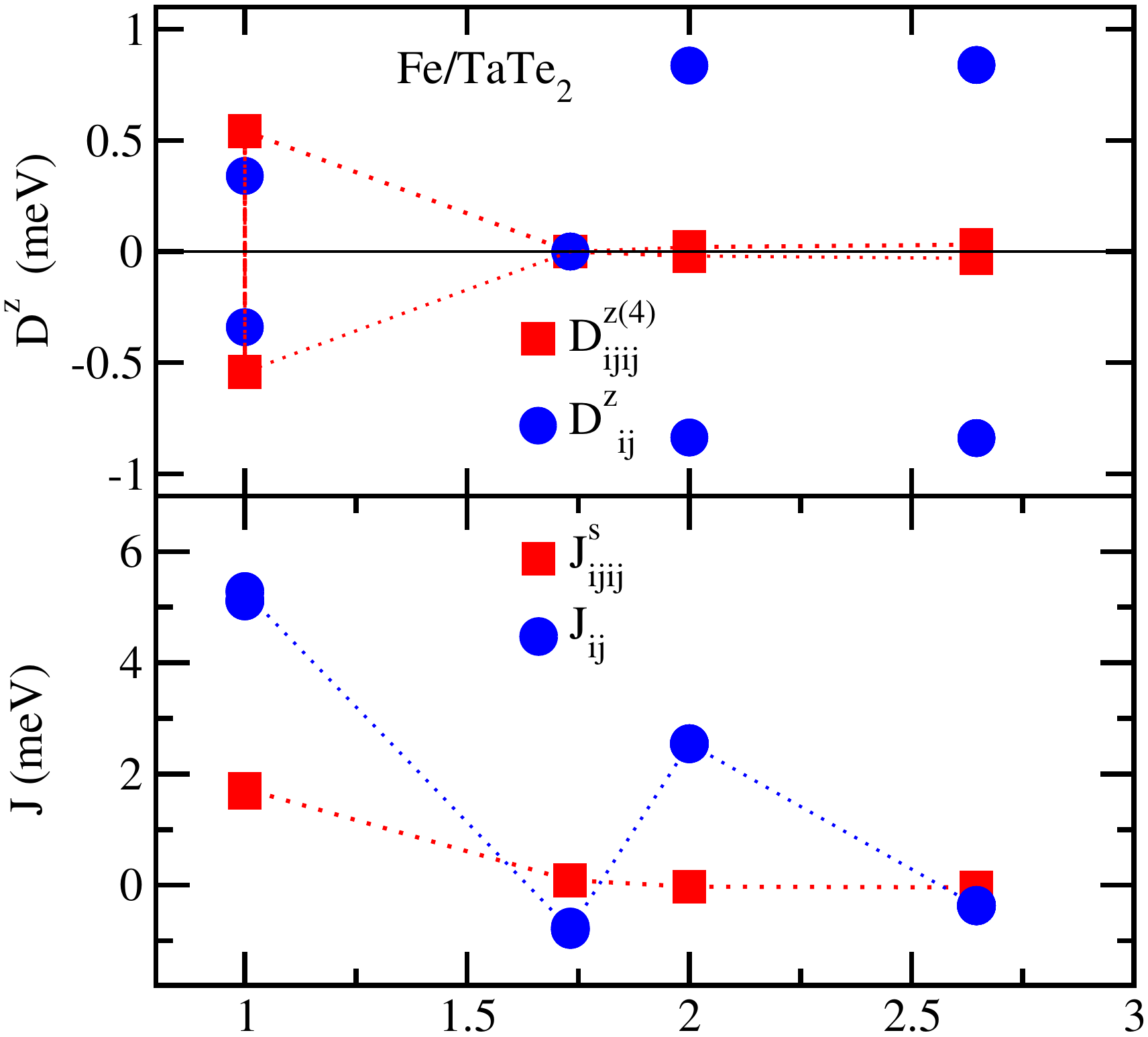}\;(a)
\includegraphics[width=0.37\textwidth,angle=0,clip]{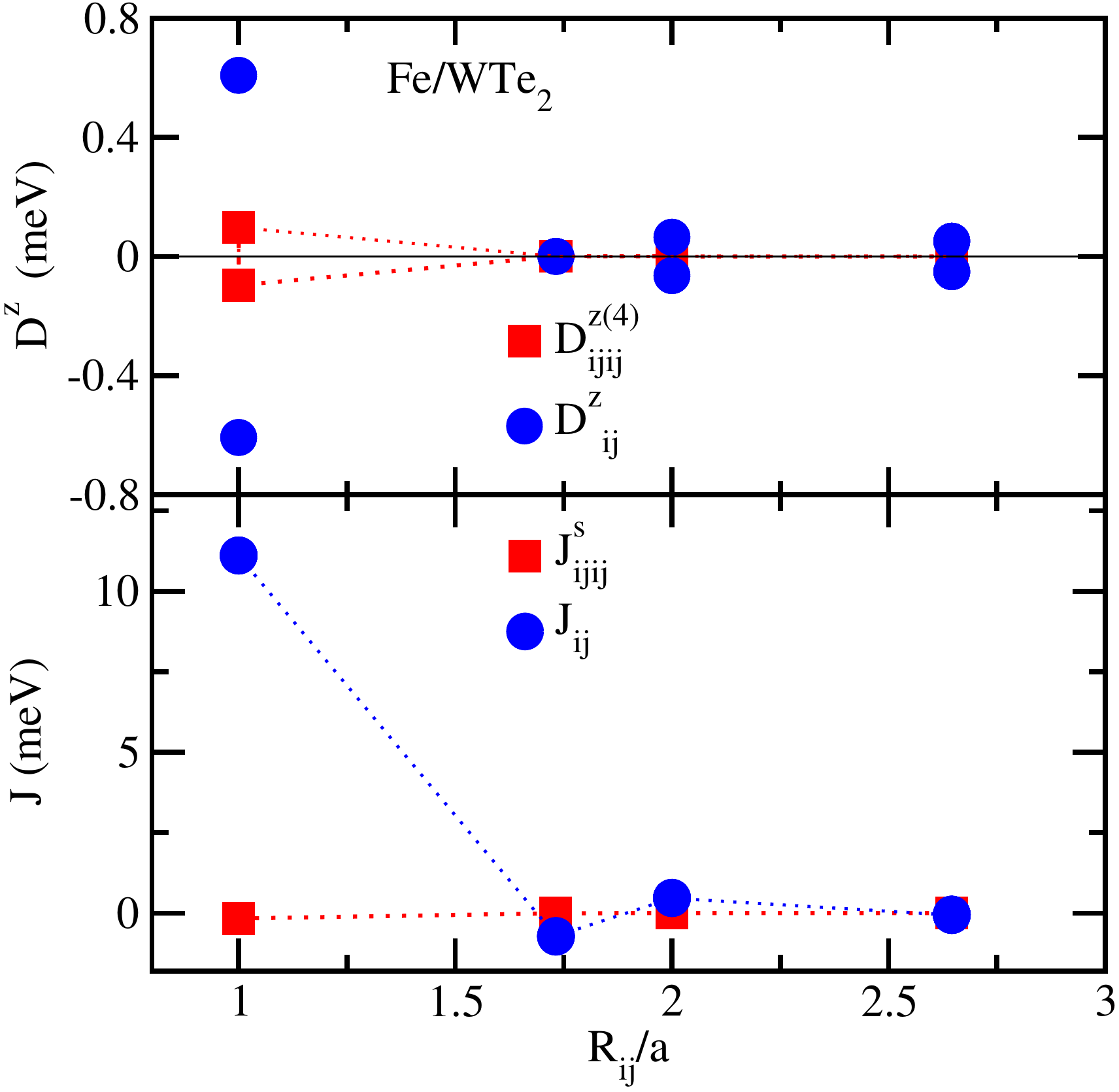}\;(b)
\caption{\label{fig:TMDC_DMI} $z$-component of the BDMI (squares, top)
  and biquadratic scalar interactions 
  (squares, bottom) as a function of the site distance $R_{ij}$ (in
  multiples of the lattice parameter $a$). Results are shown for a Fe
  overlayer deposited on a single layer of 
  TaTe$_2$ (a) and WTe$_2$ (b), in comparison with DMI and bilinear
  interaction parameters shown by circles.
    }  
\end{figure}

The calculations demonstrate similar symmetry properties of the BDMI
when compared with the conventional DMI, as 
was already pointed out recently \cite{BSL19}.
In bcc Fe having inversion symmetry, the BDMI is equal to zero, while
it is finite in the multilayer and the Fe/TMDC 
systems, following the properties of the DMI interactions.
Figure \ref{fig:TMDC_DMI} gives results for the $z$-component of the chiral
biquadratic exchange interactions, ${\cal D}^{z}_{ijij}$, calculated for a Fe
overlayer deposited on a TaTe$_2$ and WTe$_2$ single layers,
respectively, on the basis of Eq.\ (\ref{Eq_Heisenberg_D4}).
As one can see,  ${\cal D}^{z}_{ijij}$ has a significant magnitude
when compared to the bilinear DMI parameters. Interestingly, the $x$ and
$y$ components in these two materials are much smaller than the
corresponding components of the bilinear DMI. 

\begin{figure}
\includegraphics[width=0.4\textwidth,angle=0,clip]{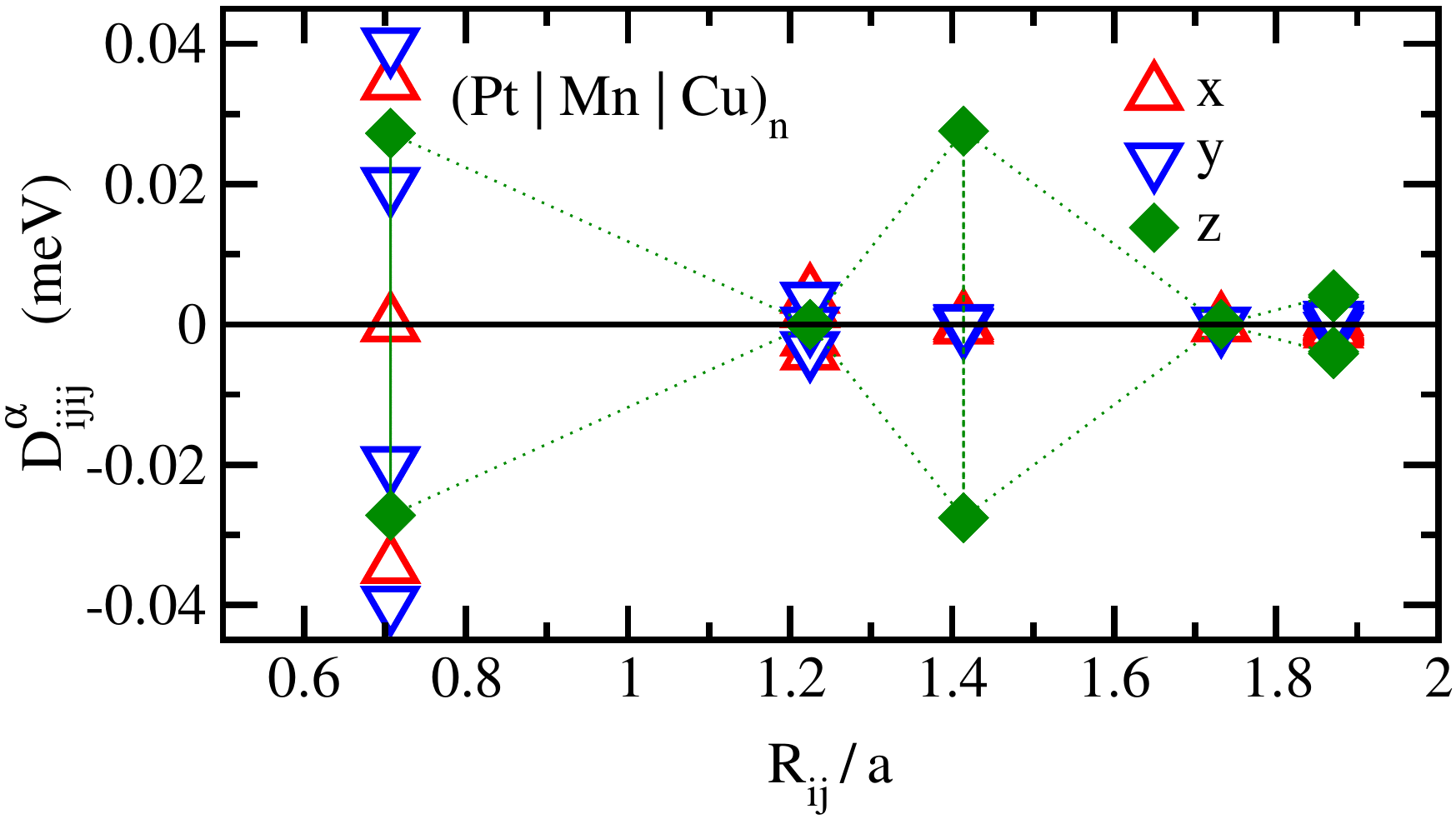}\;(a)
\includegraphics[width=0.4\textwidth,angle=0,clip]{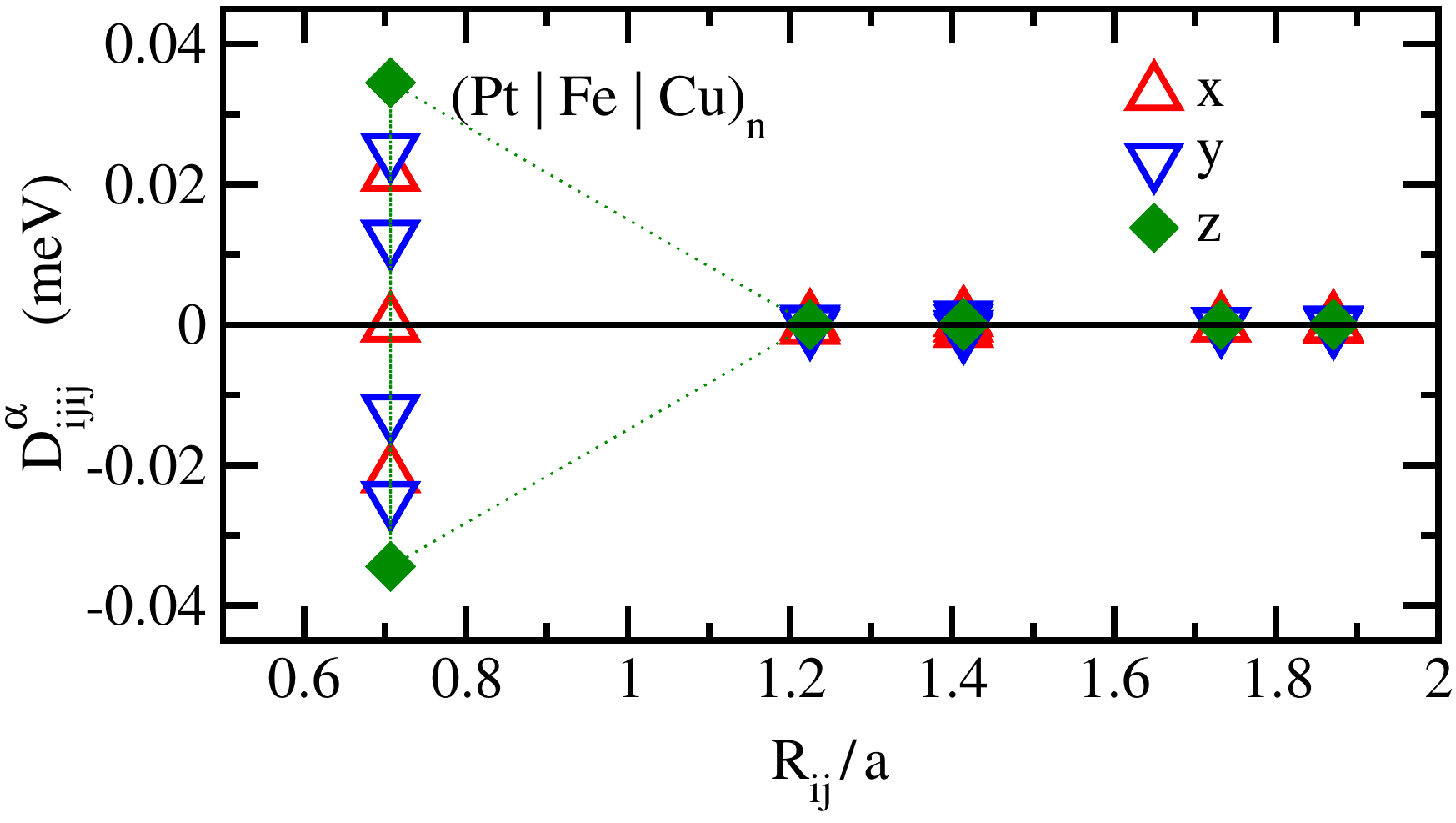}\;(b)
\includegraphics[width=0.4\textwidth,angle=0,clip]{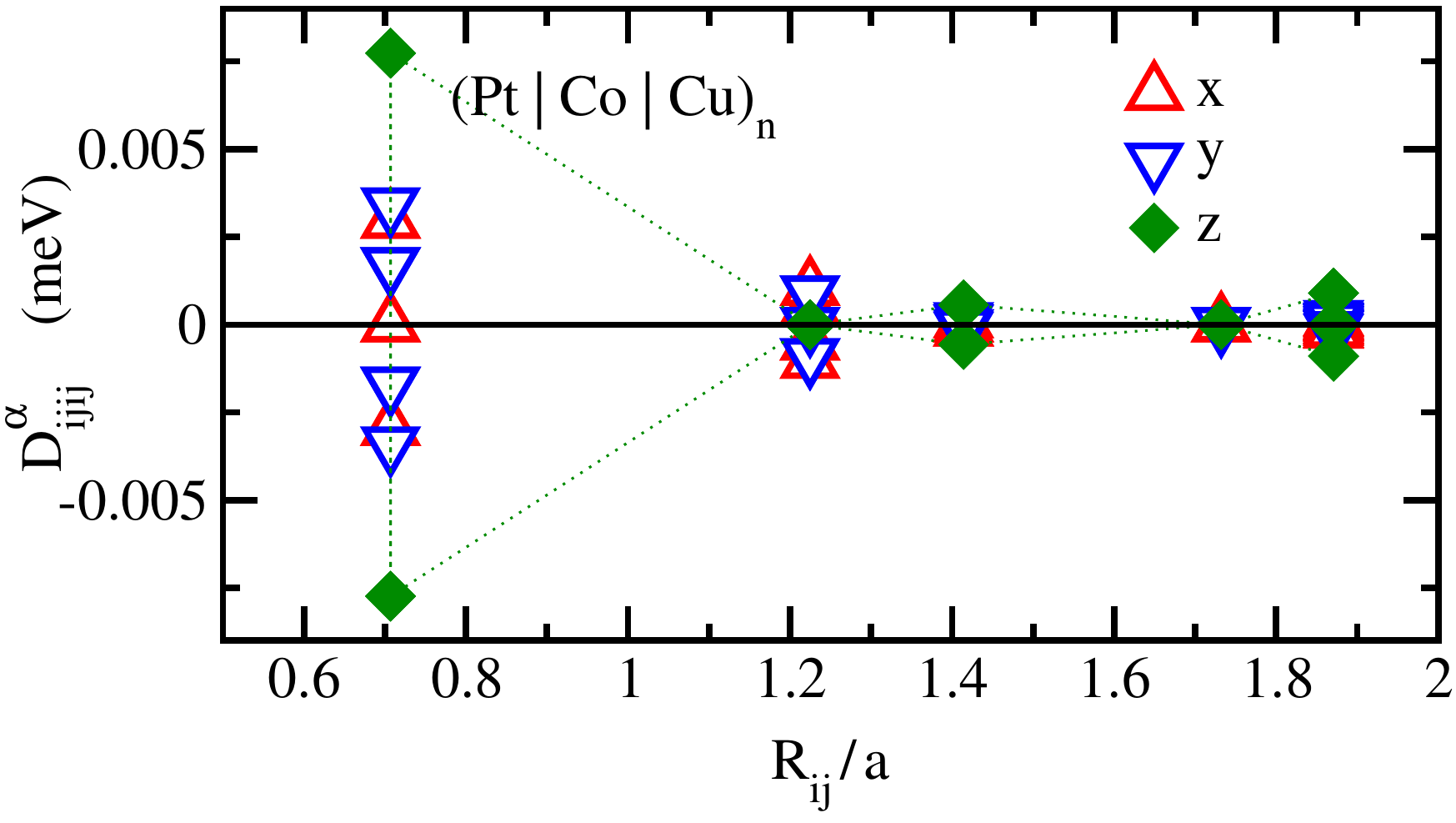}\;(c)
\caption{\label{fig:Pt/Mn/Cu_DMI-biquad}  The $x$-, $y$- and $z$-components of chiral
  biquadratic exchange interaction, $\vec{{\cal D}}_{ijij}$,
  between the magnetic 3d-metals $X$ in (Pt/$X$/Cu)$_n$ multilayers with
  $X =$ Mn (a), Fe (b), and Co (c), plotted  as a function of the interatomic distance
  $R_{ij}$. The orientation of the interaction vectors between first nearest neighbors is
  shown in Fig.\ \ref{fig:PtFeCu_3spin_inplain} 
  }  
\end{figure}

\begin{figure}
\includegraphics[width=0.13\textwidth,angle=0,clip]{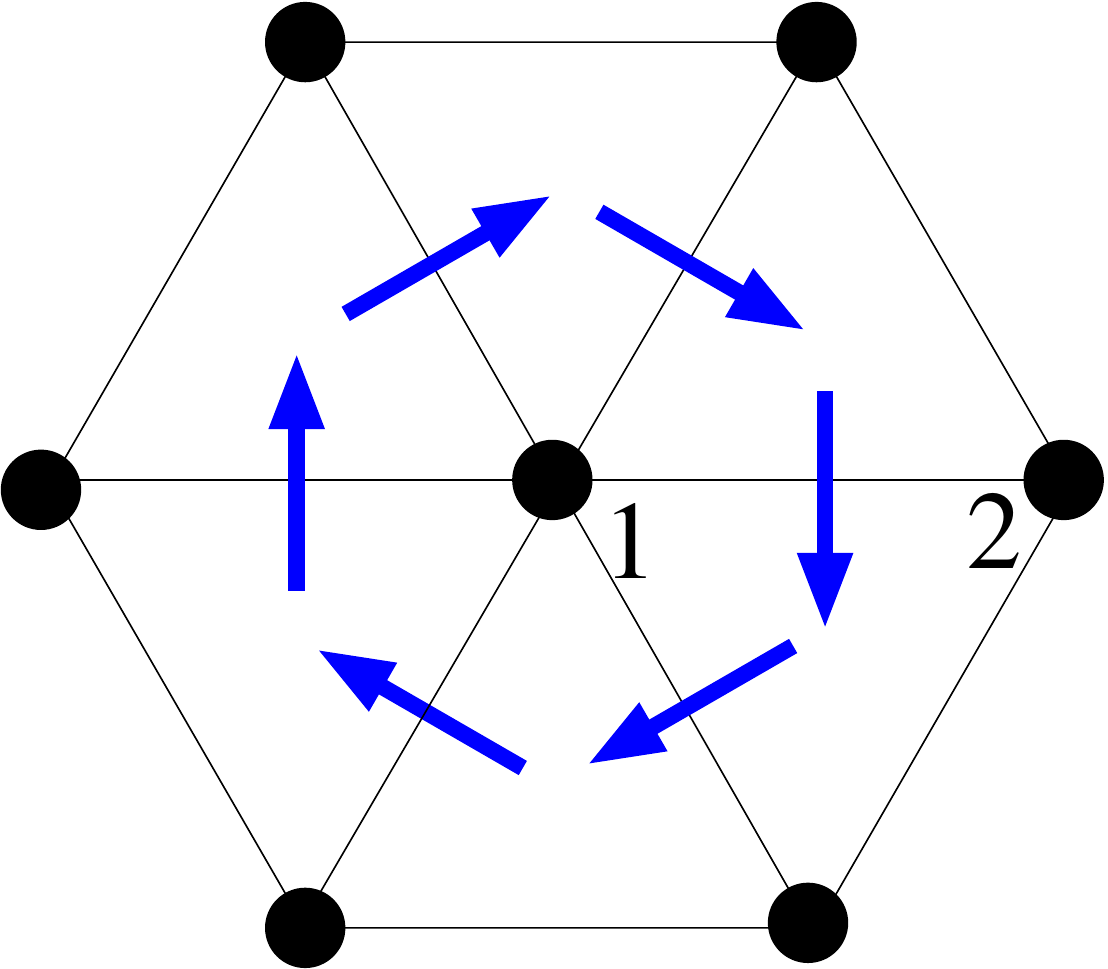}\;
\includegraphics[width=0.13\textwidth,angle=0,clip]{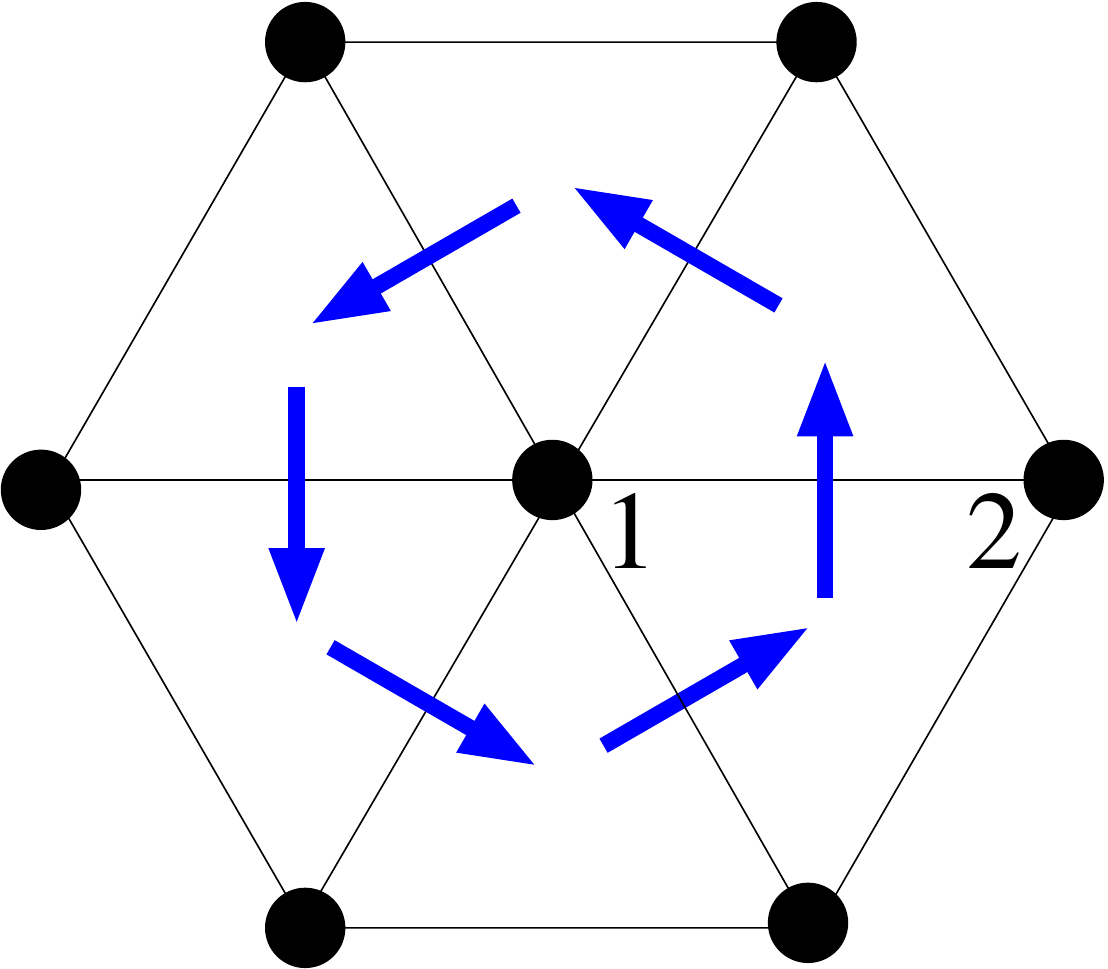}\;
\includegraphics[width=0.13\textwidth,angle=0,clip]{DMI-2_PtXCu-eps-converted-to.pdf}\;(a)
\includegraphics[width=0.13\textwidth,angle=0,clip]{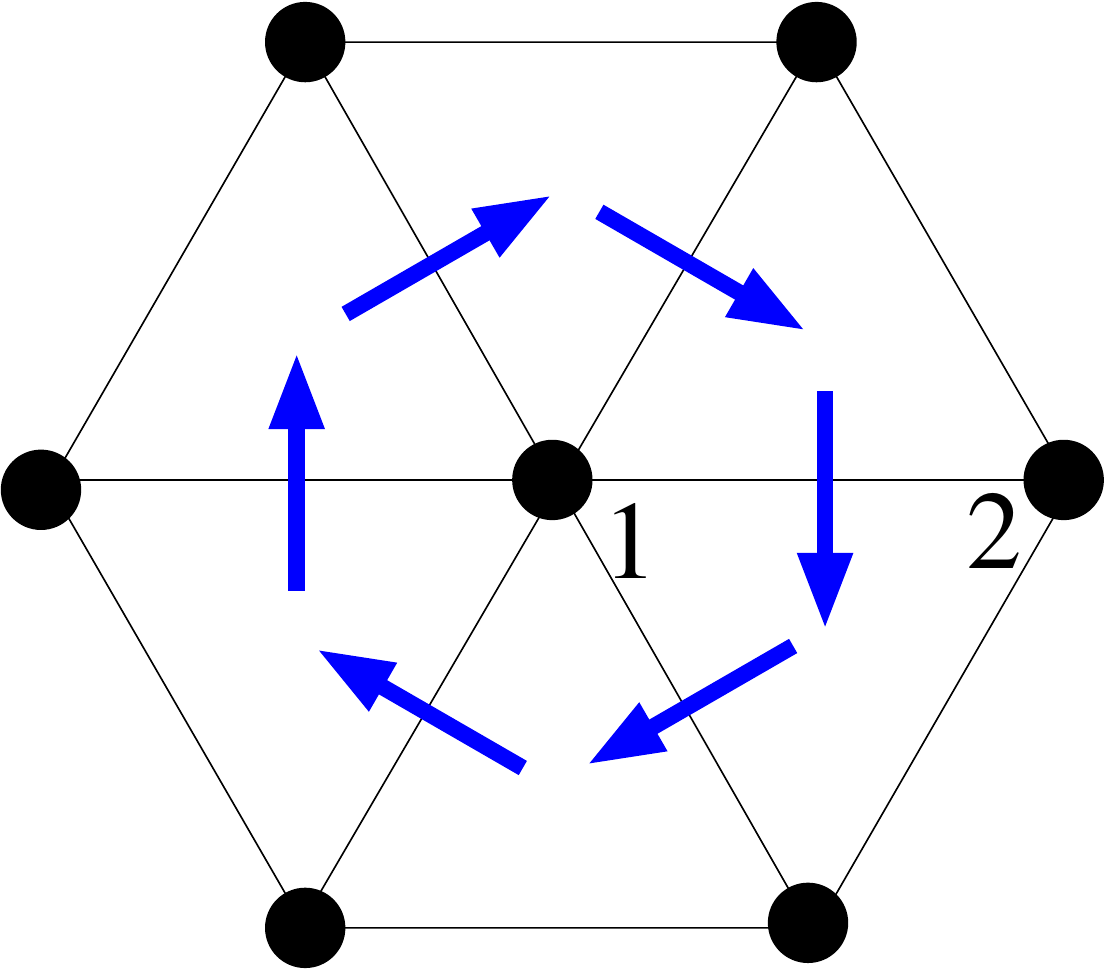}\;
\includegraphics[width=0.13\textwidth,angle=0,clip]{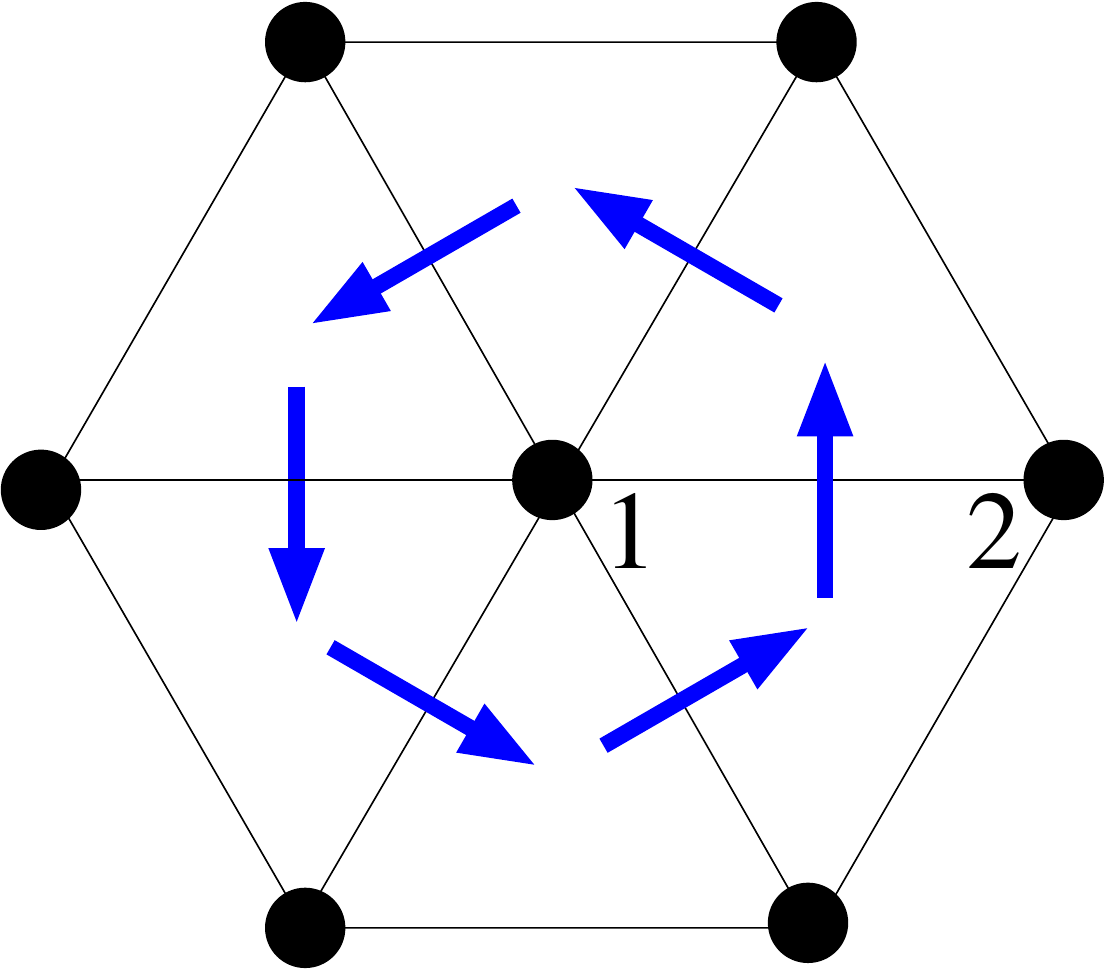}\;
\includegraphics[width=0.13\textwidth,angle=0,clip]{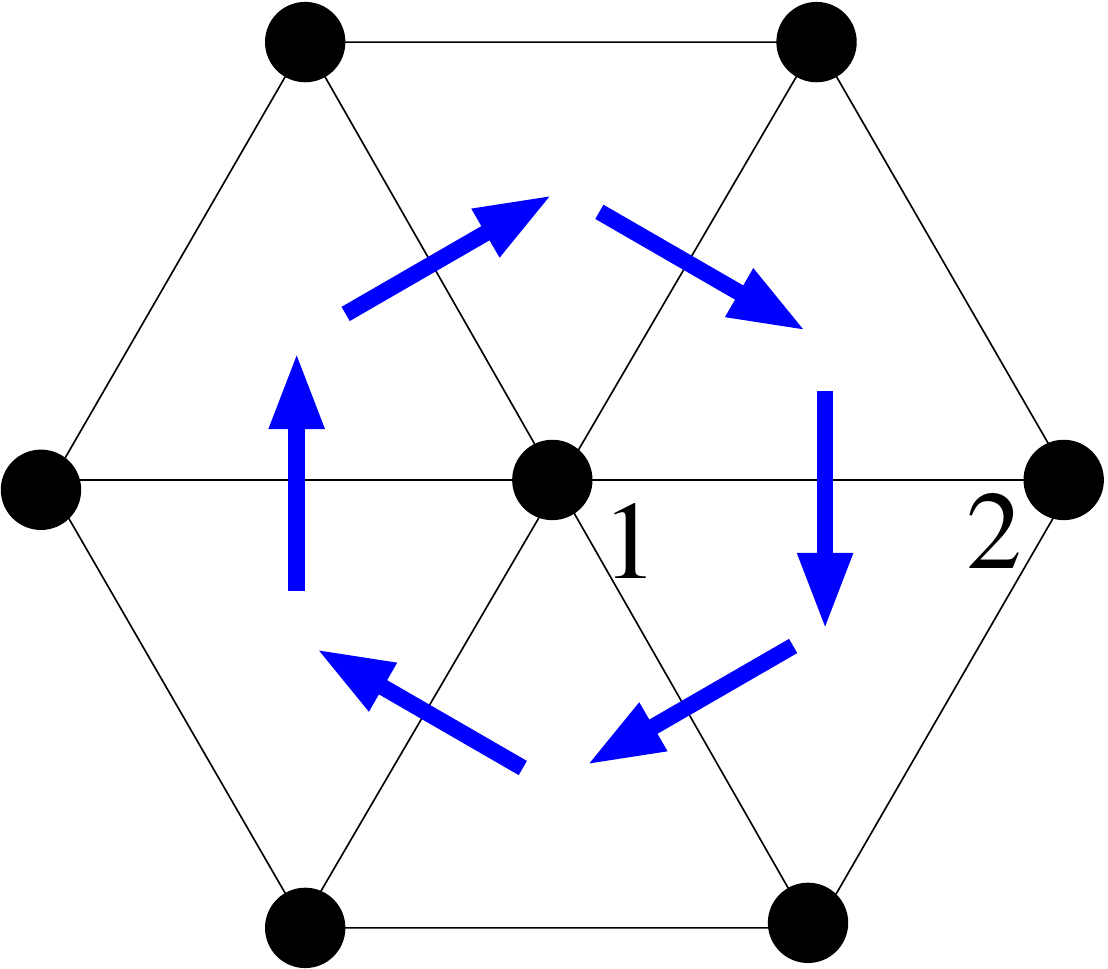}\;(b)
\caption{\label{fig:PtFeCu_3spin_inplain} 
Orientation of the DMI vector $\vec{D}_{ij}$  (a) and the vector of the
BDMI $\vec{{\cal D}}_{ijij}$ (b) corresponding to the interaction of the first 
neighbor 3d-atoms $X$ in (Pt/$X$/Cu)$_n$ multilayers: for $X =$ Mn
(left), Fe (middle), and Co (right). The magnitude of the interactions are
  given in Table \ref{TAB_MOM}.}   
\end{figure}

In the case of the multilayer systems (Pt/Fe/Cu)$_n$,   (Pt/Mn/Cu)$_n$
and (Pt/Co/Cu)$_n$ all three  components, $x, y, z$, have the same order of magnitude as
it is seen in Fig.\ \ref{fig:Pt/Mn/Cu_DMI-biquad}. The orientation of these
interactions between first nearest neighbor sites is shown in Fig.\
\ref{fig:PtFeCu_3spin_inplain}. As can be seen from Table \ref{TAB_MOM},
in contrast to the Fe/TMDC system, all components are more than one order
of magnitude smaller than the corresponding DMI components. 

Figure \ref{fig:PtFeCu_3spin_inplain} shows
schematically the in-plane components of the DMI and BDMI, which have
the same orientation for (Pt/Fe/Cu)$_n$ and  (Pt/Mn/Cu)$_n$, but not for
(Pt/Co/Cu)$_n$.
The $y$-component of ${\cal D}^y_{ijij}$  representing 
the interaction between atoms with $\vec{R}_{ij} = a(0.707,0,0)$ are
given in Table \ref{TAB_MOM}. These values give the total 
in-plane interaction as for the taken pair of atoms ${\cal
  D}^x_{ij} = 0$ and  ${D}^x_{ij} = 0$. Note also that in (Pt/Mn/Cu)$_n$
the ${\cal D}^z_{ijij}$ component has an opposite sign when compared to
${D}^z_{ij}$.

\begin{table}
\begin{tabular}{ |p{2cm}||p{1.5cm}|p{1.5cm}|p{1.5cm}|p{1.5cm}| }
 \hline
               & ${D}^y_{ij}$ & ${D}^z_{ij}$  & ${\cal D}^y_{ijij}$  & ${\cal D}^z_{ijij}$ \\
 \hline
  (Pt/Mn/Cu)$_n$      & -1.14   & -1.22  &  -0.039   &   0.031   \\ 
  (Pt/Fe/Cu)$_n$      & 0.17   & 0.35    &  0.024    &   0.034   \\ 
  (Pt/Co/Cu)$_n$      & 0.63   & 0.40    &  -0.003   &   0.008  \\
\hline                                 
\end{tabular}
\caption{\label{TAB_MOM} The $y$- and $z$-components of the DMI and chiral
  biquadratic exchange interaction (in meV)
  between 3d-metals in (Pt/$X$/Cu)$_n$ multilayers ($X =$
  Mn,~Fe,~Co). The $y$- component corresponds to the interactions between
  atoms 1 and 2 (see Fig.\ \ref{fig:PtFeCu_3spin_inplain}) with
  $\vec{R}_{12} = a(0.707,0,0)$. For this geometry ${\cal D}^y_{1212}$ and ${D}^y_{12}$
  represent the magnitude of the in-plane projection of corresponding
  interactions with the first nearest neighbors, as in this case 
  ${\cal D}^x_{1212} = 0$ and  ${D}^x_{12} = 0$. }       
\end{table}

\begin{figure}
\includegraphics[width=0.45\textwidth,angle=0,clip]{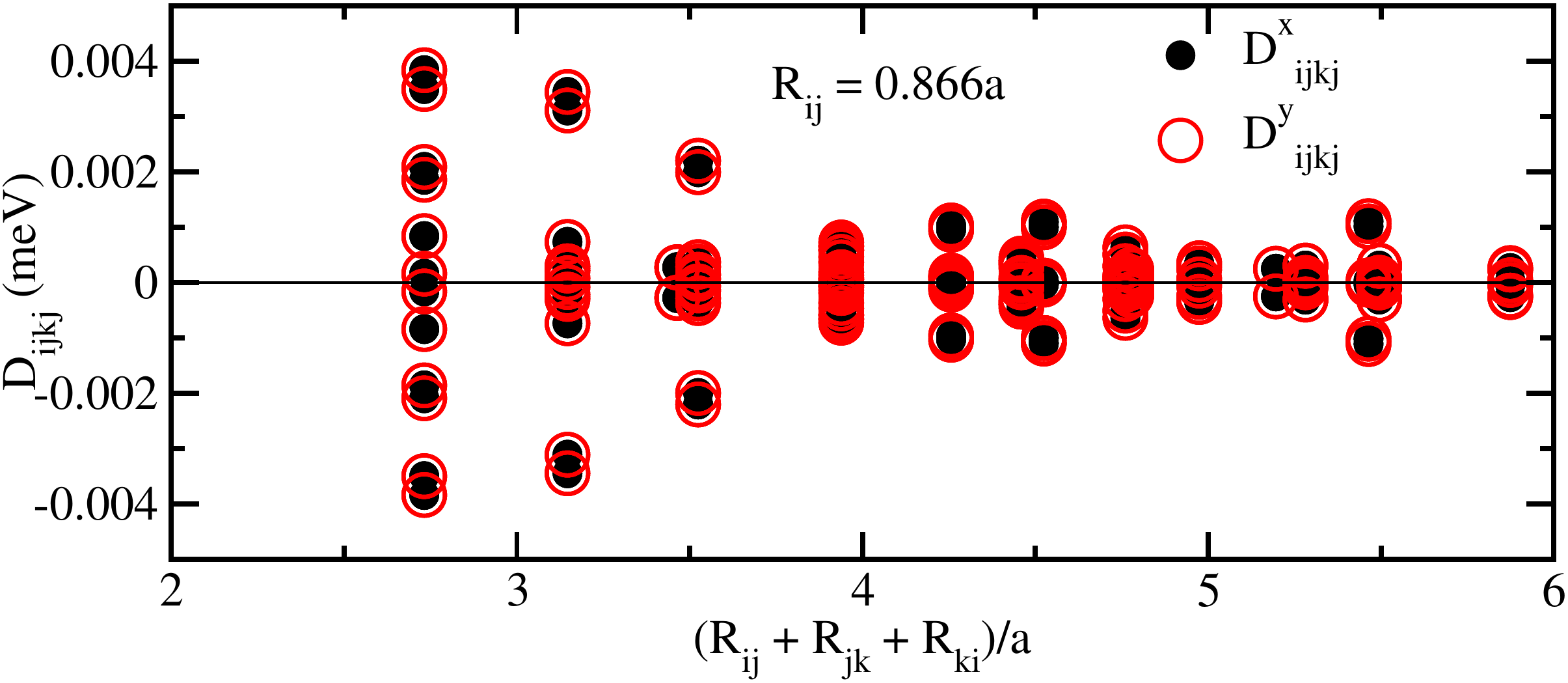}\;(a)
\includegraphics[width=0.14\textwidth,angle=0,clip]{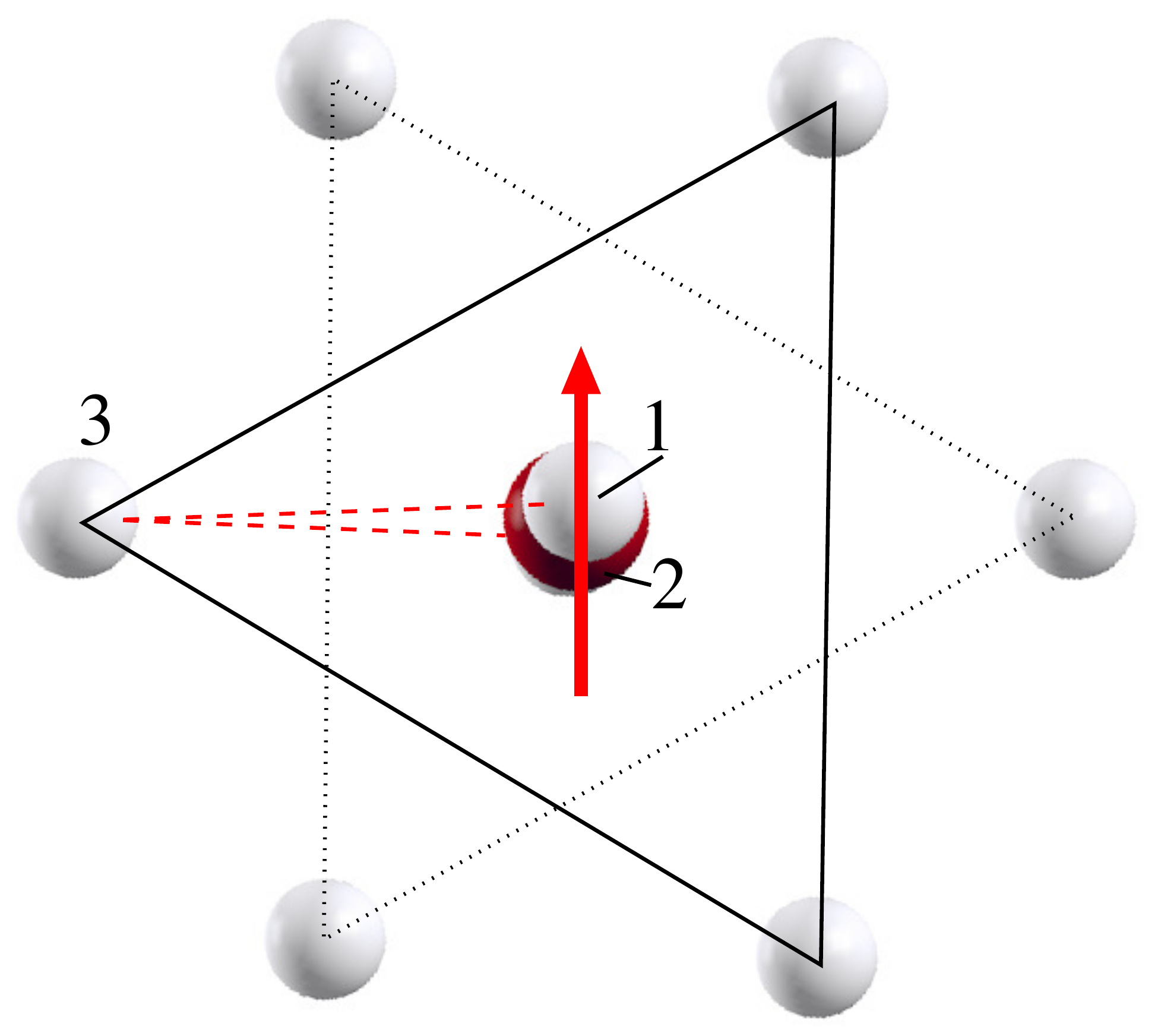}\;
\includegraphics[width=0.14\textwidth,angle=0,clip]{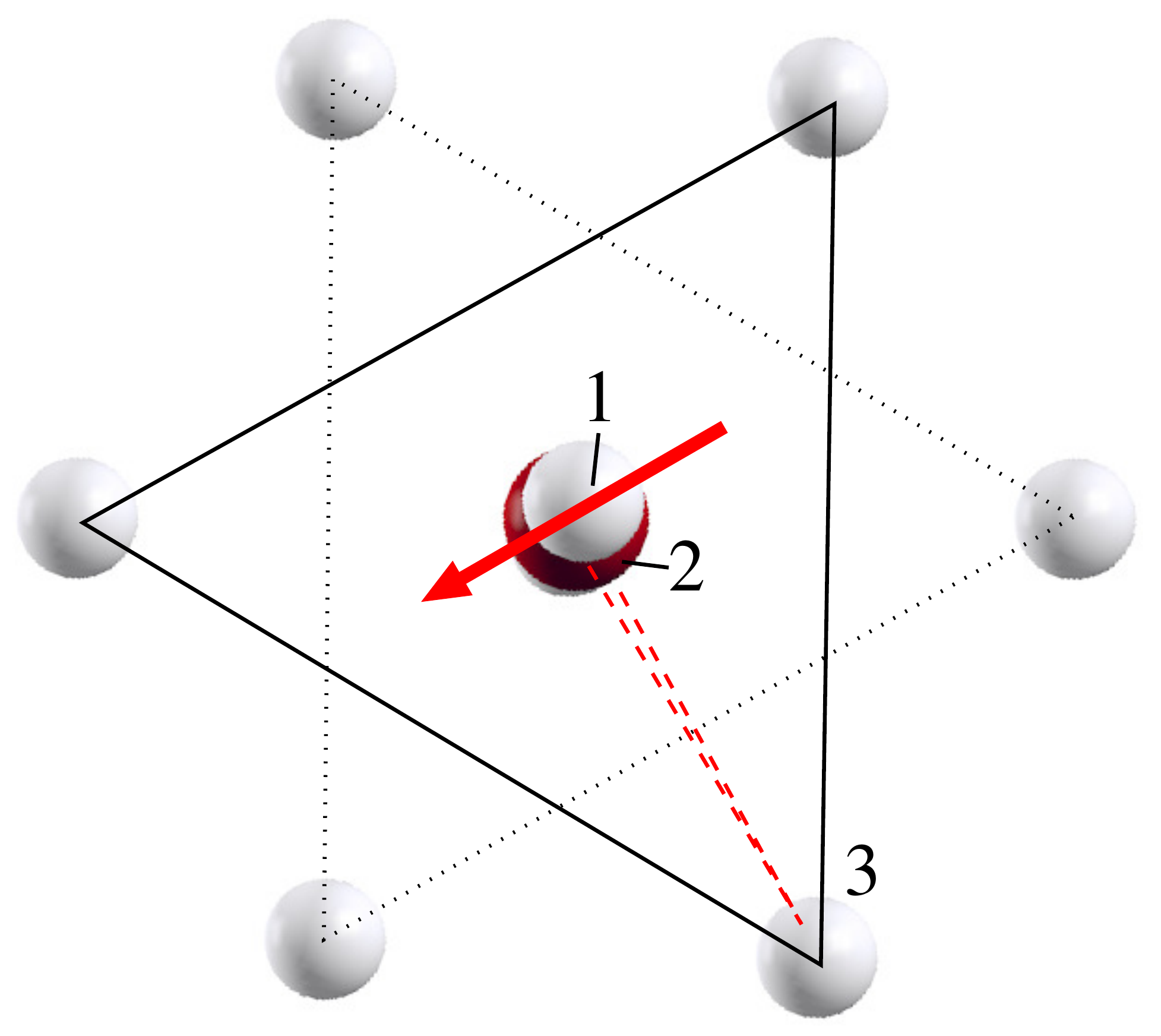}\;
\includegraphics[width=0.14\textwidth,angle=0,clip]{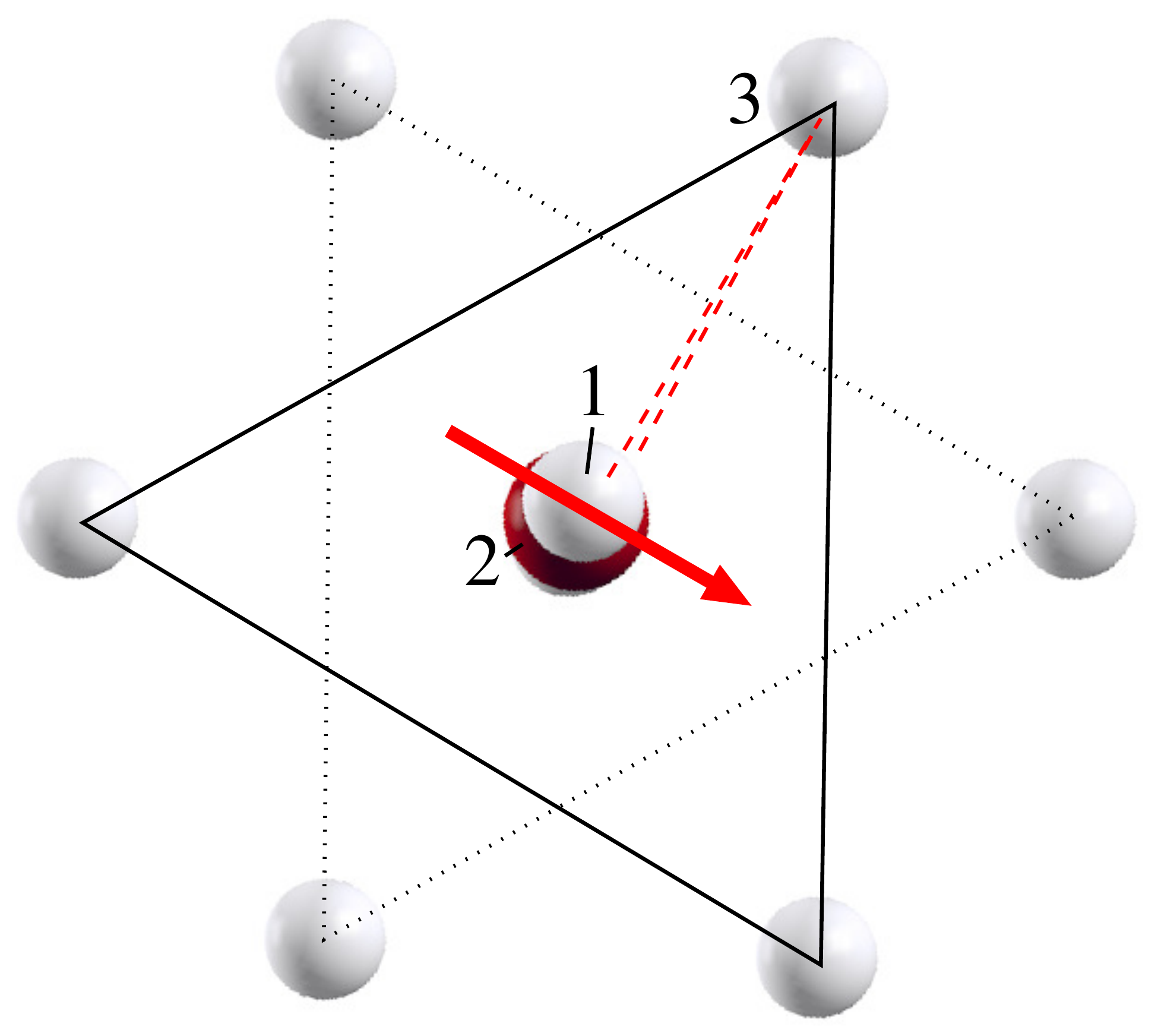}\;(b)
\caption{\label{fig:Fe_3spin-DMI} (a) TDM interactions for bcc Fe
with 'vector' coupling between the gray atom in the center (atom 1) and
the red atom behind (atom 2) with the third atom (atom 3) coupled with
atom 2 via scalar coupling. The magnitude of this three-spin interaction
energy is 0.004 meV for all three cases shown in the figure. 
}   
\end{figure}

Similar to the DMI and BDMI, the TDMI $\vec{\cal D}_{ijkj}$ is a 
SOC-induced interaction between atoms $i$ and $j$ which depends on the
relative orientation of the spin moments of the atoms $j$ and $k$. 
In contrast to the biquadratic interaction, it does not vanish for
centrosymmetric systems, as it is demonstrated by the calculations for bcc
Fe represented in Fig.\ \ref{fig:Fe_3spin-DMI}.
Let us consider the TDMI as
the DMI-like interaction between atoms $i$ and $j$ which depends on
the relative orientation of the spin moment of the atoms $j$ and $k$.  
Fig.\ \ref{fig:Fe_3spin-DMI}(a) displays the
dependence of the components ${\cal D}^x_{ijkj}$ and ${\cal D}^y_{ijkj}$
of the DMI-like interaction between the first nearest neighbors (distance
$|\vec{R}_{ij}|$) in bcc Fe as a function of the position of the third atom
$k$. One finds obviously a different sign for the various interactions for the same
value of $|\vec{R}_{ij}| + |\vec{R}_{jk}| + |\vec{R}_{ki}|$ implying a
dependence of the ${\vec {\cal D}}_{ijkj}$ interaction on the relative
position of the third atom (see Fig.\ \ref{fig:Fe_3spin-DMI}(b)).
In the case of a collinear magnetic structure this property results
in a compensation when summing over all surrounding atoms $k$, i.e.\ 
$\sum_{k}\vec{\cal  D}_{ijkj} = 0$, 
despite the finite magnitude of the individual interactions
$|\vec{\cal D}_{ijkj}| \neq 0$ for each triple of atoms. 
In other words,
the TDMI is canceled out in centrosymmetric collinear magnetic systems,
giving no contribution to the energy as the DMI and BDMI.
In the case of a non-collinear magnetic texture,
however, the sum can be non-zero leading to a 
non-vanishing contribution of the TDMI term to the 
energy that may stabilize the non-collinear magnetic structure.

To understand this behavior, one can consider once more the DMI between
two spin moments $\vec{s}_i$ and $\vec{s}_j$, 
caused by SOC seen as a perturbation (see e.g. \cite{BSL19}). 
Within a real space consideration the origin of the DMI can be
associated with the SOC of the electrons on a third atom arranged in the
vicinity of the atoms $i$ and $j$.
Following the work by Brinker et al. \cite{BSL19} this will be called 
the 'SOC carrying' atom. 
The anisotropy of the 
exchange interaction of two spin moments associated with a 
single 'SOC carrying' atom in this case is non-zero, 
while the DMI and its symmetry properties are determined by 
all surrounding 'SOC carrying' atoms and by the
crystal symmetry. In particular for a centrosymmetric system, this leads 
to a cancellation when summing all individual contributions.
In the case of the TDMI one has to make an explicit summation over the
'third' atom $k$ involved in the interaction, which can be seen  as the
'SOC carrying' one.  
 As a result, for any triple of atoms in a
centrosymmetric system  the TDMI is non-zero.
For the case of a  collinear magnetic structure, however,  
the summation over all atoms  $k$  leads to a   
canceling of the TDMI. 
In the case of a non-collinear magnetic structure, on the other hand,
this does not have to apply.

 Note that these conclusions based on the results obtained for a frame
 of reference  with the $\hat z$ axis oriented along the crystallographic [001]
  direction should hold for any other frame of reference. 
Nevertheless, it is more convenient to discuss the interactions using a
frame of reference with the $z$ axis, as well as the magnetization, oriented
along the [111] crystallographic direction, as it is shown in Fig.\
\ref{fig:Fe_3spin-DMI}(b). The arrows represent the direction of the TDM
interaction in the $(x,y)$ plane between the gray atom $1$ at the center and
the red atom $2$ behind, induced by tilting of the moment of the third
atom $(3)$ (connected in the picture by dashed lines with the atoms $1$
and $2$). One can see that the direction of this interaction
depends on the position of atom $3$.
%

\begin{figure}
\includegraphics[width=0.2\textwidth,angle=0,clip]{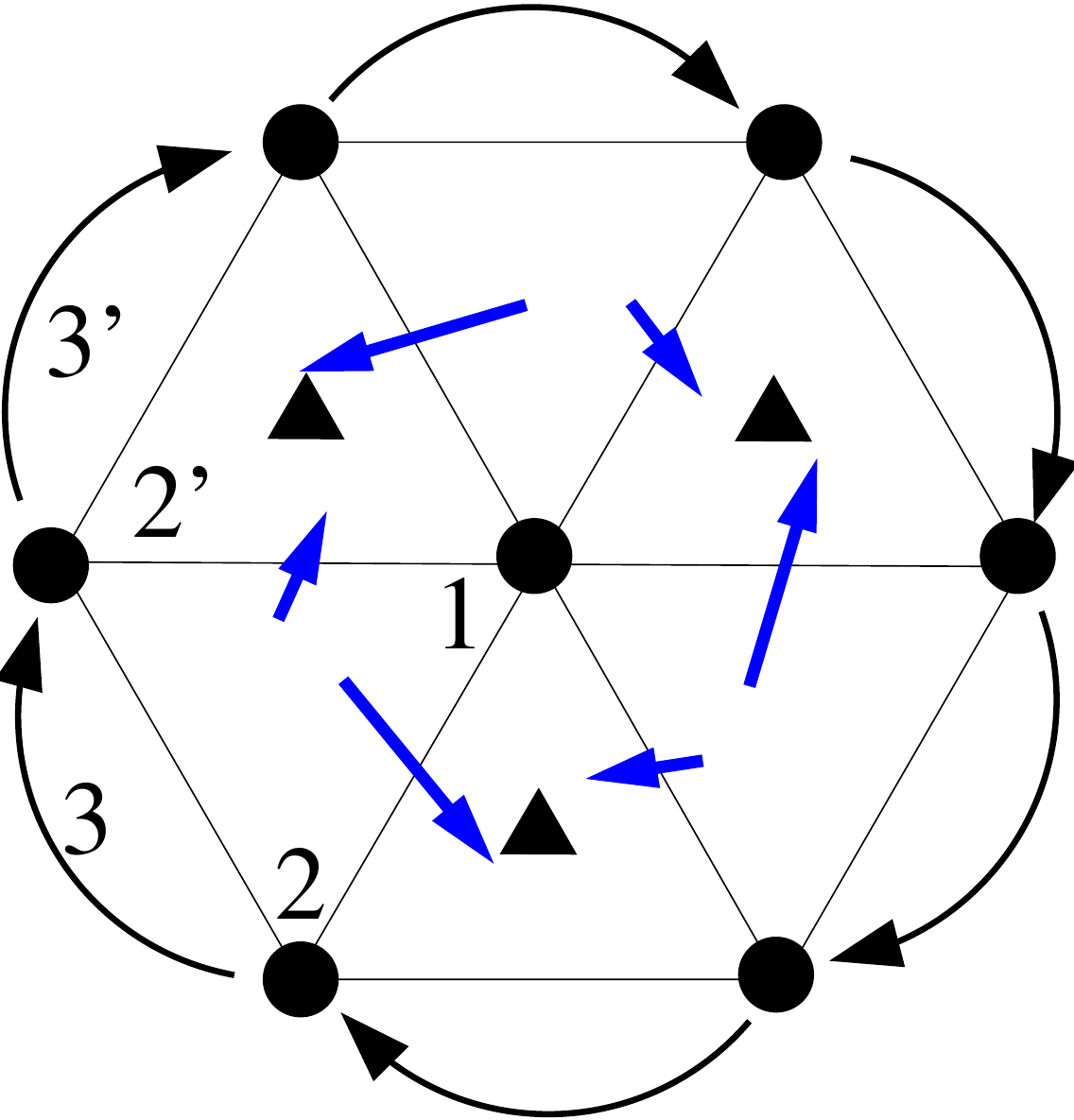}\;(a)
\includegraphics[width=0.2\textwidth,angle=0,clip]{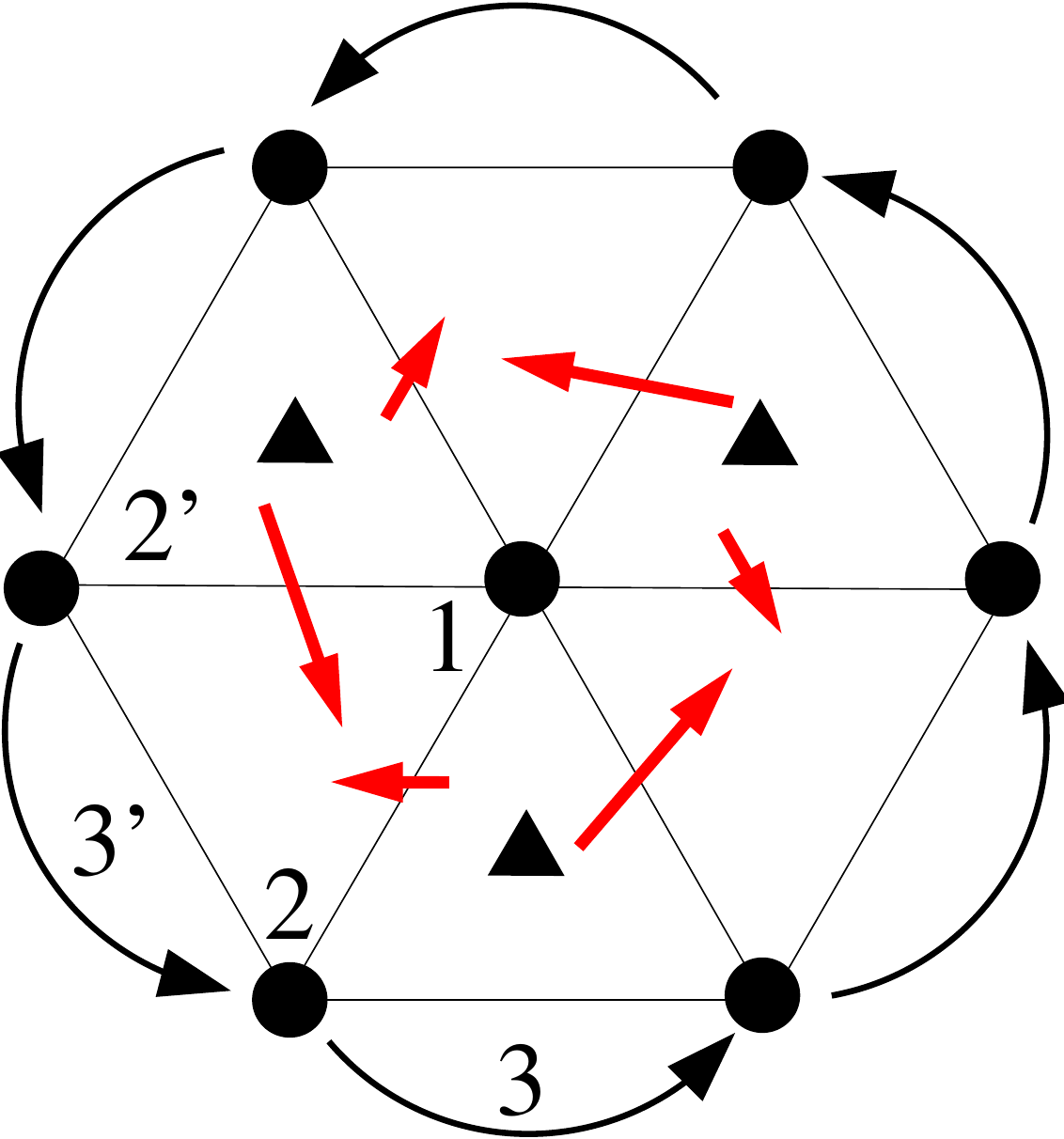}\;(b)
\caption{\label{fig:PtMnCu_3spin_DMI_inplain} The plane-projected
 three-spin DMI-like interactions (TDMI)  between atoms 1 (in the
  center) and 2 (on the hexagon) in the presence of atom 3 (relative
  position shown by a bended arrow) in the (Pt/Mn/Cu)$_n$  multilayer. (a)
  represents the vectors when the third atom follows site 2(2') on the
  hexagon in the clock-wise direction and (b) when the third atom
  follows site 2(2') in the anti-clock-wise direction. The long vectors
  represent the interactions with magnitude 0.068 meV, while the
  short ones correspond to the interactions with
  0.032 meV.  The z-component of the interactions is -0.006 meV between atoms 1 and 2 (2')
  in (a) and 0.006  between atoms 1 and 2 (2') in (b). The triangles
  show the position of Pt atoms on the neighboring layer. 
}   
\end{figure}

However, in the case of systems without inversion symmetry, the
TDM interactions do not cancel each other and can play
a certain role in the formation of the magnetic ground state
configuration. This is 
demonstrated by calculations for (Pt/$X$/Cu)$_n$  multilayer
systems. Fig.\ \ref{fig:PtMnCu_3spin_DMI_inplain} shows corresponding
results for the (Pt/Mn/Cu)$_n$ multilayer system where, using a similar representation
as before, the arrows represent the 'vector' interactions (i.e. $\sim
(\hat{s}_1 \times \hat{s}_2)$) between atoms $1$ and $2$, controlled by
the third atom $3$. Obviously, 
 the direction of this interaction depends on the position of the third
 atom as one can see in Fig.\ \ref{fig:PtMnCu_3spin_DMI_inplain} (a) and
 (b). Moreover, the magnitude of this interaction follow the 3-fold 
 in-plane symmetry of the system, and is  comparable to that of the
 biquadratic interactions and is smaller by more than one order of magnitude
 when compared to the DMI  interactions.


\subsection{Chiral exchange: Three-spin exchange interactions}
\subsubsection{First-principles calculations}

\begin{figure}
\includegraphics[width=0.2\textwidth,angle=0,clip]{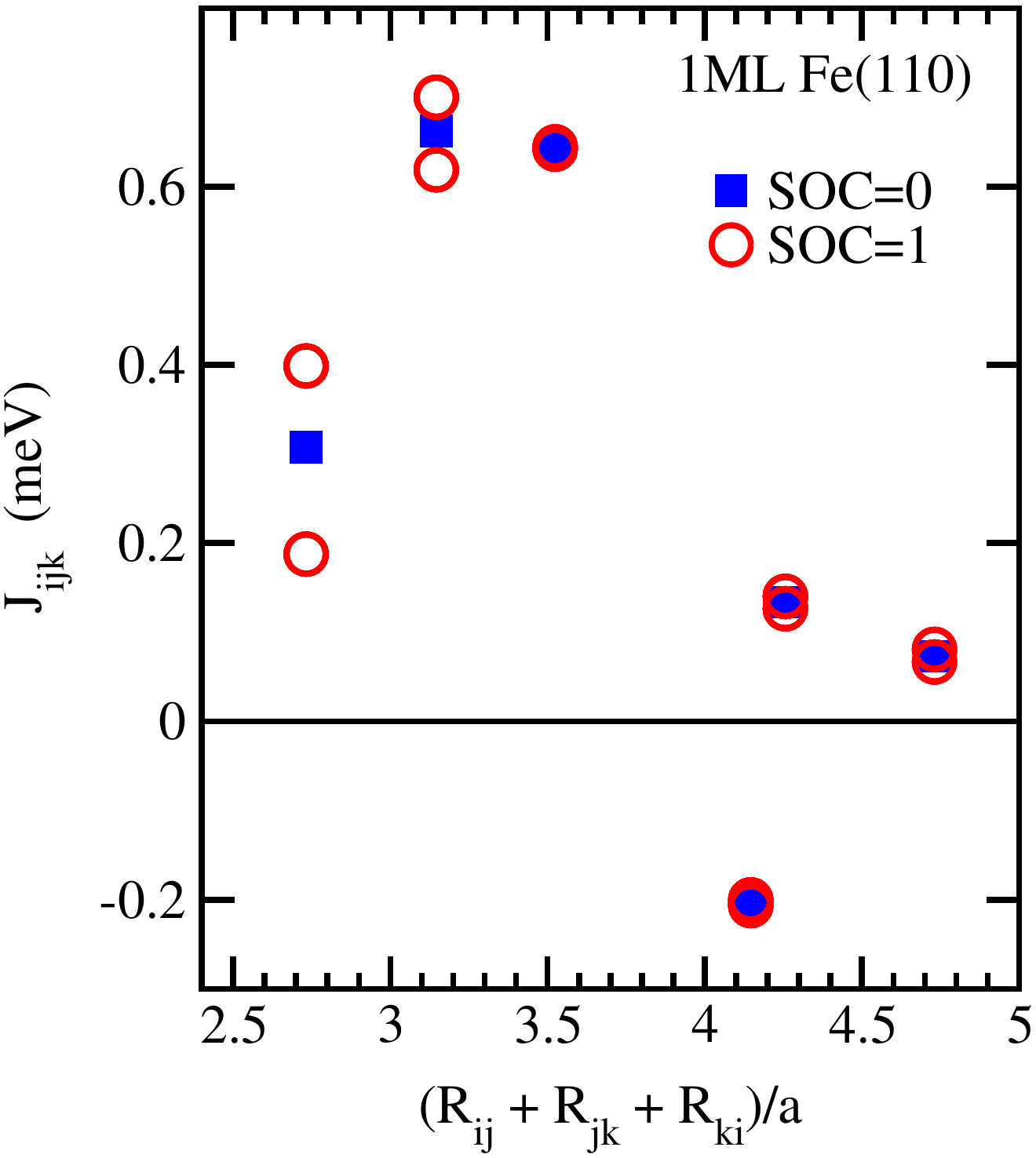}\;(a)
\includegraphics[width=0.2\textwidth,angle=0,clip]{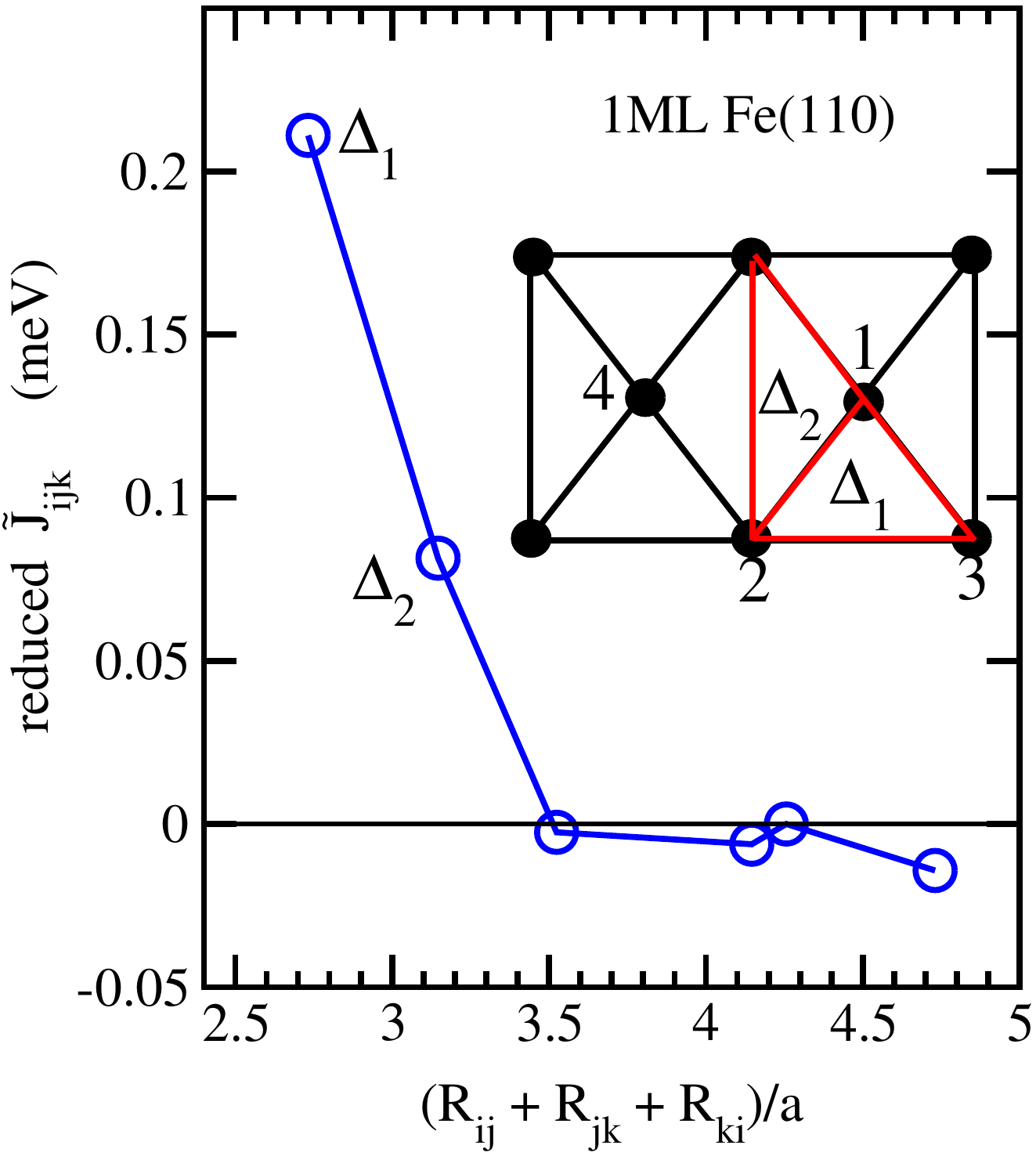}\;(b)
\caption{\label{fig:THEESPIN-Fe110} (a) Three-spin chiral exchange interaction
  parameters $J_{ijk}$ calculated for bcc Fe (110) with SOC = 0 (closed squares)
  and SOC=1 (open circles). (b) The reduced TCI $\tilde{J}_{ijk}$  for
  bcc Fe (110). Inset in (b) shows two triangles in Fe(110)
  system, having the smallest size $\Delta1$ and $\Delta2$. 
    }  
\end{figure}

Eq.\ (\ref{Eq:J_XYZ}) was used to calculate the three-spin interaction parameters for a
couple of representative 3D and 2D systems.
Fig.\ \ref{fig:THEESPIN-Fe110} (a)   represents the results on the
 TCI for 1ML bcc Fe(110). 
 The DMI and BDMI for this system vanish due to inversion
symmetry. The TCIs calculated without SOC (closed symbols) included
for various triangles of different  size do not change
 upon permutation of any two atoms, i.e.\ $J_{123} = J_{132}$. 
 As  discussed above, this leads to a cancellation of
the energy contribution due to these two terms. 
However, switching SOC on breaks the symmetry of the TCI
with respect to permutations, implying $J_{123} \neq J_{132}$.
Corresponding data are shown in Fig.\ \ref{fig:THEESPIN-Fe110} (a) by open symbols. 
As a consequence, the contribution due to the TCI
 to the magnetic energy of the system
should in general be finite. 

For further discussions it 
is convenient to introduce reduced TCI parameters defined as
$\tilde{J}_{ijk} = J_{ijk} - J_{ikj}$ for
counter-clock-wise sequences of atoms $i, j, k$. Corresponding
results for $\tilde{J}_{ijk} $  for 1ML Fe(110) are shown in 
Fig.\ \ref{fig:THEESPIN-Fe110} (b). 
In this case the energy term $ H^{(3)} $ given in
 Eq.\ (\ref{Eq_Heisenberg_3-spin})
associated with $\tilde{J}_{ijk}$ can be written as follows 
\begin{eqnarray}
  H^{(3)} &=&  - \sum_{i \neq j\neq k}
              \tilde{J}_{ijk} \chi_{ijk} 
\label{Eq_Heisenberg_TCI-reduced}
\end{eqnarray}
with the scalar spin chirality $\chi_{ijk}$, accounting only the
contributions due to the counter-clock-wise sequence of atoms $i, j,
k$. As one can see in Fig.\ \ref{fig:THEESPIN-Fe110} 
(b), the magnitude of the 
TCI decreases quickly with an increasing perimeter
of a triangle. 
As a consquence one may restrict the sum in  
Eq.\ (\ref{Eq_Heisenberg_TCI-reduced}) to the two smallest triangles.
 In this case, making use of the symmetry of $\tilde{J}_{ijk}$ with respect to cyclic permutation,
 i.e.\ accounting for 
  $\tilde{J}_{123} = \tilde{J}_{312} = \tilde{J}_{231} =
\tilde{J}_{\Delta}$, the expression for $H^{(3)}$
 can be further simplified to 
\begin{eqnarray}
  H^{(3)} &=&  -  \tilde{J}_{\Delta1} \sum_{(i,j,k)\in\Delta1}
              \chi_{ijk}  - \tilde{J}_{\Delta2} \sum_{(i,j,k)\in\Delta2} \chi_{ijk} \; .
\label{Eq_Heisenberg_TCI-reduced2}
\end{eqnarray}

The TCI  parameters calculated for 1ML Fe(110) can be compared to the
TDMI parameters, as both are non-vanishing in centrosymmetric
systems. Considering the smallest triangle $\Delta1$, we have for
the  TCI
$\tilde{J}_{\Delta1} = 0.21$ meV, while the
z-component of the TDMI (the only non-vanishing one) 
between spin moments 1 and 2  is found to be
$D^z_{1213} = 0.0014$ meV and $D^z_{1214} = -0.039$
(the positions of the third  atoms are shown in
Fig.\ \ref{fig:THEESPIN-Fe110})
 demonstrating that the  TDMI is much weaker  when compared to the  TCI. 

On the other hand, the origin of the TCI can be discussed in more detail
on the basis of the 
spin-chiral interaction introduced by Grytsiuk et al. \cite{GHH+20}.
According to this approach, the  TCI can be associated with
a topological orbital moment $\vec{L}^{TO}_{ijk}$ induced on the atoms of each triangle
\cite{LFBM18a} due to the non-coplanar orientation of
the  spin magnetic moments.
According to   Refs.\ \cite{GHH+20,FHZ+20,TOY+01}, 
one has $\vec{L}^{TO}_{ijk} = \kappa^{TO}_{ijk} 
\chi_{ijk} \vec{n}_{ijk}$, where  $\kappa^{TO}_{ijk}$ is the topological
orbital susceptibility, and $\vec{n}_{ijk}$ is the normal to the
triangle $\Delta_{i,j,k}$. 
Accounting for the SOC, the
 interaction energy  between spin moments on the atoms with
corresponding topologically induced orbital moments can   
be written as $~\sim \sum_{i} \xi \vec{L}^{TO}_{i}\cdot \vec{s}_1$ where
$\xi$ is the spin-orbit interaction parameter for atom $i$ having the spin
moment $\vec{s}_i$. 
In the case of all  atoms being equivalent, 
the sum  can be written as $~\sim
\xi \kappa^{TO}_{ijk} \chi_{ijk} (\vec{n}_{ijk}\cdot \langle
\vec{s} \rangle )$, with $\langle \vec{s} \rangle = \vec{s}_i +
\vec{s}_j + \vec{s}_k $. 
This expression shows that the dependence
of the three-spin interactions on the orientation of spin magnetic moments
is given by their projection on the normal vector of a triangle
that is proportional to the
flux of local spin magnetization through the triangle area.

\begin{figure}
\includegraphics[width=0.2\textwidth,angle=0,clip]{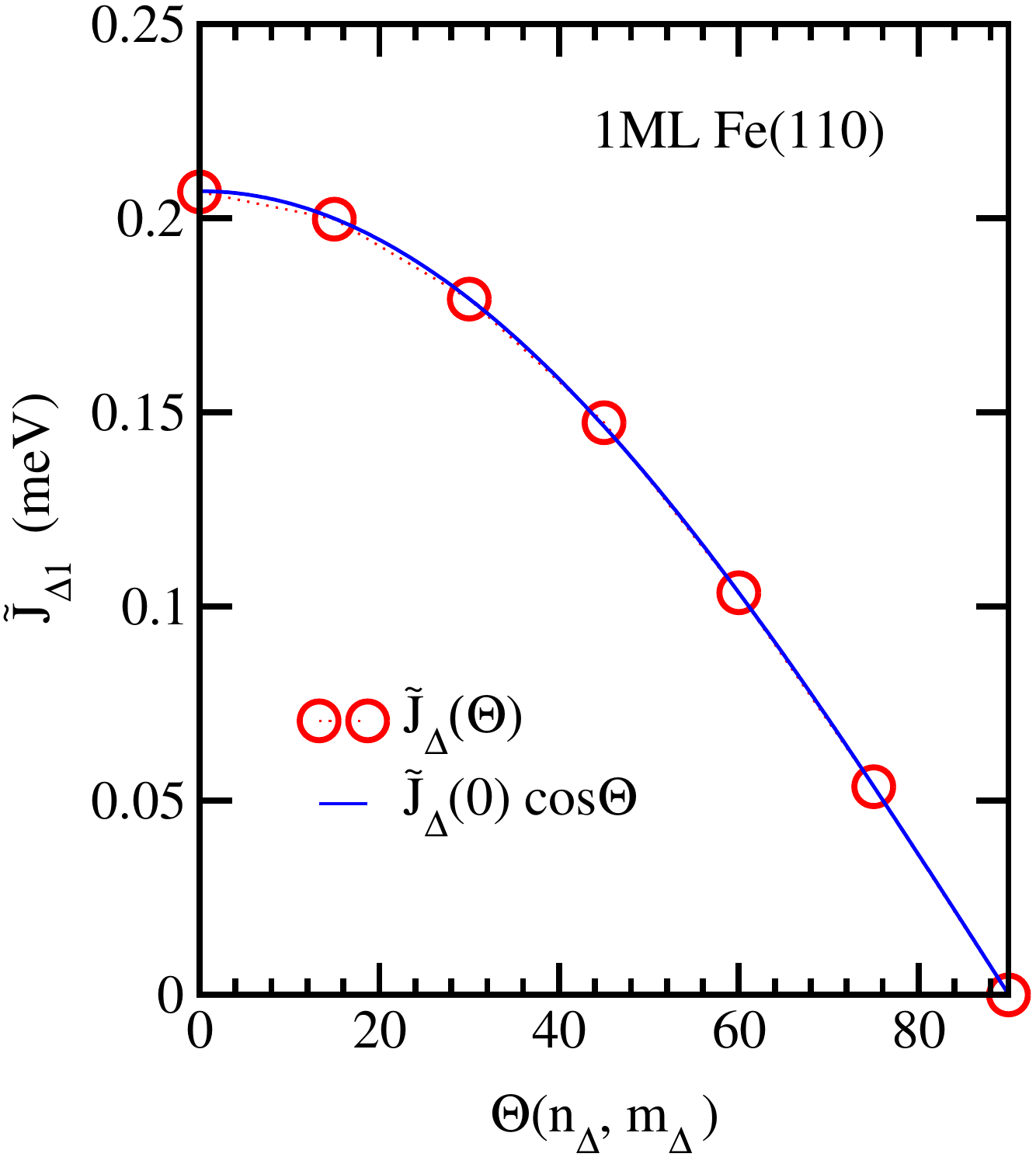}\;(a)
\includegraphics[width=0.2\textwidth,angle=0,clip]{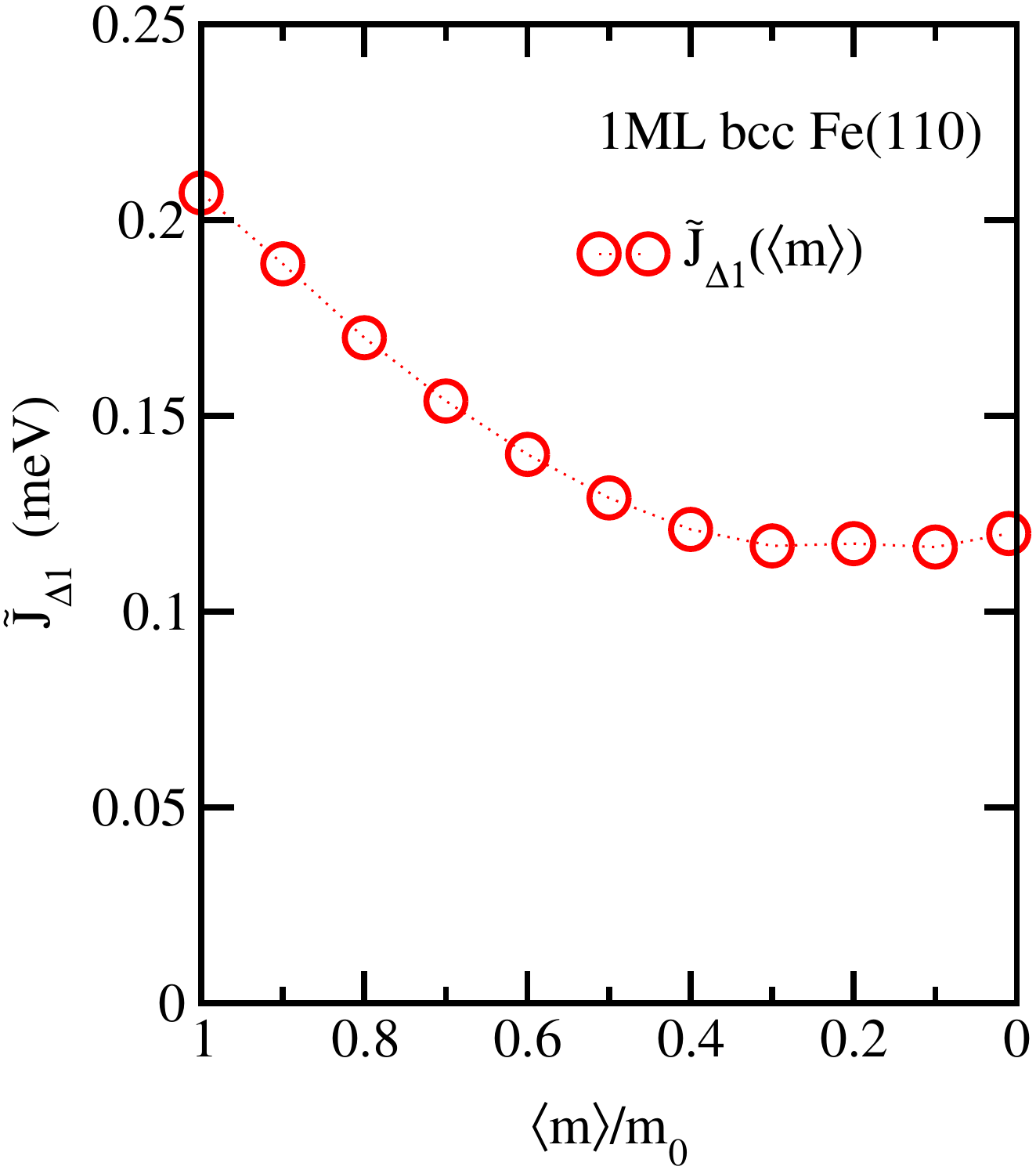}\;(b)
\caption{\label{fig:THEESPIN-Fe110-Theta} Three-spin chiral exchange interaction
  parameters $\tilde{J}_{\Delta}$ calculated for the smallest triangle $\Delta1$ in 1 ML of
  Fe(110): (a) as a function of the orientation of magnetization with
  respect to normal vector of triangle, and (b) as function of average
  reduced magnetization $|\langle \vec{m} \rangle|/m_0$. 
    }  
\end{figure}

The dependence of the  TCI on the orientation of the
 magnetization has been
  investigated also for 1ML bcc Fe (110).
The parameters calculated for the smallest triangle,
$\tilde{J}_{\Delta1}(\theta)$ are plotted in Fig.\
\ref{fig:THEESPIN-Fe110-Theta}(a) as a function of the angle $\theta$
between the  magnetization and the triangle normal. 
As one can see, the results given by circles are in  perfect
agreement with the  function $\tilde{J}_{\Delta1}(0) cos\theta$, in line
with the discussion above.

An increase in
 temperature results in general in an increase of magnetic disorder in
the system leading to a corresponding decrease of the net magnetization
that finally vanishes at the critical temperature $T_C$. 
In order to investigate the
dependence of the TCI parameter $\tilde{J}_{ijk}$ 
on the normalized magnetization seen as the order parameter, 
calculations have been performed for the partially ordered state described
by means of  the relativistic disorder local moment (RDLM) approach
\cite{GPS+85, EMC+15}. In these calculations the exchange parameters
$\tilde{J}_{ijk}$ are associated with the energy change due to a tilting
of the spin magnetic moments of a triple of atoms with respect to the
magnetization direction of the reference system, accounting for partial
(as well as full) magnetic disorder of all surrounding atoms. Fig.\
\ref{fig:THEESPIN-Fe110-Theta}(b) represents the results for the smallest
triangle, showing a decrease of $\tilde{J}_{\Delta1}$ by about twice
approaching $|\langle \vec{m} \rangle|/m_0  
\approx 0.5$, and staying nearly unchanged for larger disorder
i.e.\ higher temperature.
 The non-vanishing behavior of the TCI parameters can be
 understood by keeping in mind their dependence  on the 
 local magnetic order.

\begin{figure}
  \includegraphics[width=0.4\textwidth,angle=0,clip]{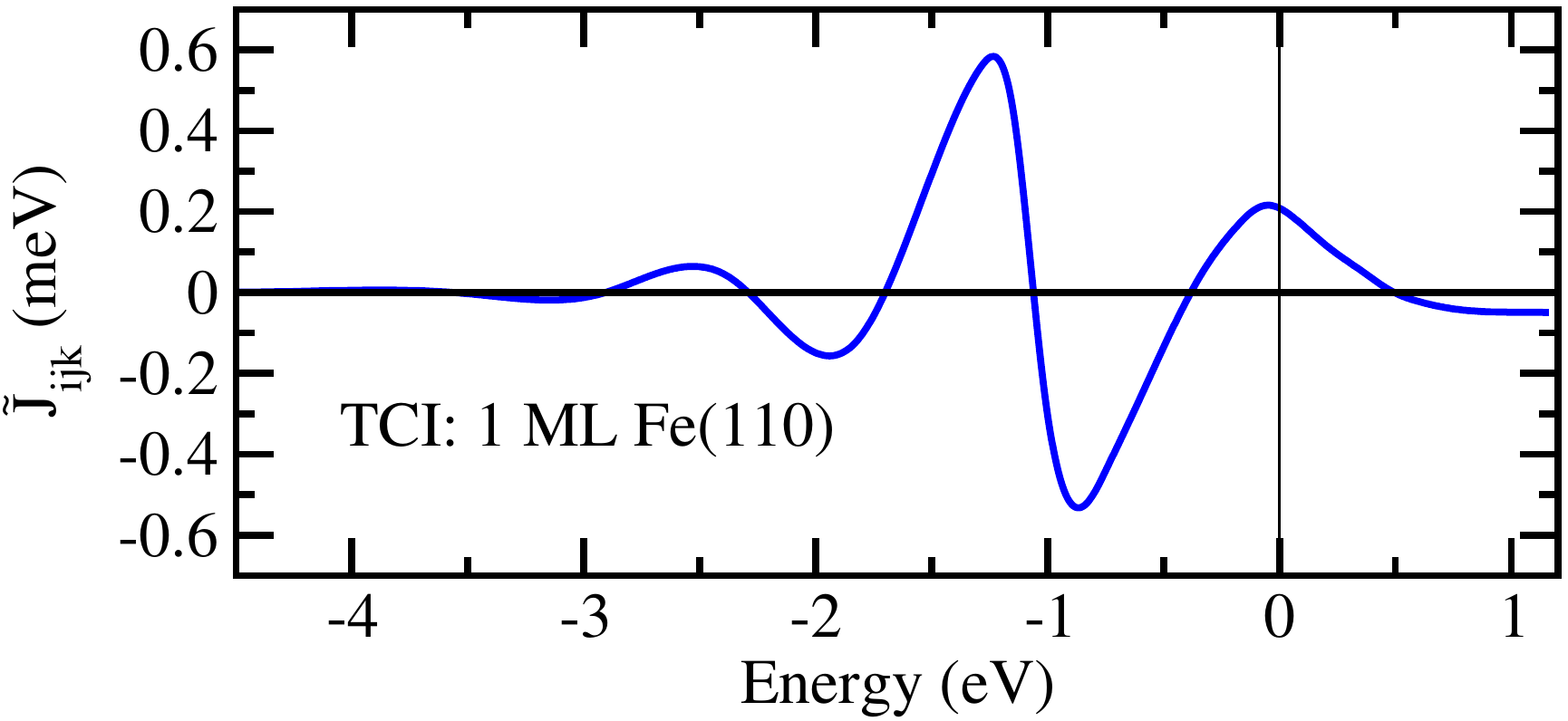}  
\caption{\label{fig:Fe110_TCI}  The reduced three-spin chiral
  interactions $\tilde{J}_{ijk}(E)$ for the smallest triangle $\Delta1$ in 1 ML of
  Fe(110) as a function of  occupation. 
  }  
\end{figure}

 In Fig.\ \ref{fig:Fe110_TCI} one can see in addition
  a strong dependence of the 
  TCI on the occupation of the  valence states, with the magnitude reaching
  its maximum  below the true Fermi energy.  

\begin{figure}
\includegraphics[width=0.44\textwidth,angle=0,clip]{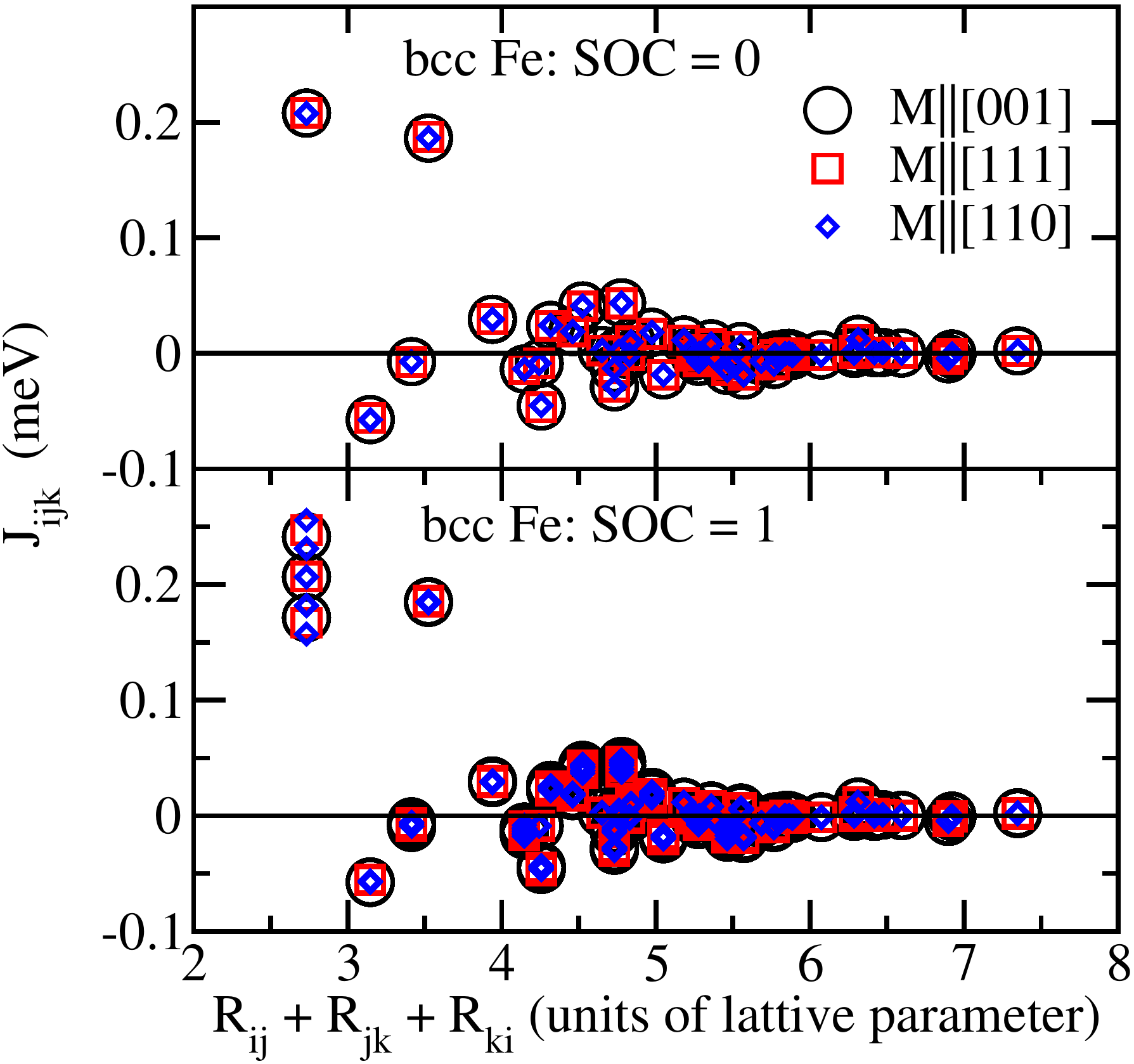}\;(a)
\includegraphics[width=0.2\textwidth,angle=0,clip]{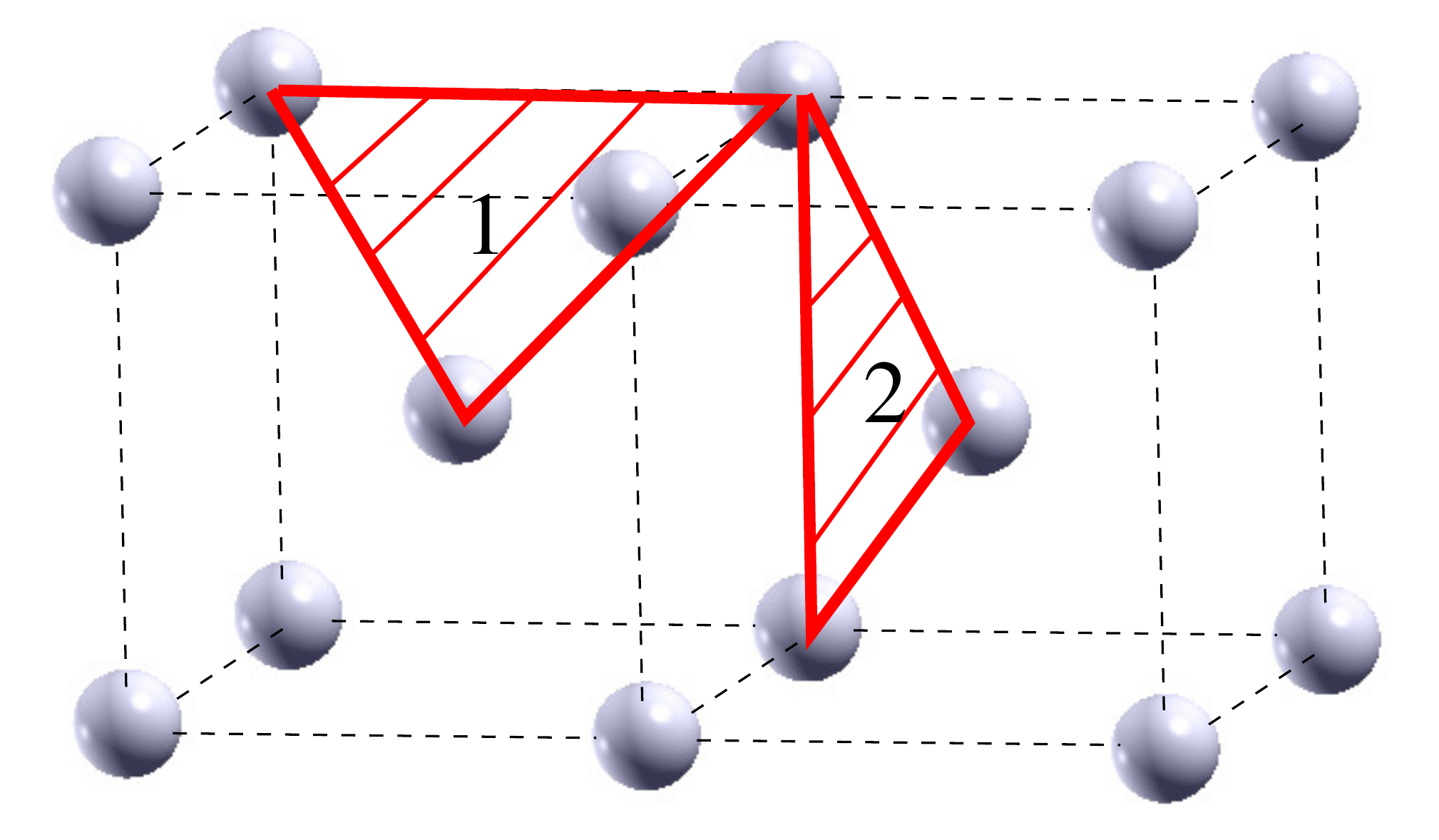}\;
\includegraphics[width=0.2\textwidth,angle=0,clip]{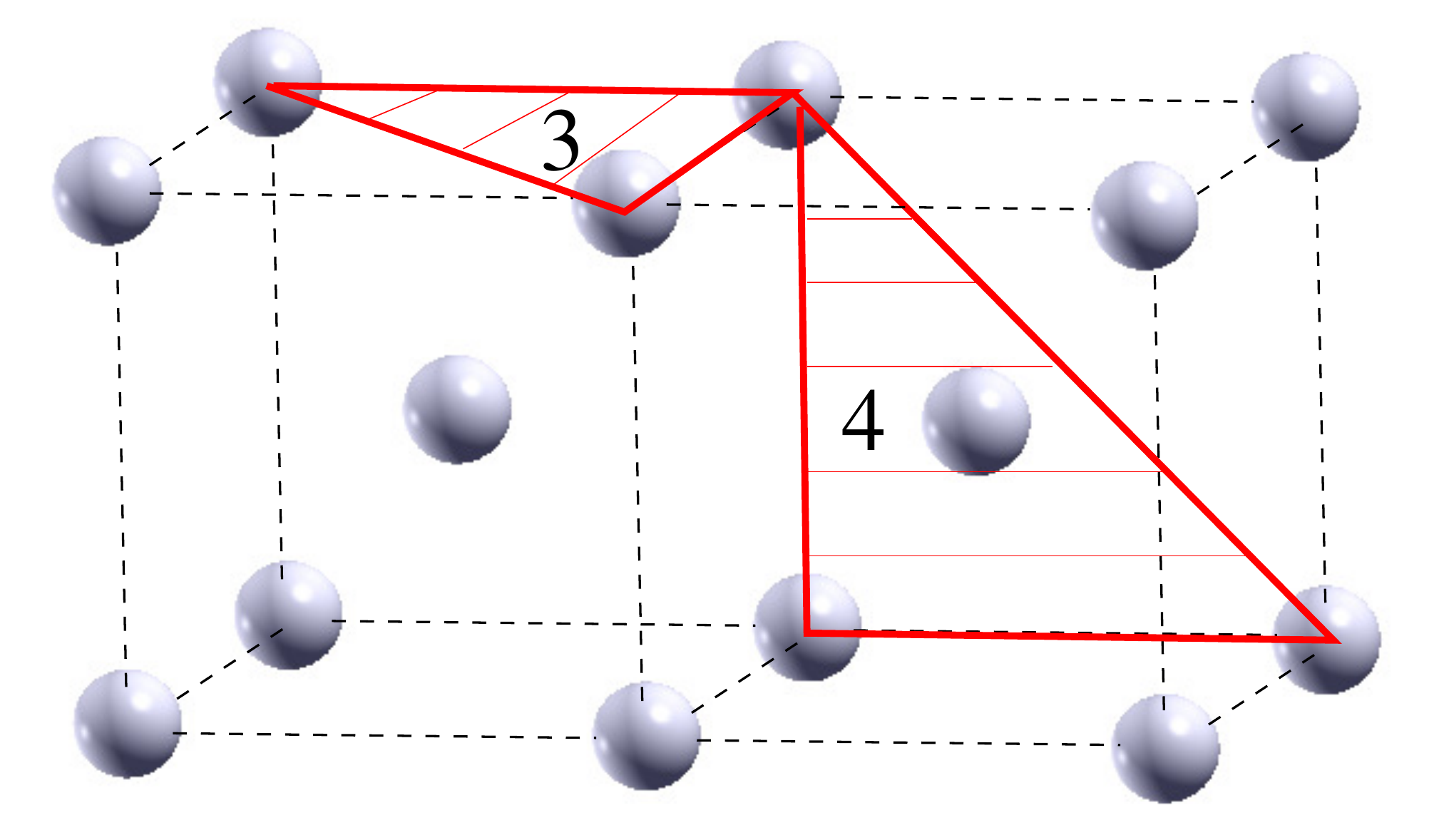}\;(b)
\caption{\label{fig:THEESPIN-Fe} (a) Three-spin chiral exchange interaction
  parameters calculated for bcc Fe without  (top panel) and with SOC   (bottom panel), respectively, for three different
  magnetization   directions: [001] {circles}, [111] (squares) and [110]
  (diamonds). (b) The reduced TCI calculated for the triangles 1,2,3,4,
  assuming magnetization along z direction:
  $ \tilde{J}^1_{\Delta} = 0.07$ meV,  $ \tilde{J}^2_{\Delta} = 0.0$
  meV,   $ \tilde{J}^3_{\Delta} = 0.005$ meV,   $ \tilde{J}^4_{\Delta} = 0.0$ meV.    }  
\end{figure}

 Fig.\ \ref{fig:THEESPIN-Fe}(a) shows the TCI parameters calculated for
 centrosymmetic bcc Fe, as a function of the perimeter of the considered
 triangles for three different orientations of the magnetic moment: [001]
 (circles), [111] (squares) and [110] (diamonds).
  The results given in Fig.\ \ref{fig:THEESPIN-Fe}(a) (top panel) have 
  been   obtained
 without including the SOC.
 As one can see, the TCI parameters  do not depend on the
  orientation of the   magnetization. 
  However, according to the discussions above,
 this leads to a cancellation of their contribution to the energy upon
 summation over all sites in the lattice. 
 In the presence of SOC, on the other hand, the
 interactions shown in Fig.\ \ref{fig:THEESPIN-Fe}(a) (bottom panel)
 change their magnitude upon permutation, i.e.\ $J_{123} \neq J_{132}$.
 The corresponding 
 dependence on the relative orientation of the magnetization and
 the triangle normal is also given 
  in Fig.\ \ref{fig:THEESPIN-Fe}(b) for four
 different triangles, assuming the  magnetization oriented along 
 the $z$ axis. In the
 case of triangles 1 and 3 the TCI parameters $\tilde{J}_{ijk}$ are
 non-zero. This is not the case for the triangles 2 and 4 as the
 flux of the magnetization through their area is equal to zero.
This changes however due to change of the magnetization toward the 
[111] and [110] crystallographic directions (see
Fig.\ \ref{fig:THEESPIN-Fe}(a), bottom panel).

\begin{figure}
\includegraphics[width=0.2\textwidth,angle=0,clip]{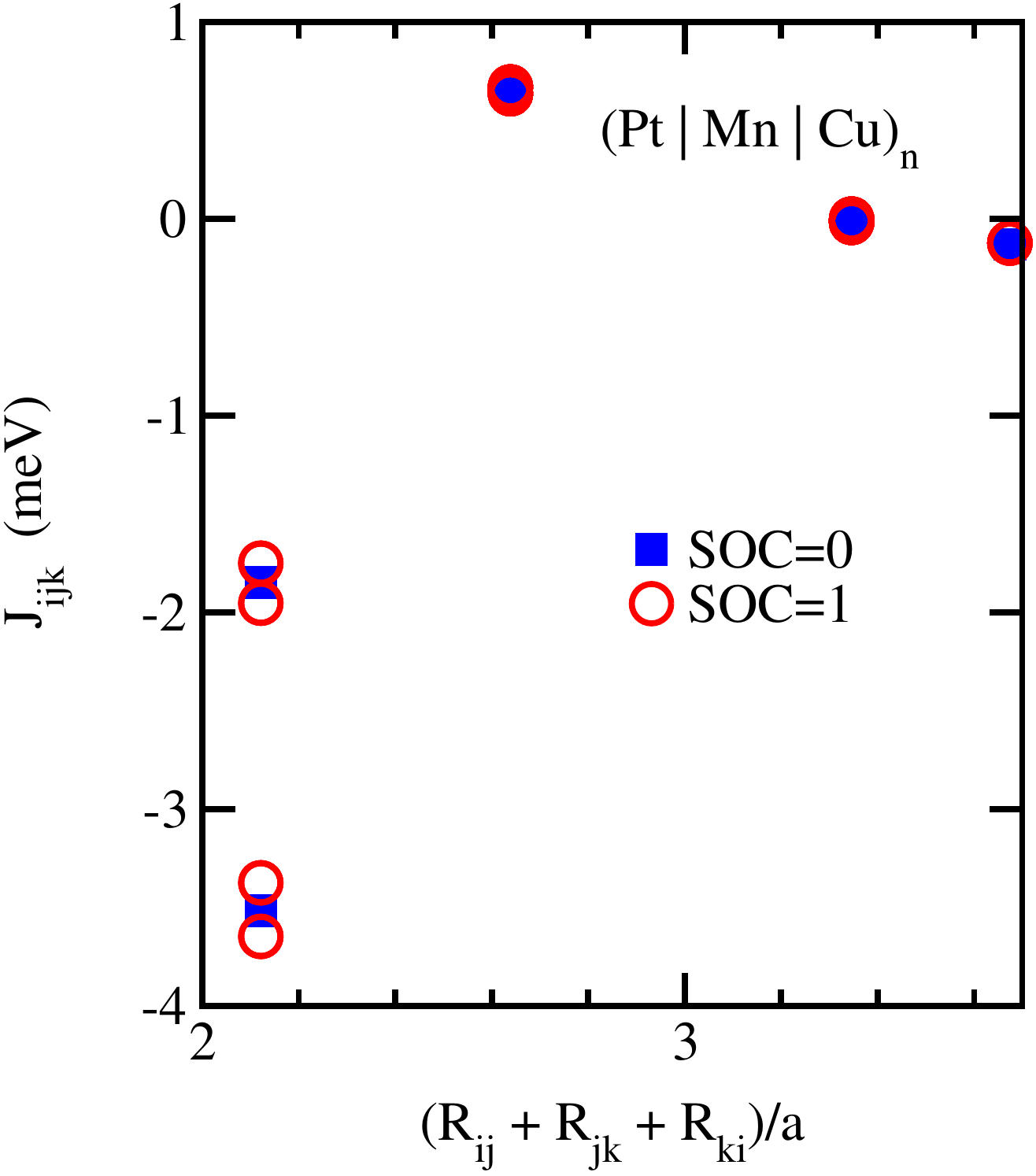}\;(a)
\includegraphics[width=0.2\textwidth,angle=0,clip]{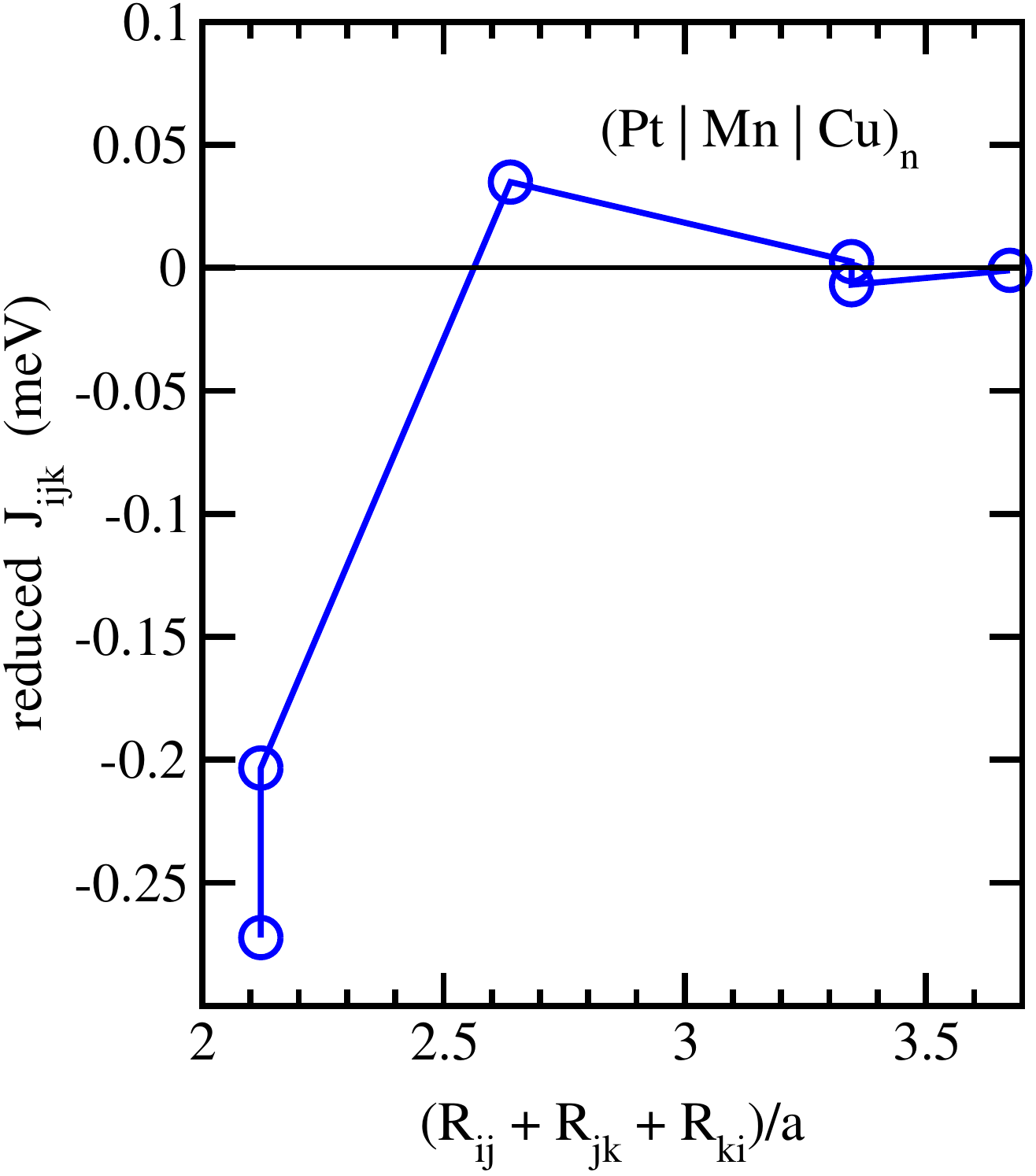}\;(b)
\includegraphics[width=0.2\textwidth,angle=0,clip]{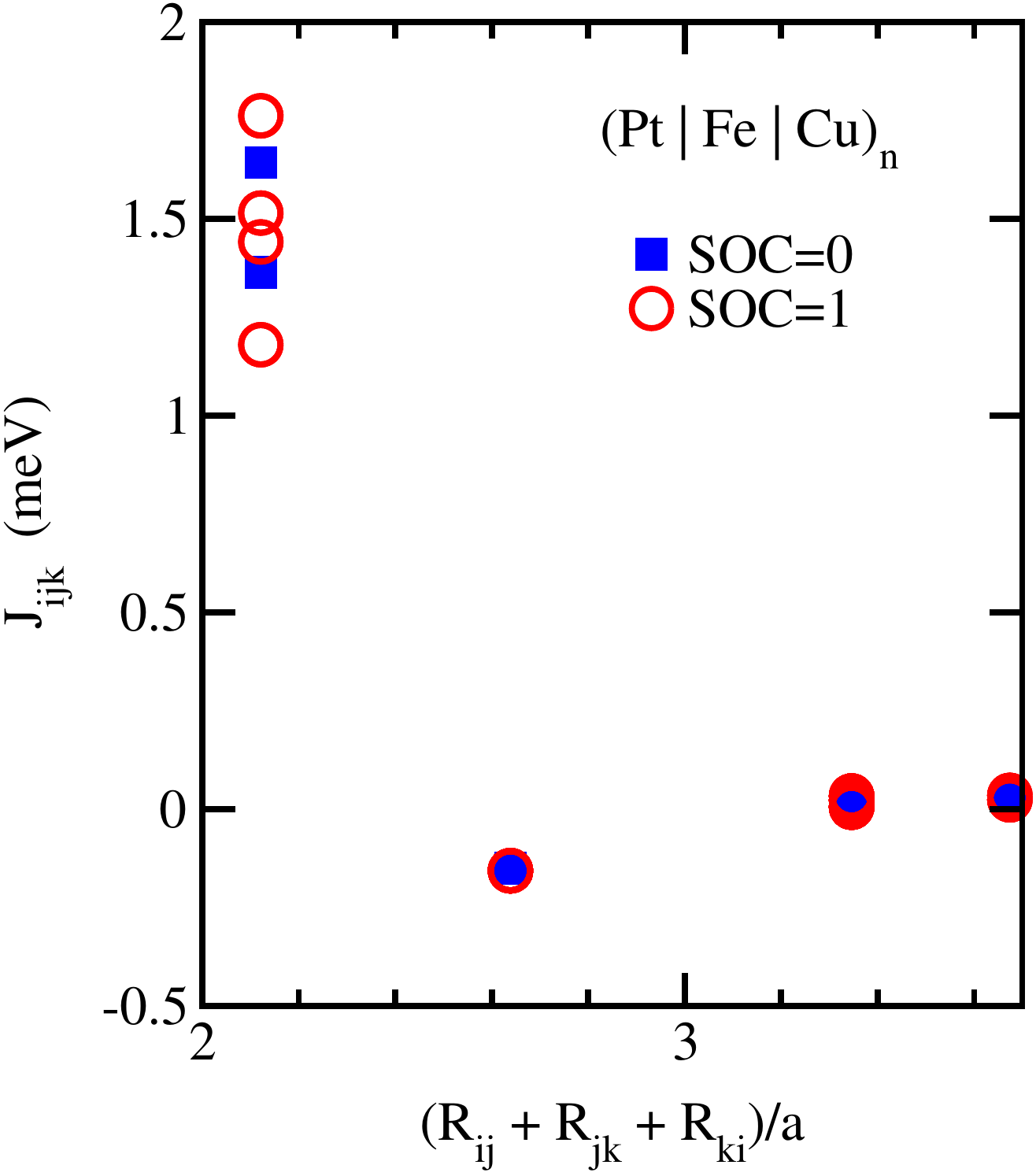}\;(c)
\includegraphics[width=0.2\textwidth,angle=0,clip]{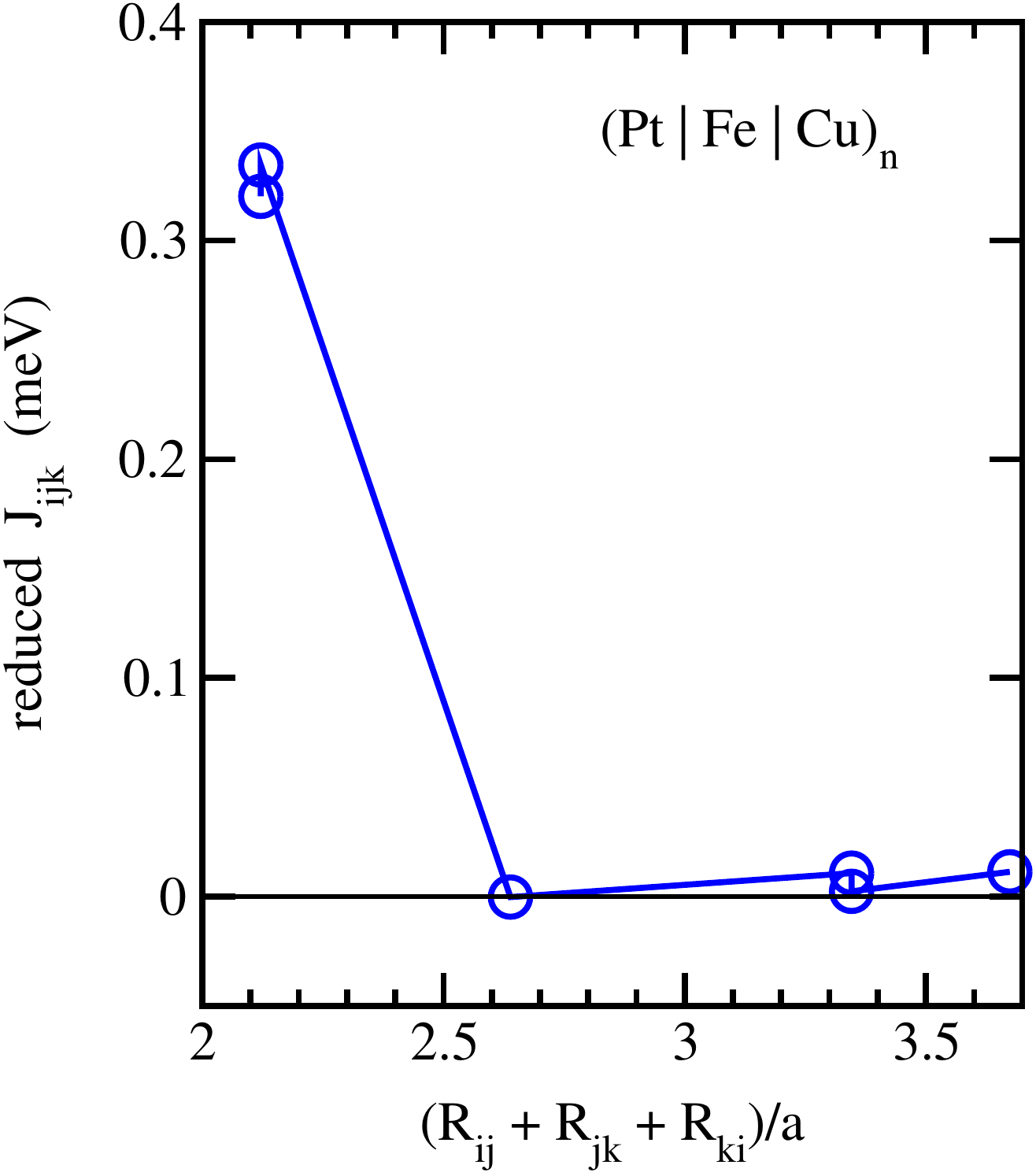}\;(d)
\includegraphics[width=0.2\textwidth,angle=0,clip]{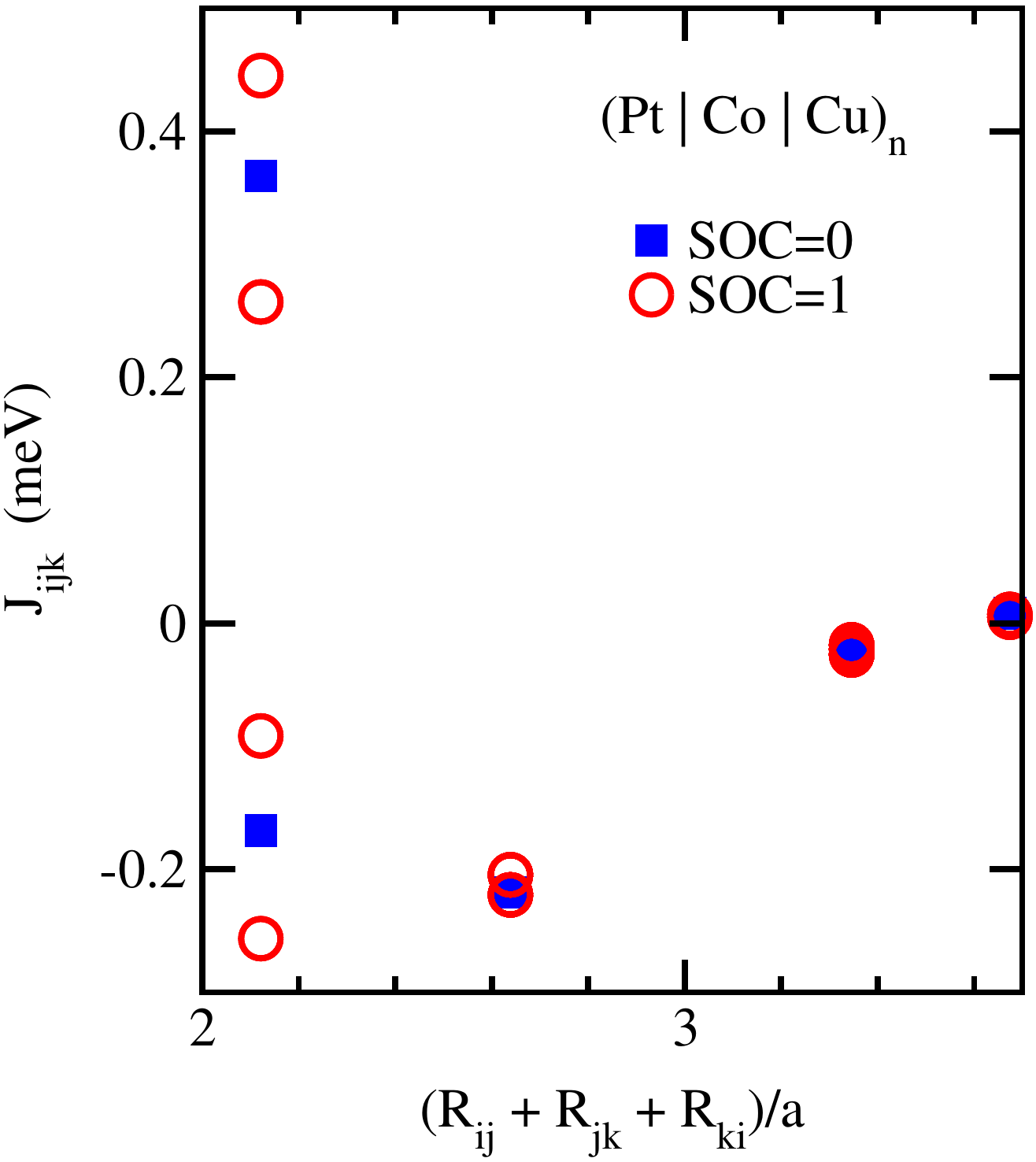}\;(e)
\includegraphics[width=0.2\textwidth,angle=0,clip]{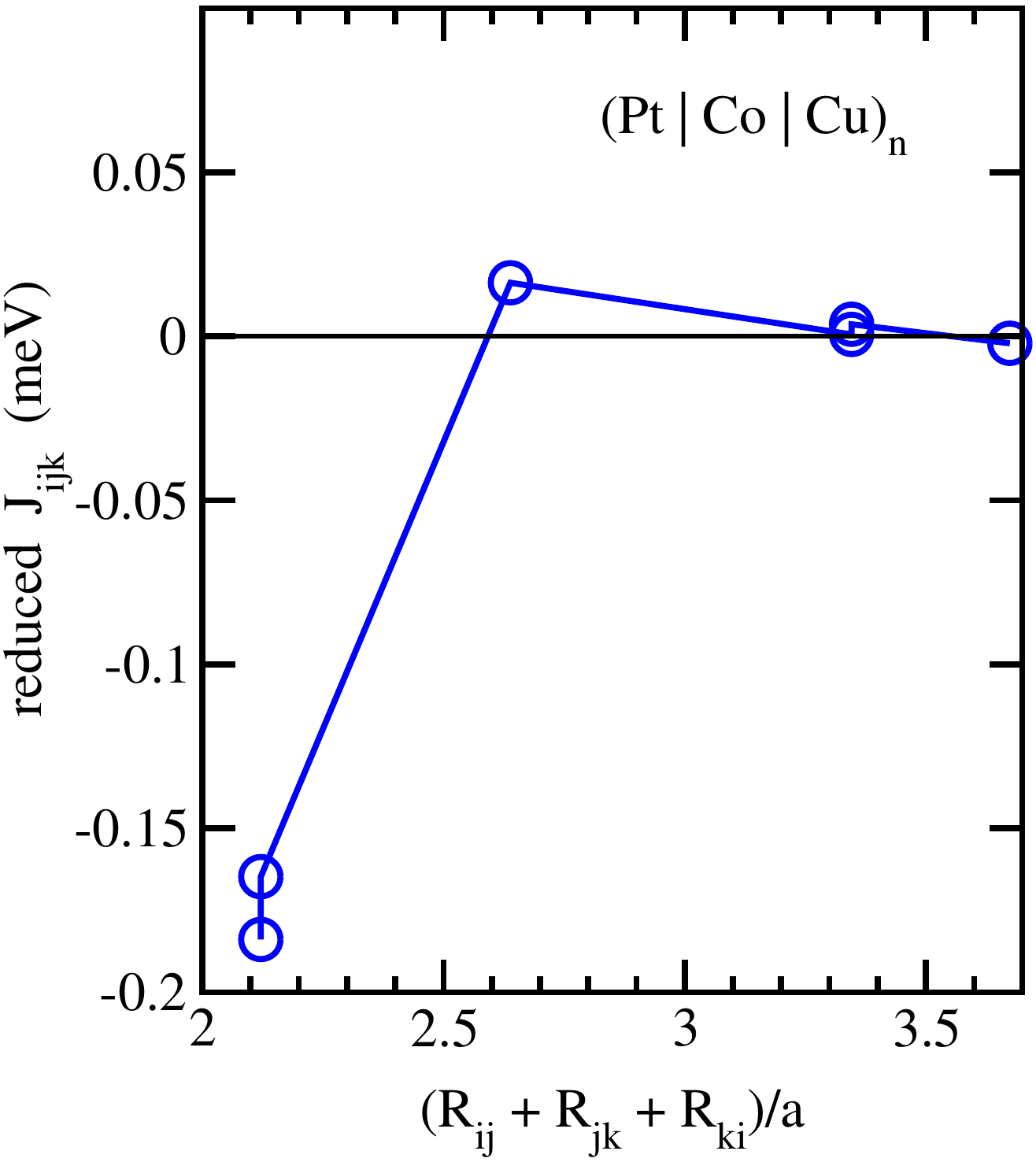}\;(f)
\caption{\label{fig:PtFeCu_DMI} (a),(c),(e) Three-spin exchange interaction
  parameters $J_{ijk}$ between the 3d-atoms $X$ in (Pt/$X$/Cu)$_n$ multilayer,
  calculated using Eq.\ (\ref{Eq:J_XYZ}) without 
  (closed symbols) and with SOC (open symbols) included, respectively,
  plotted as a function of
  the total length for a  3-atomic cluster, $R_{ij} + R_{jk} +  R_{ki}$,
  created by the coupling atoms. (b), (d) and (f) represent the
  corresponding reduced three-spin interaction parameters
  $\tilde{J}_{ijk} = J_{ijk} -J_{ikj}$.
 }  
\end{figure}

Fig.\ \ref{fig:PtFeCu_DMI} shows the three-spin chiral
interaction between 3$d$ atoms in the (Pt/$X$/Cu)$_n$ multilayer system,
with $X$ = Mn (a), Fe (c) and Co (e), calculated without (closed
squares) and with SOC  (open circles), respectively. 
The reduced parameters $
\tilde{J}_{ijk}$ are plotted in Fig. \ref{fig:PtFeCu_DMI} (b), (d) and
(f) for  Mn, Fe and Co, respectively. As one can see, the dominating
exchange parameters $\tilde{J}_{\Delta1}$ 
are associated with  the smallest
triangle. Their magnitude is rather close for all three systems, while the
sign of $\tilde{J}_{\Delta1}$ in the case of (Pt/Fe/Cu)$_n$ is opposite
to that for the  two other systems.
In addition, one can see a weak dependence of the TCI on the arrangement, as 
they are slightly different for the triangles with neighboring Pt atom,
$\tilde{J}^{Pt}_{\Delta1}$, and Cu $\tilde{J}^{Cu}_{\Delta1}$ atoms.

\begin{figure}
\includegraphics[width=0.32\textwidth,angle=0,clip]{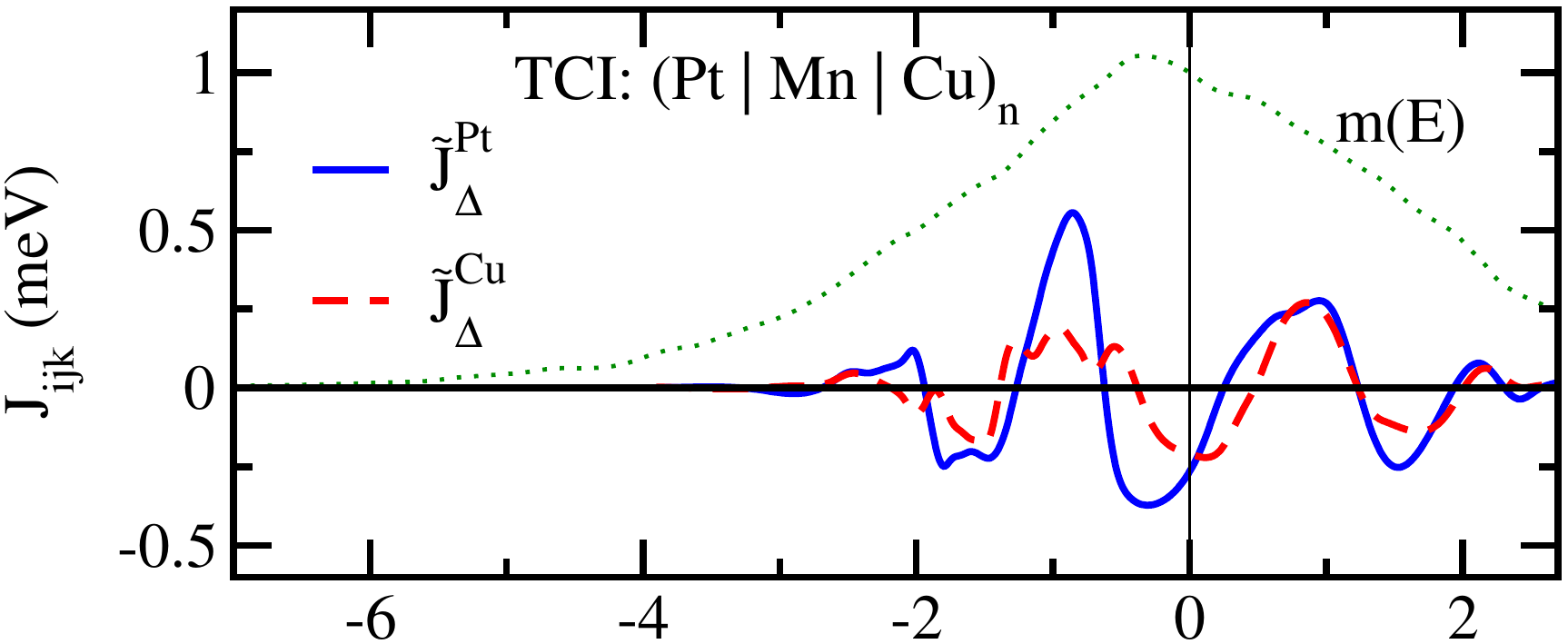}\;(a)
\includegraphics[width=0.32\textwidth,angle=0,clip]{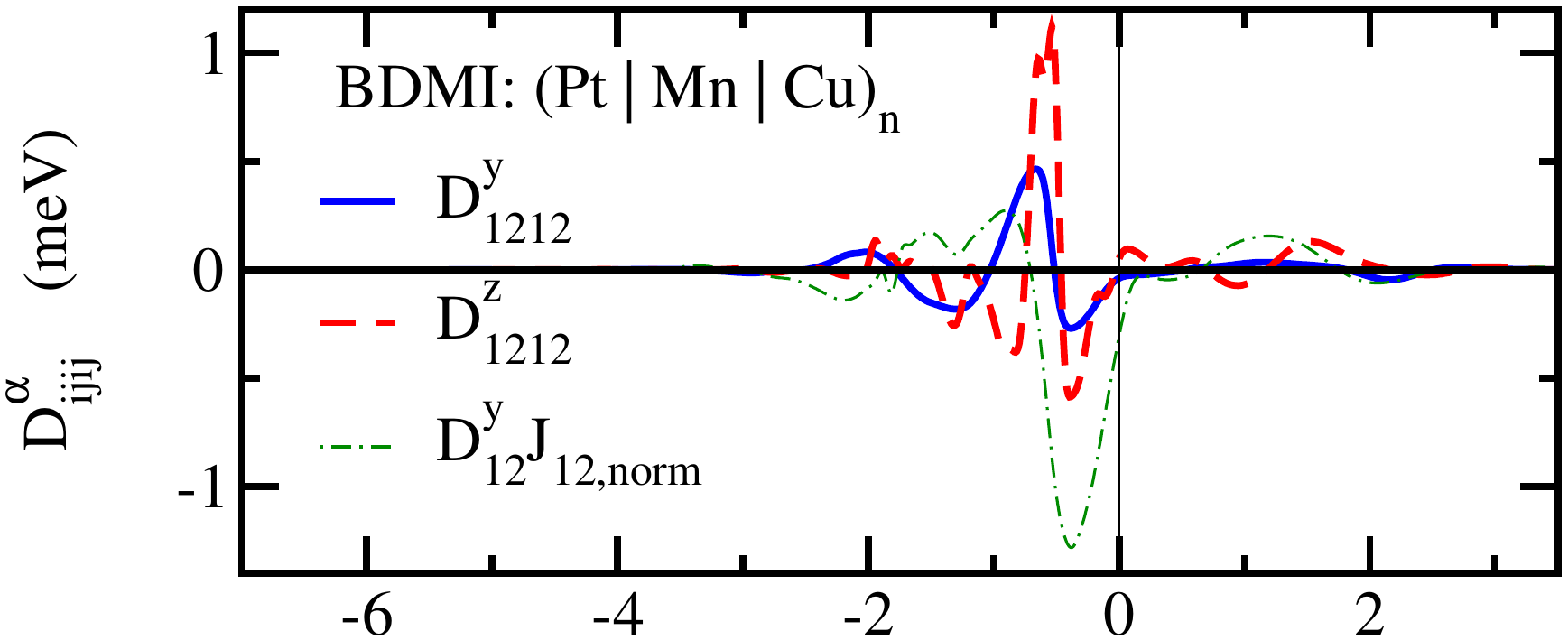}\;(b)
\includegraphics[width=0.32\textwidth,angle=0,clip]{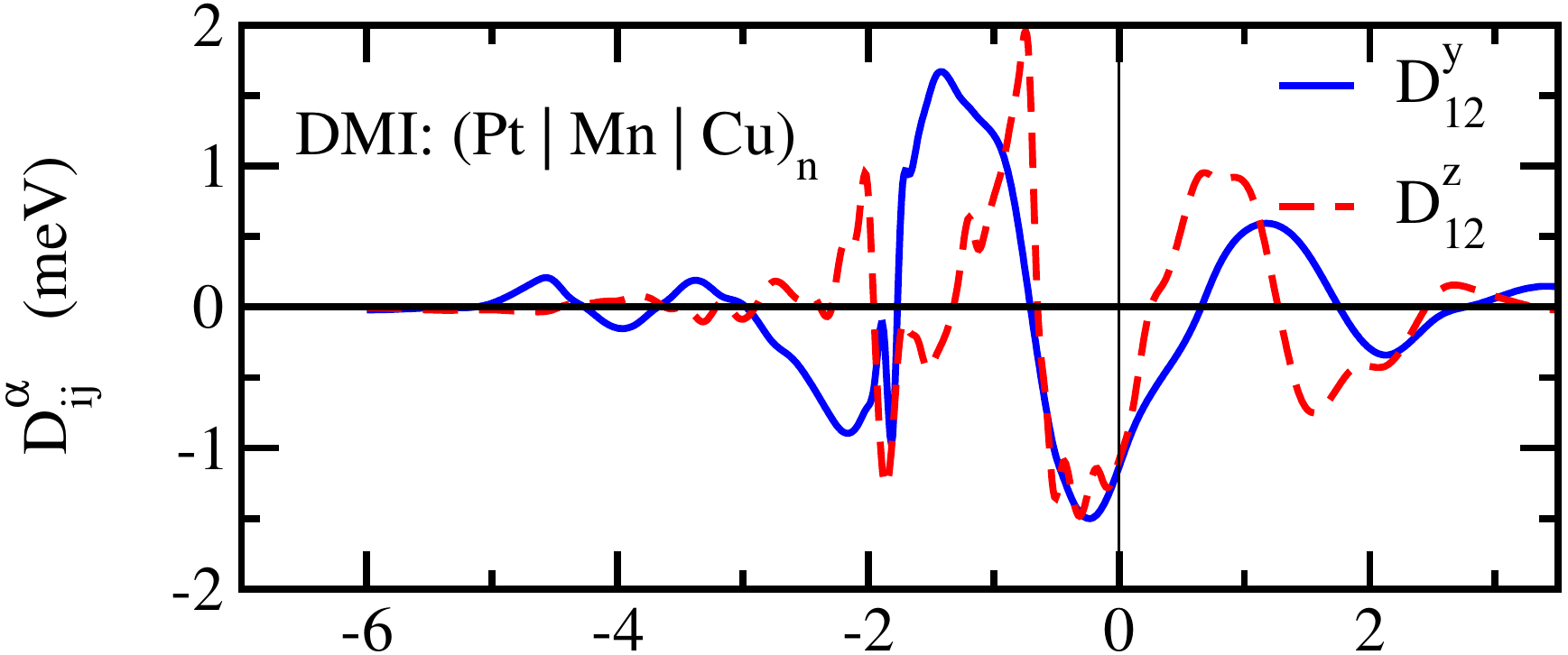}\;(c)
\includegraphics[width=0.32\textwidth,angle=0,clip]{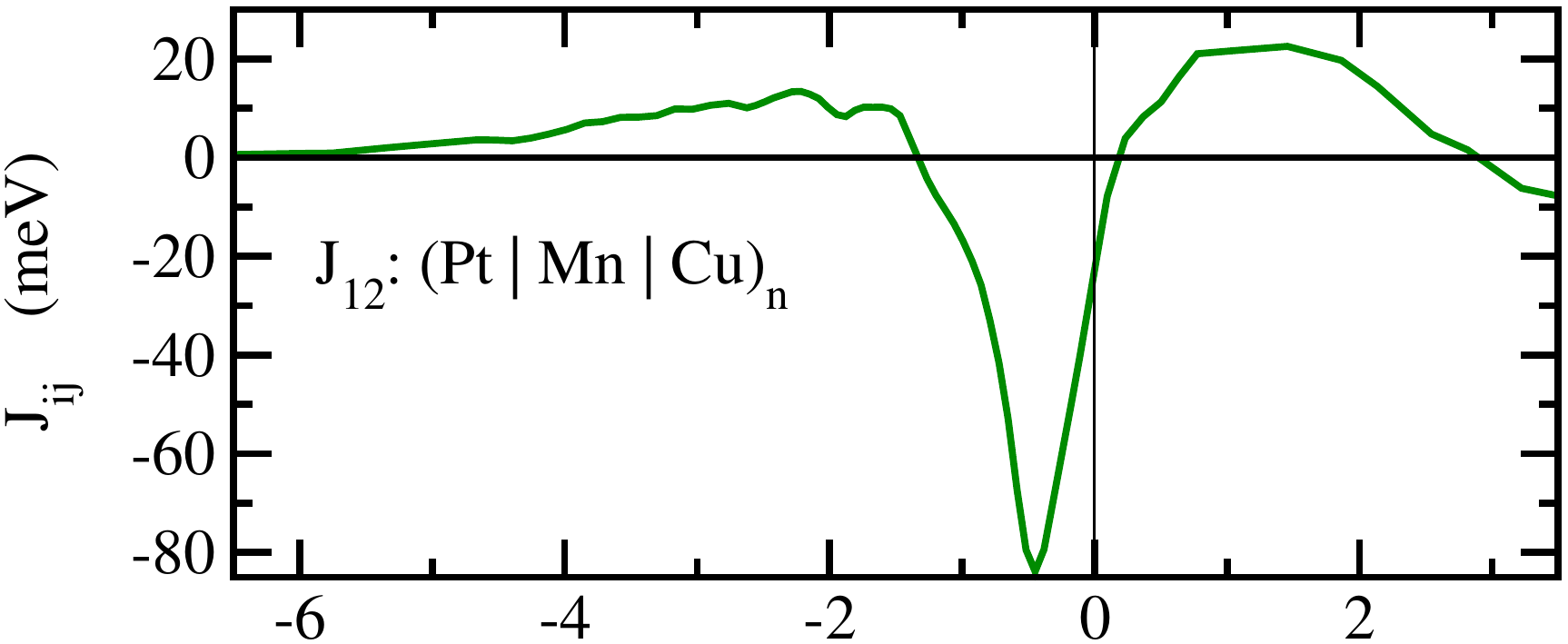}\;(d)
\includegraphics[width=0.32\textwidth,angle=0,clip]{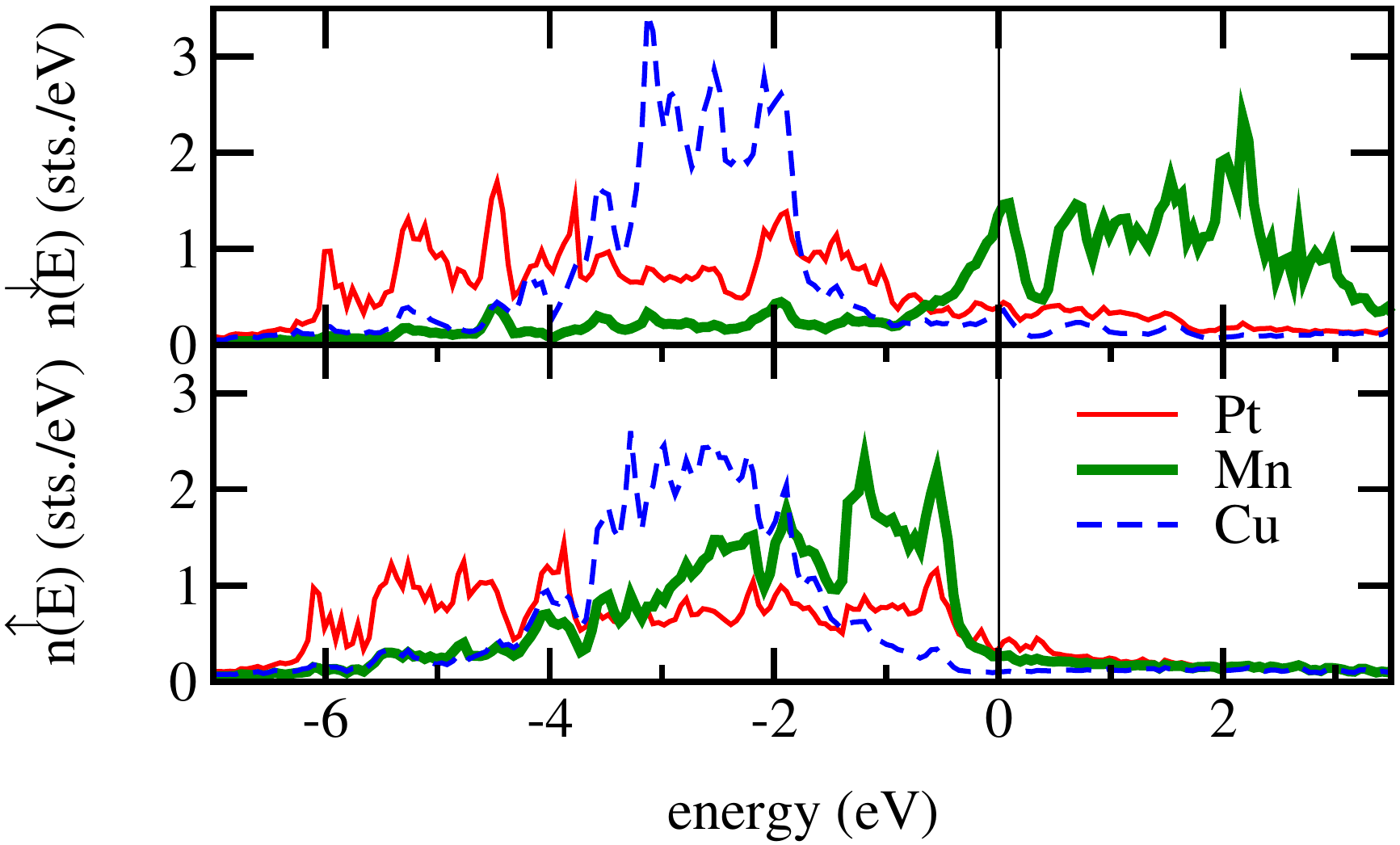}\;(e)
\caption{\label{fig:Pt/Mn/Cu_3spin-E}  Three-spin interaction parameters
  ${\tilde J}_{ijk}(E)$ for the smallest Mn-triangles in (Pt/Mn/Cu)$_n$
  multilayer, centered at Cu (${\tilde J}^{Cu}_{\Delta}$) and Pt
  (${\tilde J}^{Pt}_{\Delta}$); the dotted line represents the Mn spin 
  magnetic moment $m(E) = M(E)/M_{Mn}$ ($M_{Mn} = 3.7 \mu_B$) as a function of the occupation (a);
  $y$-, $z$-components of the chiral biquadratic exchange interaction,
  $\vec{{\cal D}}_{1212}$ (b) and DM interactions, $\vec{D}_{ij}$ (c)
  between Mn atoms as a function of the occupation; (d) the isotropic exchange
  $J_{ij}(E)$, and (e) the element-projected DOS(E).
  }  
\end{figure}

\begin{figure}
\includegraphics[width=0.32\textwidth,angle=0,clip]{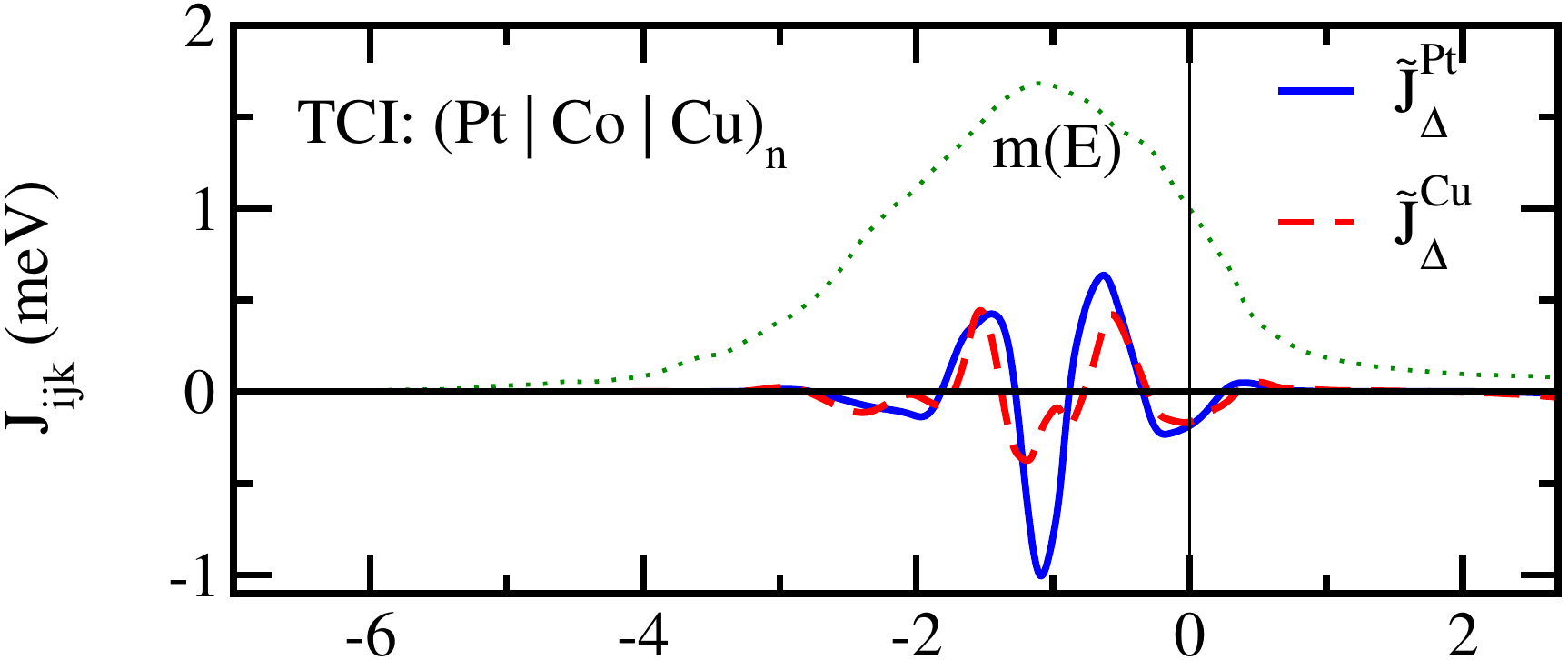}\;(a)
\includegraphics[width=0.32\textwidth,angle=0,clip]{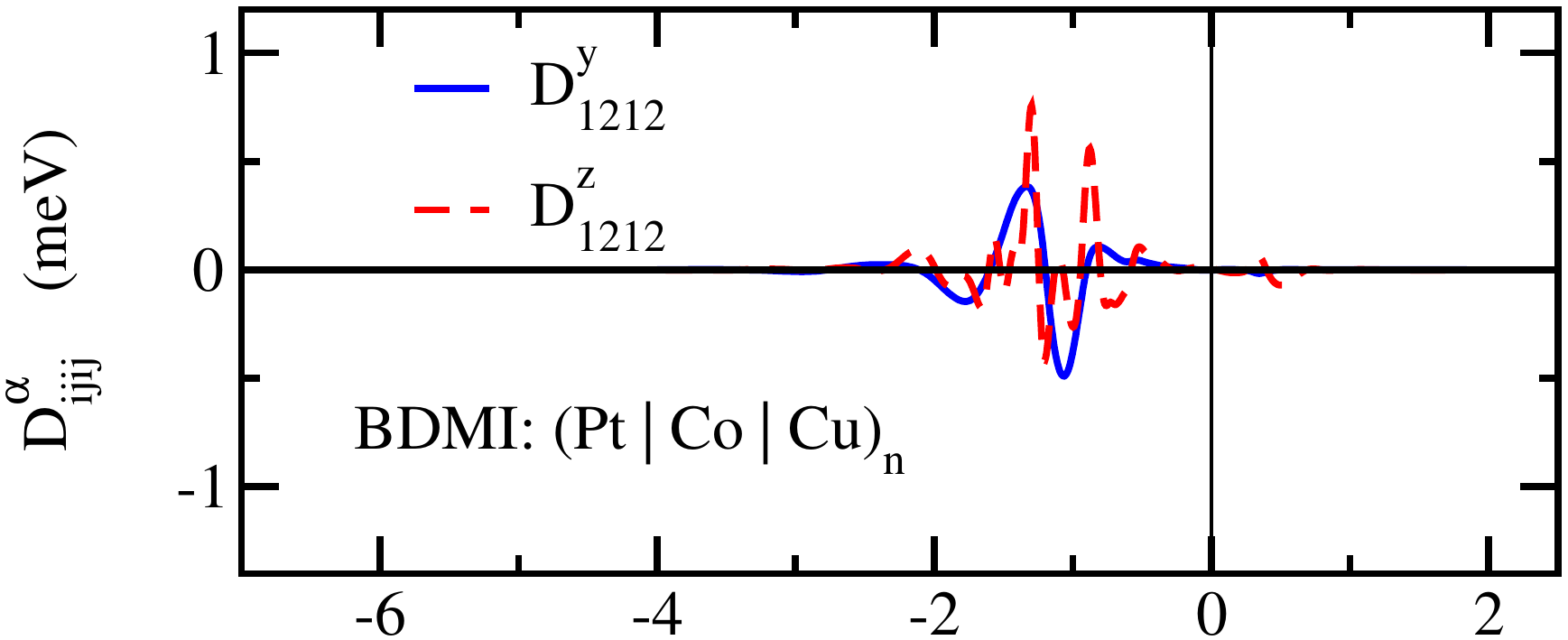}\;(b)
\includegraphics[width=0.32\textwidth,angle=0,clip]{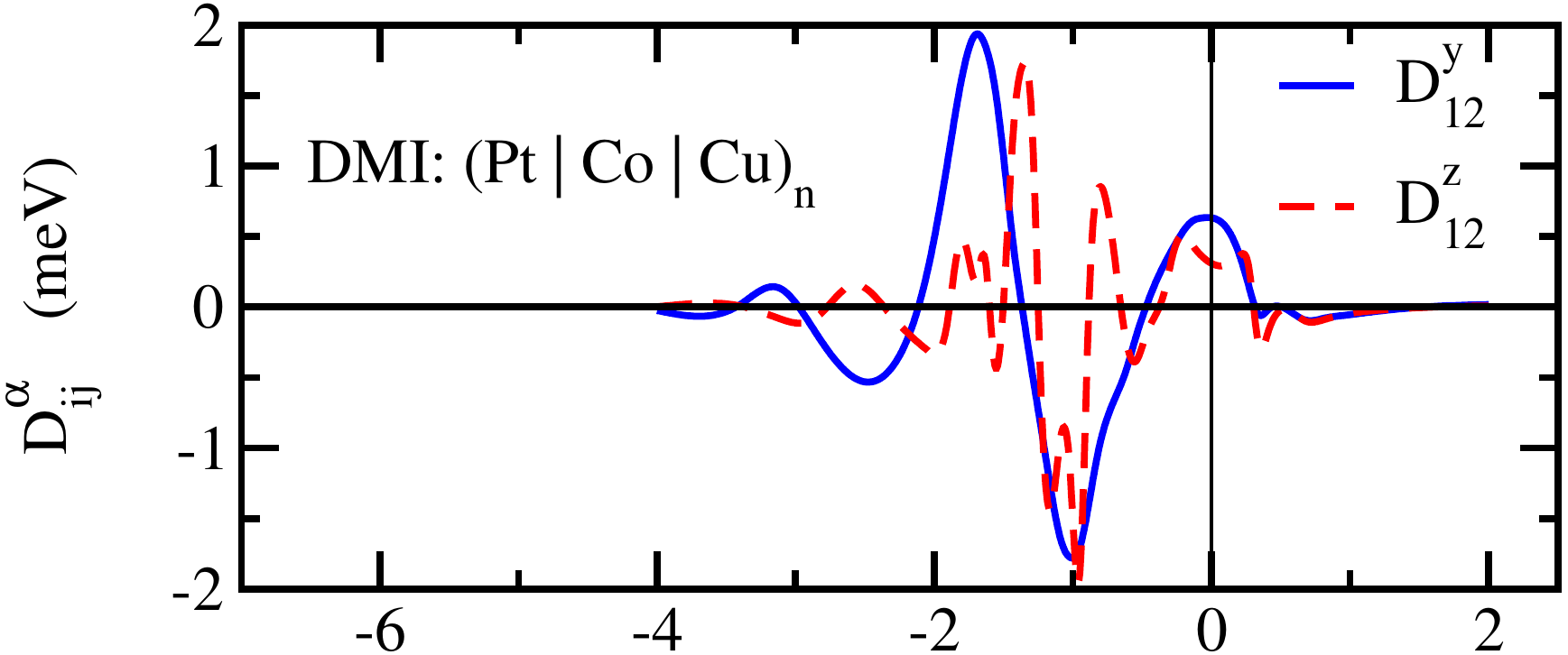}\;(c)
\includegraphics[width=0.32\textwidth,angle=0,clip]{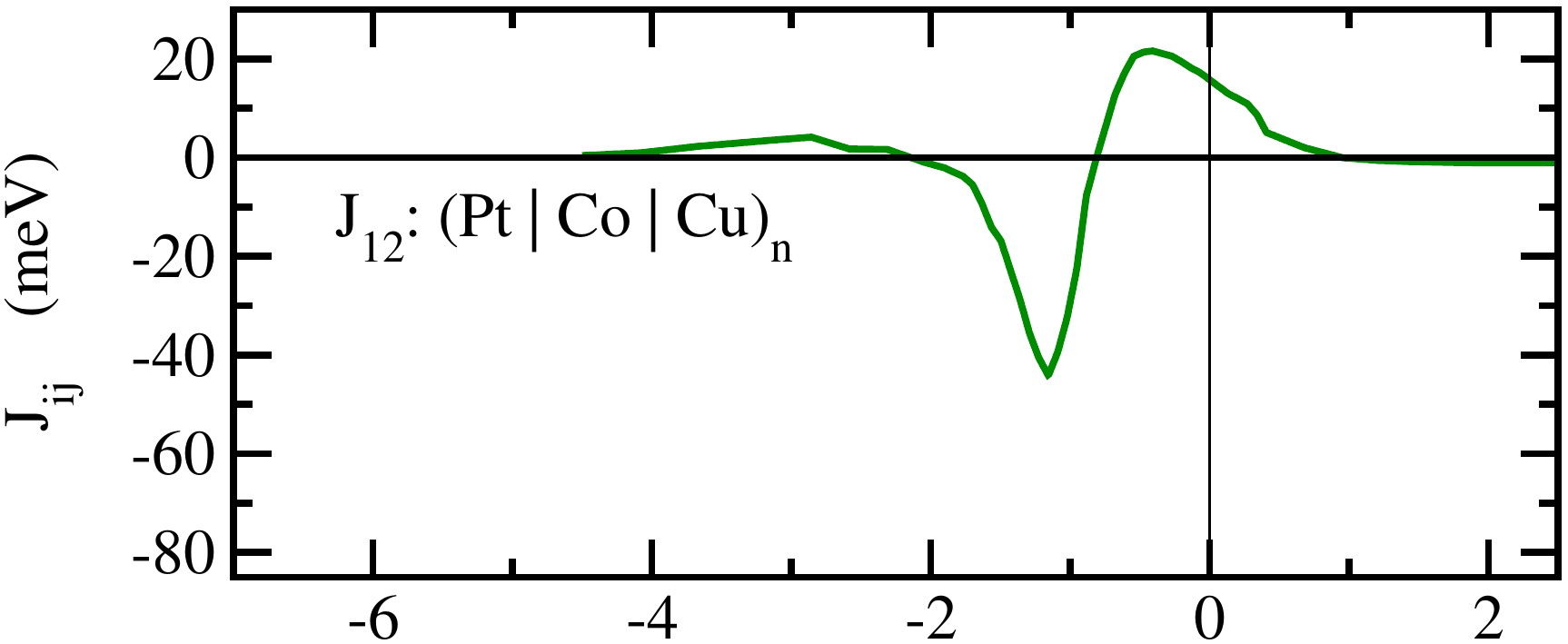}\;(d)
\includegraphics[width=0.32\textwidth,angle=0,clip]{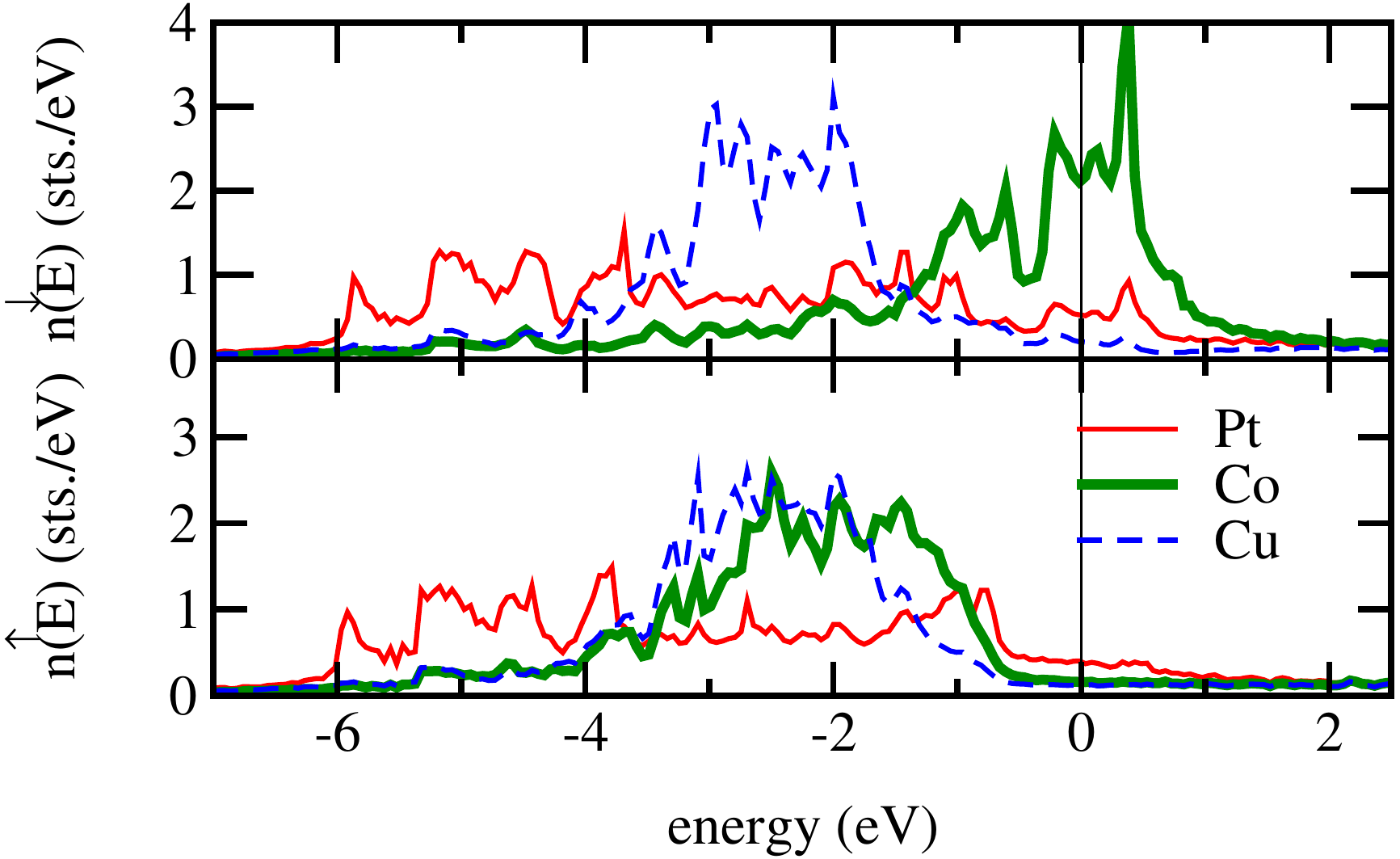}\;(e)
\caption{\label{fig:Pt/Co/Cu_3spin-E} Three-spin interaction parameters
  ${\tilde J}_{ijk}(E)$ for the smallest Co-triangles in (Pt/Co/Cu)$_n$
  multilayer, centered at Cu (${\tilde J}^{Cu}_{\Delta}$) and Pt
  (${\tilde J}^{Pt}_{\Delta}$); the dotted line represents the Co spin 
  magnetic moment  $m(E) = M(E)/M_{Co}$ ($M_{Co} = 1.9 \mu_B$) as a function of the occupation (a);
  $y$-, $z$-components of the chiral biquadratic exchange interaction,
  $\vec{{\cal D}}_{1212}$ (b) and the DM interactions, $\vec{D}_{ij}$ (c)
  between Co atoms as a function of the occupation; (d) the isotropic exchange
  $J_{ij}(E)$, and (e) the element-projected DOS, n(E).
  }  
\end{figure}

Figs.\ \ref{fig:Pt/Mn/Cu_3spin-E}  and \ref{fig:Pt/Co/Cu_3spin-E}
represent the TCI (a) in comparison with BDMI
 (b) DMI (c) and isotropic exchange interactions, 
plotted as a function of energy characterizing the occupation of the valence
band (i.e. an artificial Fermi energy position). One can see an
oscillating behavior for all parameters when the occupation 
increases, with their sign changing at different energies
because of different origin of these interactions. Note, however, that all
quantities shown in  Figs.\ \ref{fig:Pt/Mn/Cu_3spin-E} and
\ref{fig:Pt/Co/Cu_3spin-E} have a maximum at approximately half
occupation of the Mn(Co) $d$-band (see Figs.\
\ref{fig:Pt/Mn/Cu_3spin-E}(e)  and \ref{fig:Pt/Co/Cu_3spin-E}(e)), that 
correlates also with the maximum of the spin magnetic moment (Figs.\
\ref{fig:Pt/Mn/Cu_3spin-E}(a)  and 
\ref{fig:Pt/Co/Cu_3spin-E}(a)) and  maximum of antiferromagnetic
exchange interactions (Figs.\ \ref{fig:Pt/Mn/Cu_3spin-E}(d)  and
\ref{fig:Pt/Co/Cu_3spin-E}(d)).

Comparing the $y$- and $z$-components of the BDMI and DMI shown in Figs.\
\ref{fig:Pt/Mn/Cu_3spin-E} (a) and (b), 
respectively, one can see a more narrow energy region,
in which the former quantity has a significant magnitude.
Note, however, that the biquadratic interaction is a higher
order term in the energy expansion and should represent simultaneously the
features of vector and scalar interactions of two spin moments.
Thus, plotting in  Fig.\ \ref{fig:Pt/Mn/Cu_3spin-E} (a) (thin
lines) the function $D^{\alpha}_{ij}(E)J_{ij}(E)/max(J_{ij}(E))$ for the
nearest-neighbor interactions, one can see a
 localization in energy of this function similar to the one seen for the BDMI. 

 \subsubsection{Monte Carlo simulations}

 In order to demonstrate a possible impact of the higher-order chiral
 interactions on the magnetic structure, Monte Carlo simulations have
 been performed for model systems. We focus here on the
 effect of the three-spin chiral interactions having rather different
 properties  when compared to the DMI-like interactions.
 In particular, they depend on the orientation of the coupling spin moments.
 The calculations have illustrative character, therefore we present only
 few results showing a non-vanishing impact such interactions, seen here  
 as free parameter, in the formation of the magnetic texture.  

 Monte Carlo (MC) simulations are performed for 2D lattice having a
 triangular structure, on the basis of the model 
 Hamiltonian
\begin{eqnarray}
  H &=&  - J_1\sum_{(i,j)_1}  \hat{s}_i \cdot \hat{s}_j \nonumber \\
  &&    - J_2\sum_{(i,j)_2}  \hat{s}_i \cdot \hat{s}_j
        - \frac{1}{3} \tilde{J}_{\Delta1} \sum_{(i,j,k)\in\Delta1}
        \hat{s}_i\cdot (\hat{s}_j \times \hat{s}_k) \; .        
\label{Eq_Heisenberg_TCI-reduced2}
\end{eqnarray}
Dealing with this expression, the three-spin contribution is evaluated accounting for the
counter-clock-wise sequence of atoms $i, j$ and $k$ with $\hat{s}_i$,
$\hat{s}_j$ and $\hat{s}_k$ the orientation of the
corresponding spin moments respectively.
In the model Hamiltonian only the first- (positive) and second-neighbor (negative)
isotropic exchange interactions, $J_1$ and $J_2$,
respectively, are taken into account, while the three-spin chiral 
interactions $\tilde{J}_{\Delta1}$ are accounted for the smallest possible triangles.
To take into account the dependence of the 3-spin interactions on the
magnetic configuration we use an algorithm similar to that
used to calculate the exchange interactions mediated by non-magnetic
components in alloys (e.g. FePd, FeRh, etc.), which also depends on the
local magnetic configuration \cite{PMS+10,PMK+16}. 
As $\tilde{J}_{\Delta1}$ depends on the magnetization flux through the
triangle, at each MC step they have been calculated according to the
discussions above, using the expression  
$\tilde{J}_{\Delta1} = \tilde{J}^0_{\Delta1} \hat{n}\cdot(\hat{s}_i +
\hat{s}_j + \hat{s}_k)$, where $\hat{n}$ is the normal to the film. 
Here $\tilde{J}^0$ is the maximal value of three-spin interaction
corresponding to a small deviation of the spin magnetic moments from the
collinear direction.

\begin{figure}
\includegraphics[width=0.12\textwidth,angle=0,clip]{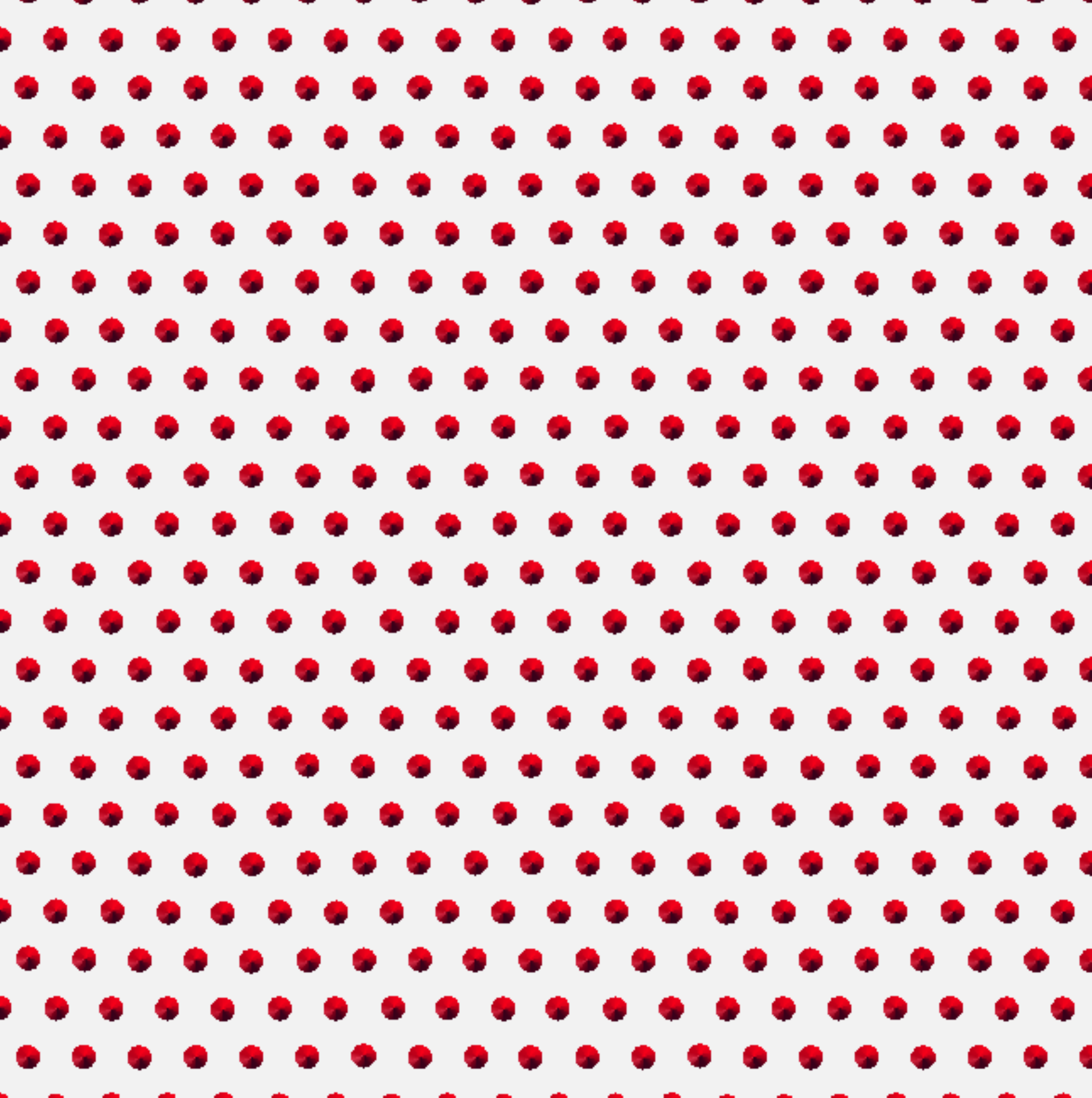}(a)
\includegraphics[width=0.12\textwidth,angle=0,clip]{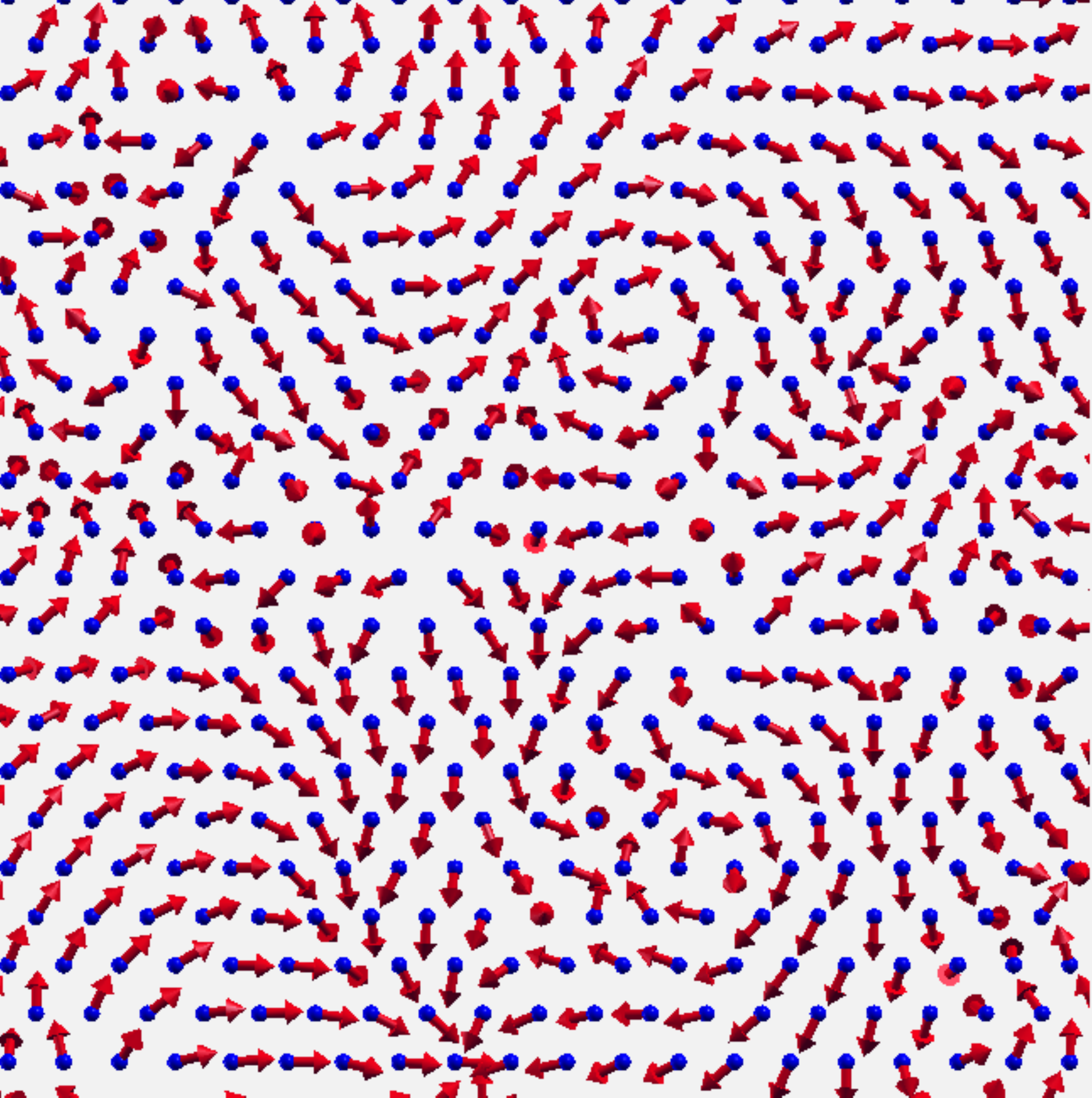}(b)
\includegraphics[width=0.12\textwidth,angle=0,clip]{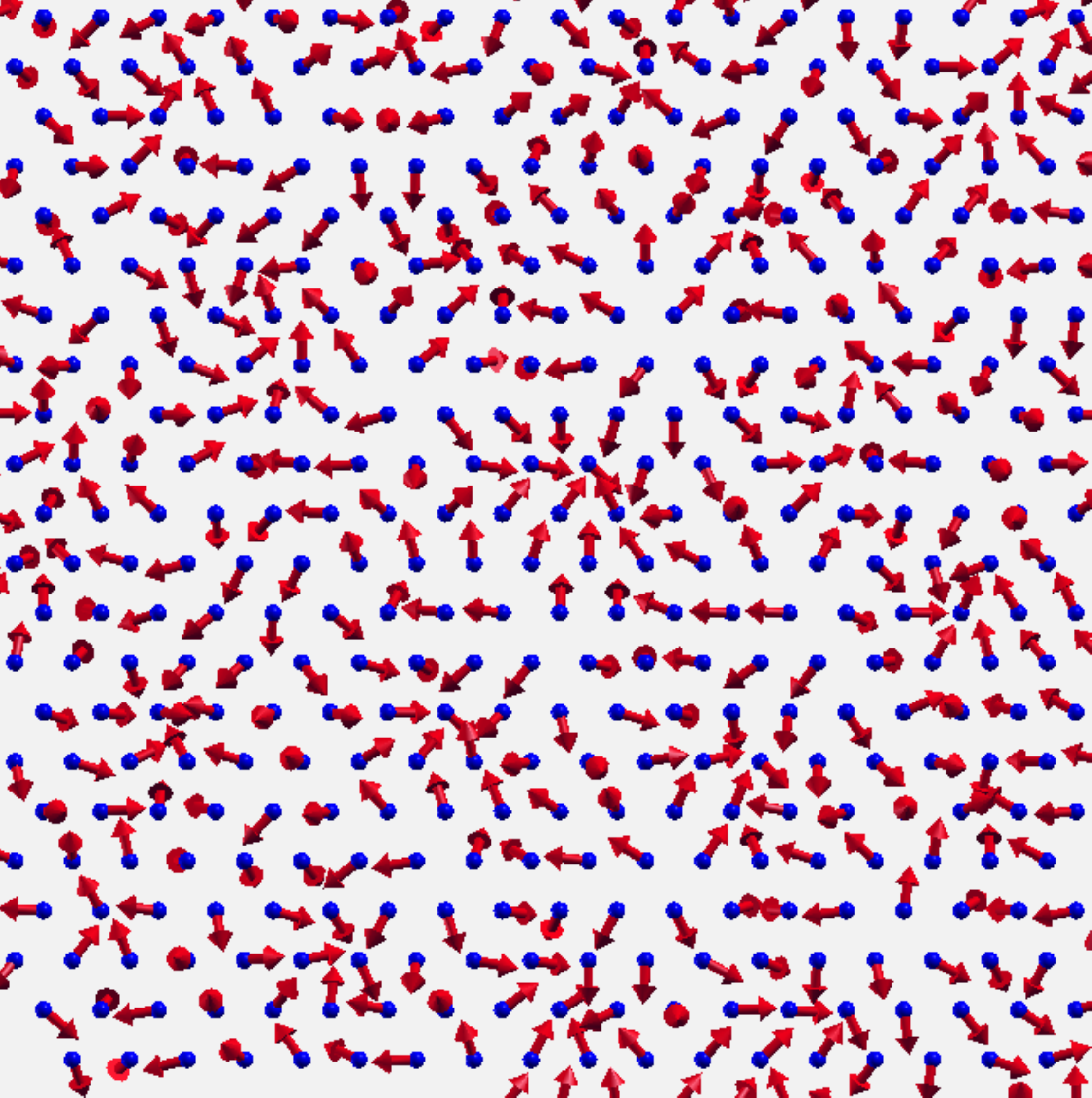}(c)\\
\caption{\label{fig:MC} Snapshot ($30 \times 30$ atoms) of the magnetic
  structure obtained by MC simulations for $T = 0.1$~K, accounting for
  first-neighbor $J_1$ and second-neighbor
  $J_2$ and three spin $J_{\Delta1}$ interactions ($J_{\Delta1}/J_1 =
  1.0$) only: (a) $J_2 = 0.0$, (b) $J_2/J_1 = -0.25$, and (c)
  $J_2/J_1 = -0.5$. }  
\end{figure}

Periodic boundary conditions have been used with the MC cell
having $60 \times 60$ atomic sites. For the sequential update the Metropolis
algorithm has been used. Up to 5000 MC steps have been used to reach the
equilibrium.
The results represented in Fig. \ref{fig:MC} correspond to the
temperature $T = 0.1$ K. In all cases we used the first-neighbor
parameter $J_1 = 3$ meV and rather large value for three-spin
interaction parameter $J_{\Delta1}/J_1 = 1.0$. Fig.  \ref{fig:MC}
shows the magnetic structures for $J_2/J_1 = 0$ (a),
$J_2/J_1 =-0.25$ (b) and $J_2/J_1 = -0.5$ (c).
One can see no impact of three-spin interactions on the magnetic
structure when the negative second-neighbor interactions are not taken
into account. But for systems having competing FM and AFM interactions,
the three-spin interactions results in the formation of a vortex
structure with the size of the vortexes dependent 
on relative magnitude of the exchange parameters.

\section{Summary}

To summarize, in the present work we present a general approach to
calculate the multispin exchange interactions in order to extend
the classical Heisenberg Hamiltonian. This approach allows first principles
calculations of multispin interactions in real-space within the
framework of the multiple scattering Green function formalism.
We discussed some properties of different types of chiral
interactions, with the main focus on the three-spin exchange
interactions. A  specific feature of TCI is its  topological origin in
contrast to the three-spin DMI-like interactions. We demonstrated by
means of MC simulations that this term can lead to a stabilization of
vortex-like magnetic texture.

 \section{Acknowledgement}
Financial support by the DFG via SFB 1277 (Emergent Relativistic Effects
in Condensed Matter - From Fundamental Aspects to Electronic
Functionality).

\section{Appendix A}

\subsection{Computational details}
\label{SEC:Computational-scheme}

The numerical results for the coupling parameters are based on
first-principles electronic structure calculations 
performed using the spin-polarized relativistic KKR (SPR-KKR) Green
function method  \cite{SPR-KKR7.7,EKM11}. The calculations were done
in a fully-relativistic mode, except for  some special cases pointed
out in the manuscript, where scaling of the spin-orbit interaction
was applied.
All calculations have been performed using the atomic sphere
approximation (ASA) within the framework of the local 
spin density approximation (LSDA) to spin density 
functional theory (SDFT), using a parametrization for the exchange and
correlation potential as given by Vosko {\em et  al.} \cite{VWN80}.
For the angular momentum expansion of the Green function the angular
momentum cutoff $l_{max} = 3$ was used.

To demonstrate the properties of multispin interactions, several
reference systems have been considered. These are the 3d-metals bcc Fe (lattice
parameters $a = 5.40$ a.u), fcc Ni ($a = 6.65$ a.u.) and hcp Co ($a =
4.72$ a.u., $c/a = 1.62$), $L1_0$ compounds FePt ($a = 7.36$ a.u., $c/a
= 0.95$) and FePd ($a = 7.28$ a.u., $c/a = 0.96$),  as well as
the multilayer model systems
(Cu/Fe/Pt)$_n$  (Cu/Co/Pt)$_n$ and (Cu/Mn/Pt)$_n$ having fcc
structure with (111) orientation of the layers. The lattice parameter $a
=7.407$ a.u. was used for all systems. 
The k-mesh $98 \times 98
\times 98$ was used for the integration over the BZ of the considered
3d-metals, while $41 \times 41 \times 41$ was used for the
(Cu/$X$/Pt)$_n$ multilayers, and $38 
\times 38 \times 11$ for 1ML Fe(110). 

The calculations for 1ML of bcc Fe ($a = 5.40$ a.u) have been performed
in the supercell geometry with Fe layers separated by three vacuum
layers. This decoupling that was sufficient to demonstrate the properties of the exchange
interaction parameters for the 2D system. 

Another system dealt with is an Fe overlayer on the top of a
TMDC compound Fe/1H-TaTe$_2$ and Fe/1H-WTe$_2$, with space group
$P6_3/mmc$ for the bulk TMDC compounds. These
calculations have been performed in supercell geometry with the Fe/TMDC
films separated by vacuum layers. The lattice parameters are $a =
6.82$ a.u. and  $c = 26.29$ a.u. for 1H-TaTe$_2$ and $a = 6.69$ a.u.and
$c = 25.71$ a.u. for 1H-WTe$_2$. 
More structure information about TMDC monolayers one can find for
example in Ref. \cite{ASC12}.

\begin{widetext}
  \section{Appendix B}

  In the following we present some more details concerning the mapping
  of the ab-initio magnetic energy to the Heisenberg Hamiltonian. The
  main idea is demonstrated first dealing with the bilinear interatomic exchange
  interactions. In order to derive higher-order interactions, rather
  lengthy transformations are required, which follow similar scheme as
  that outlined to get the expressions for bilinear interactions.
  
  The expressions for the exchange interaction parameters can be
  obtained by comparing the energy change due to spin modulations characterized by
  a wave vector $\vec{q}$, described either
  in the first-principles formulation or based on the extended
  Heisenberg model. Obviously, one has to identify the terms having the
  same 
  dependence on the interatomic distance, in particular by comparing 
  the derivatives with respect to $\vec{q}$-vector in the limit of $q = 0$. 
  It may be useful also to select the elements giving zero
  contributions to the corresponding energy derivatives.

  \subsection{Bilinear terms}
  \subsubsection{Spin modulation 1}
  Let's start with the bilinear terns in the Heisenberg Hamiltonian:  
 \begin{eqnarray}
   H^{(2)} &=&  - \sum_{i,j} J_{ij} (\hat{s}_i \cdot \hat{s}_j) -
               \sum_{i,j}  \vec{{D}}_{ij} \cdot (\hat{s}_i \times
               \hat{s}_j)  \; .
 \label{Eq_deriv_Heisenberg_2-spin}
 \end{eqnarray}
Depending on the considered interaction parameters different types of
spin modulation are used to simplify the derivation.
In the present case we use the geometry with the magnetization along the $z$
axis, and the spin spiral having the form 
\begin{eqnarray}
  \hat{s}_i &=&
(\sin\theta \cos(\vec{q}\cdot\vec{R}),\sin\theta \sin(\vec{q}\cdot\vec{R}), \cos\theta)
\label{Eq:spin-spiral-stand}
\end {eqnarray}

This leads to the the following energy change for the Heisenberg model
 \begin{eqnarray}
   \Delta E^{(2)}(\vec{q}) &=&  - \sum_{i,j}
                                J_{ij}  (\sin^2\theta
                               \cos \vec{q}\cdot(\vec{R}_j - \vec{R}_i)
                              + \cos^2\theta  - 1)
                                - \sum_{i,j}
                                D^z_{ij} \sin^2\theta
                               \sin \vec{q}\cdot(\vec{R}_j - \vec{R}_i)
                               \nonumber \\
                       && -  \sum_{i,j} D^x_{ij} \sin\theta \cos\theta
                               (\sin(\vec{q}\cdot\vec{R}_i) -
                          \sin(\vec{q}\cdot\vec{R}_j))
                          -  \sum_{i,j} D^y_{ij} \sin\theta \cos\theta
                               (\cos(\vec{q}\cdot\vec{R}_j) -
                          \cos(\vec{q}\cdot\vec{R}_i))       \,.           
 \label{Eq_deriv_Heisenberg_2-spin-SW}
 \end{eqnarray}

Using the relation for the sum over the lattice sites $\vec{R}_i$   
 \begin{eqnarray}
  \frac{1}{N} \sum_{i}^N e^{-i \vec{q} \cdot \vec{R}_i} &=&  \delta_{\vec{q},0}
 \label{Eq_lattice-sum}
 \end{eqnarray}
 one can show that the terms due to $D^x_{ij}$ and $D^y_{ij}$ in the
present case do not give contributions to the derivatives with respect
to the  $\vec{q}$ vector:
 \begin{eqnarray}
   \sum_{i,j}^N   D^{y}_{ij}  (\cos(\vec{q}\cdot\vec{R}_j) -  \cos(\vec{q}\cdot\vec{R}_i) ) &=& 
   \sum_{j}^N \bigg[ \sum_{i}^N   D^{y}_{ij} \bigg]                                                                                                \cos(\vec{q}\cdot\vec{R}_j)- \sum_{i}^N \bigg[ \sum_{j}^N D^{y}_{ij} \bigg]\cos(\vec{q}\cdot\vec{R}_i)  = N D^{y}_{0} (\delta_{\vec{q},0} - \delta_{\vec{q},0}) = 0
 \end{eqnarray}
and
 \begin{eqnarray}
   \sum_{i,j}^N   D^{x}_{ij}  (\sin(\vec{q}\cdot\vec{R}_j) -  \sin(\vec{q}\cdot\vec{R}_i) ) &=&
   \sum_{j}^N \bigg[ \sum_{i}^N   D^{x}_{ij}\bigg]  \sin(\vec{q}\cdot\vec{R}_j) - \sum_{i}^N \bigg[ \sum_{j}^N D^{x}_{ij}\bigg] \sin(\vec{q}\cdot\vec{R}_i)    = 0 \,,
 \end{eqnarray}
taking into account that the sum over $i(j)$ in square brackets in these
expressions gives the same value for each site $j(i)$.

Taking the first- and second-order derivatives of the energy change
$\Delta E(\vec{q})$ (Eq.\ (\ref{Eq_deriv_Heisenberg_2-spin-SW})) with
respect to $\vec{q}$, one obtains the expressions 
contributed either by the DMI  interactions
 \begin{eqnarray}
   \frac{\partial}{\partial \vec{q}}\Delta E(\vec{q})\bigg|_{q \to 0} &=&  
    - \sin^2\theta \sum_{i \neq j}^N
         D^{z}_{ij}   \hat{q}\cdot(\vec{R}_i - \vec{R}_j) 
 \label{Eq_deriv_Heisenberg_2-derivativ1}
 \end{eqnarray}

or by the isotropic exchange interaction
 \begin{eqnarray}
   \frac{\partial^2}{\partial \vec{q}^2}\Delta E(\vec{q})\bigg|_{q \to 0} &=&  
     \sin^2\theta \sum_{i,j}
      J_{ij} (\hat{q}\cdot(\vec{R}_i - \vec{R}_j))^2
 \label{Eq_deriv_Heisenberg_2-derivativ2}
 \end{eqnarray}
 with $\hat{q} = \vec{q}/|\vec{q}|$ the unit vector giving the direction of
 the wave vector $\vec{q}$.

 Now let's consider the first-principles energy change due to a spin
 spiral, evaluated in terms of the Green function
\begin{eqnarray}
\Delta {\cal E}^{(2)} &=& -\frac{1}{\pi} \mbox{Im}\,\mbox{Tr} \int^{E_F}
                    dE\, \Delta V G_0 \Delta V G_0 \;.
\label{Eq_Free_Energy-4}
\end{eqnarray}
Substituting the multiple-scattering representation for the
Green function into this equation together with the perturbation  
 \begin{eqnarray}
 \Delta V(\vec{r}) &=&  \sum_i \beta \big( \vec{\sigma}\cdot\hat{s}_i
  -  \sigma_z\big) B_{xc}(\vec{r}) \;,
\label{Eq_perturb_stiff}
\end{eqnarray}
for a spin spiral according to Eq.\ (\ref{Eq:spin-spiral-stand}), and introducing
the definition 
\begin{eqnarray}
{\cal J}^{\alpha\beta}_{ij} &=& \frac{1}{2\pi} \mbox{Im}\,\mbox{Tr} \int^{E_F}
                          dE\, T^{\alpha}_{i} \tau_{ij}T^{\beta}_{j}
                                \tau_{ji} \, ,
\label{Eq_first_principles_J-tensor}
\end{eqnarray}
we obtain the following symmetrized expressions for the energy change
%
\begin{eqnarray}
\Delta {\cal E}^{(2)} &=& 
  \sum_{i,j} \sin^2 \theta  \bigg[\frac{1}{2} ({\cal J}^{xx}_{ij}  +
                          {\cal J}^{yy}_{ij}) \cos
      \vec{q}\cdot(\vec{R}_i - \vec{R}_j ) + {\cal J}^{zz}_{ij} (\cos
                          \theta -1)^2  \bigg]  \nonumber\\
                      &&+ \sin^2 \theta \frac{1}{2} ({\cal J}^{xx}_{ij}  - {\cal J}^{yy}_{ij}) \cos
      \vec{q}\cdot(\vec{R}_i + \vec{R}_j )\nonumber\\
                      && 
                         + \sin^2\theta \frac{1}{2} ({\cal J}^{xy}_{ij}  + {\cal
                         J}^{yx}_{ij})\sin  \vec{q}\cdot(\vec{R}_i + \vec{R}_j )\nonumber
                         \\
                     &&    + \sin^2\theta \frac{1}{2} ( {\cal J}^{xy}_{ij}  - {\cal J}^{yx}_{ij}) \sin
      \vec{q}\cdot(\vec{R}_j - \vec{R}_i ) \nonumber\\  
                      &&+  \sin \theta (\cos \theta -1) [ {\cal
                         J}^{xz}_{ij} \cos(\vec{q}\cdot\vec{R}_i) +
                         {\cal J}^{zx}_{ij} \cos(\vec{q}\cdot\vec{R}_j)]
                         \nonumber \\ 
                      &&+
  \sin \theta (\cos \theta -1) [ {\cal J}^{yz}_{ij} \sin(\vec{q}\cdot\vec{R}_i)   +
                        {\cal J}^{zy}_{ij} \sin(\vec{q}\cdot\vec{R}_j)]                    \,.
\label{Eq_Free_Energy_bilin-2}
\end{eqnarray}
In Eq.\ (\ref{Eq_first_principles_J-tensor}) we use $T^\alpha_i  =
\langle Z_i| \beta \sigma^\alpha B_{xc,i}|Z_i \rangle$ and the prefactor
$1/2$  prevents a double counting of contribution to the energy
associated with a pair of atoms.

Using again Eq.\ (\ref{Eq_lattice-sum}) for the sum over the lattice
 sites $\vec{R}_i$    
 one can show that the derivatives with respect to  $\vec{q}$ of the
 off-diagonal terms associated with $J^{\alpha z}_{ij}$ and $J^{z
   \alpha}_{ij}$ parameters vanish.
 
For instance, one finds:
 \begin{eqnarray}
   \sum_{i,j}^N {\cal J}^{yz}_{ij} \sin(\vec{q}\cdot\vec{R}_i) &=&
 -\frac{i}{2} \sum_{i}^N   \bigg[ \sum_{j} {\cal J}^{yz}_{ij} \bigg] (e^{i(\vec{q}\cdot\vec{R}_i)}
                             - e^{-i(\vec{q}\cdot\vec{R}_i)})  \nonumber
   \\
 &=&
  -\frac{i}{2} {\cal J}^{yz}_{0} \sum_{i}^N  (e^{i(\vec{q}\cdot\vec{R}_i)}
                             - e^{-i(\vec{q}\cdot\vec{R}_i)}) 
 =
 -\frac{i}{2} N  J^{yz}_{0}   (\delta_{\vec{q},0} - \delta_{\vec{q},0}) = 0 
 \end{eqnarray}
since the sum  $ \sum_{j} J^{\alpha\beta}_{ij} =
J^{\alpha\beta}_{0} $ is the same for all sites $i$. For the other term
one finds

 \begin{eqnarray}
  \sum_{i,j}^N   {\cal J}^{xz}_{ij}  \cos(\vec{q}\cdot\vec{R}_i) &=& \frac{N}{2} {\cal J}^{xz}_{0}\delta_{\vec{q},0}
 \end{eqnarray}
This value does not depend on $q$. As a result, its derivative with
respect to $\vec{q}$, used to derive the expression for the exchange
parameters, vanishes.  The same concerns also the exchange parameters ${\cal J}^{zy}$
and   ${\cal J}^{zx}$.
 
Transforming the term proportional to  $ ({\cal J}^{xy}_{ij} + {\cal J}^{yx}_{ij})$
and using again Eq.\ (\ref{Eq_lattice-sum}), one obtains
 \begin{eqnarray}
   \sum_{i,j}^N  ({\cal J}^{xy}_{ij} + {\cal J}^{yx}_{ij}) \sin \vec{q}\cdot(\vec{R}_i + \vec{R}_j) &=&
   \sum_{i,j}^N ({\cal J}^{xy}_{ij} + {\cal J}^{yx}_{ij}) (\cos(\vec{q}\cdot(\vec{R}_i- \vec{R}_j) \sin(2\vec{q}\cdot\vec{R}_j)
            +  \sin(\vec{q}\cdot(\vec{R}_i- \vec{R}_j)
                                                                            \cos(2\vec{q}\cdot\vec{R}_j))
   \nonumber \\
                                                                                     &=&
   \sum_{i}^N  \bigg[ \sum_{j}^N  ({\cal J}^{xy}_{ij} + {\cal J}^{yx}_{ij})
 \cos(\vec{q}\cdot(\vec{R}_i- \vec{R}_j)) \bigg]
                                                                                         \sin(2\vec{q}\cdot\vec{R}_i)  \nonumber \\
                                                                                     &&+
  \sum_{i}^N \bigg[ \sum_{j}^N  ({\cal J}^{xy}_{ij} + {\cal J}^{yx}_{ij})\sin(\vec{q}\cdot(\vec{R}_i- \vec{R}_j))  \bigg]
        \cos(2\vec{q}\cdot\vec{R}_i)) = const \, ,
 \end{eqnarray}
where the first sum vanishes due to the summation over $i$, sum does not
depend on $q$.
The same behavior is shown by the term proportional to $ ({\cal J}^{xx}_{ij} - {\cal J}^{yy}_{ij})$

Taking the first- and second-order derivatives of
the energy change $\Delta {\cal E}^{(2)}(\vec{q})$ with respect to $\vec{q}$, Eq.\
(\ref{Eq_Free_Energy_bilin-2}) leads to the expressions contributed
either by the DMI  interactions 
 \begin{eqnarray}
   \frac{\partial}{\partial \vec{q}}\Delta {\cal E}^{(2)}(\vec{q})\bigg|_{q \to 0} &=&  
  -  \sin^2\theta \sum_{i \neq j}^N
        \frac{1}{2} ({\cal J}^{xy}_{ij}  - {\cal J}^{yx}_{ij}) \hat{q}\cdot(\vec{R}_i - \vec{R}_j) 
 \label{Eq_deriv_Free_energy_2-deriv1}
 \end{eqnarray}
 and the isotropic exchange
 \begin{eqnarray}
   \frac{\partial^2}{\partial \vec{q}^2}\Delta {\cal E}^{(2)}(\vec{q})\bigg|_{q \to 0} &=&  
     \sin^2\theta \sum_{i \neq j}^N
        \frac{1}{2} ({\cal J}^{xx}_{ij}  + {\cal J}^{yy}_{ij})
                                                                                    (\hat{q}\cdot(\vec{R}_i - \vec{R}_j))^2 \,.
 \label{Eq_deriv_Free_energy_2-deriv2}
 \end{eqnarray}

Comparing these equations with the Eqs.\
(\ref{Eq_deriv_Heisenberg_2-derivativ1})-(\ref{Eq_deriv_Heisenberg_2-derivativ2}),
one obtains for the DMI
 \begin{eqnarray}
   D^z_{ij} &=&  \frac{1}{2} ({\cal J}^{xy}_{ij}  - {\cal J}^{yx}_{ij})
 \label{Eq_DMI_z}
 \end{eqnarray}
 and for the isotropic exchange
 \begin{eqnarray}
  J_{ij} &=&  \frac{1}{2} ({\cal J}^{xx}_{ij}  + {\cal J}^{yy}_{ij}) \,.
 \label{Eq_Js}
 \end{eqnarray}
The two others DMI components are discussed in detail in previous works
\cite{ME17,MPE19}.

\subsubsection{Spin modulation 2}

For comparison, we use here also the spin modulation within the
$xy$-plane, characterized by two q-vectors:   
\begin{eqnarray}
  \hat{s}_i &=&
 (\mbox{sin}(\vec{q}_1 \cdot \vec{R}_i) \;\mbox{cos}(\vec{q}_2 \cdot \vec{R}_i)\;,
                \mbox{sin}(\vec{q}_2 \cdot \vec{R}_i) \;, \mbox{cos}(\vec{q}_1 \cdot
 \vec{R}_i)\mbox{cos}(\vec{q}_2 \cdot \vec{R}_i) ) \;.
\label{eq:spiral2-2}
\end {eqnarray}
The corresponding results can be used later to derive the three-spin and biquadratic
DMI-like interactions.

Within the Heisenberg model the second order terms associated with the isotropic exchange
interaction are given by 
\begin{eqnarray}
\Delta {E}^{(2)} &=&  - \sum_{i,j}  J_{ij} [(\hat{s}_i \cdot \hat{s}_j)
                          - 1]  \nonumber \\
                          &=& - \sum_{i,j} J_{ij} \bigg[
                        \cos \vec{q}_1\cdot (\vec{R}_j - \vec{R}_j)
                        \cos(\vec{q}_2\cdot \vec{R}_i)
                          \cos(\vec{q}_2\cdot \vec{R}_j)
                          + \sin(\vec{q}_2\cdot \vec{R}_i)
                          \sin(\vec{q}_2\cdot \vec{R}_j) - 1 \bigg]  
  \label{Eq_Free_Energy-DX1}
\end{eqnarray}
Taking $q_1 = 0$, the second-order derivative of the energy
with respect to $q_2$ is given by
\begin{eqnarray}
\frac{\partial^2}{\partial \vec{q}_2^2} \Delta {\cal E}^{(2)} &=&  \sum_{i,j}  J_{ij}
              (\hat{q} \cdot (\vec{R}_j - \vec{R}_i))^2
  \label{Eq_Free_Energy-J-D-4}
\end{eqnarray}

On the other hand, taking $q_2 = 0$, the second-order derivative of
the energy with respect to $q_1$ has the form
\begin{eqnarray}
\frac{\partial^2}{\partial \vec{q}_1^2} \Delta {\cal E}^{(2)} &=&  \sum_{i,j} 
              J_{ij}  (\hat{q} \cdot (\vec{R}_j - \vec{R}_i))^2  \,.
  \label{Eq_Free_Energy-J-D-4}
\end{eqnarray}

In the first-principles approach, substituting the perturbation due to
the spin modulation according to Eq.\ (\ref{eq:spiral2-2}) and using the definition
in Eq.\ (\ref {Eq_first_principles_J-tensor}) one obtains
\begin{eqnarray}
\Delta {\cal E}^{(2)} &=&  - \sum_{i,j} {\cal J}^{xx}_{ij} 
                        \sin(\vec{q}_1\cdot \vec{R}_i)\cos(\vec{q}_2\cdot \vec{R}_i)
                         \sin(\vec{q}_1\cdot
                          \vec{R}_j)\cos(\vec{q}_2\cdot \vec{R}_j) +
                       {\cal J}^{yy}_{ij}  
                        \sin(\vec{q}_2\cdot \vec{R}_i)
                         \sin(\vec{q}_2\cdot \vec{R}_j)       \nonumber \\
                      && +                     
                          {\cal J}^{zz}_{ij}                      
                        (\cos (\vec{q}_1\cdot \vec{R}_i) \cos
                          (\vec{q}_2\cdot \vec{R}_i) - 1)        
                        (\cos (\vec{q}_1\cdot \vec{R}_i) \cos
                          (\vec{q}_2\cdot \vec{R}_i) - 1)  \nonumber \\
                      && + {\cal J}^{xy}_{ij} 
                        \sin(\vec{q}_1\cdot \vec{R}_i)\cos(\vec{q}_2\cdot \vec{R}_i)
                         \sin(\vec{q}_2\cdot \vec{R}_j) +
                       {\cal J}^{yx}_{ij}  
                        \sin(\vec{q}_2\cdot \vec{R}_i)
                         \sin(\vec{q}_2\cdot \vec{R}_j)\cos(\vec{q}_2\cdot \vec{R}_j)  \nonumber \\
                      && +    {\cal J}^{xz}_{ij}
                        \sin(\vec{q}_1\cdot \vec{R}_i)\cos(\vec{q}_2\cdot \vec{R}_i)
                         (\cos (\vec{q}_1\cdot \vec{R}_j) \cos
                         (\vec{q}_2\cdot \vec{R}_j) - 1)   \nonumber \\
                      && +                        
                         {\cal J}^{zx}_{ij}
                        (\cos (\vec{q}_1\cdot \vec{R}_i) \cos
                          (\vec{q}_2\cdot \vec{R}_i) - 1)        
                        \sin(\vec{q}_1\cdot \vec{R}_j)\cos(\vec{q}_2\cdot \vec{R}_j)  \nonumber \\
                      && +  {\cal J}^{yz}_{ij}
                        \sin(\vec{q}_2\cdot \vec{R}_i)
                         (\cos (\vec{q}_1\cdot \vec{R}_j) \cos
                         (\vec{q}_2\cdot \vec{R}_j) - 1)+                          
                         {\cal J}^{zy}_{ij}
                        (\cos (\vec{q}_1\cdot \vec{R}_i) \cos
                          (\vec{q}_2\cdot \vec{R}_i) - 1)        
                        \sin(\vec{q}_2\cdot \vec{R}_j) \;.
  \label{Eq_Free_Energy-modulation-2}
\end{eqnarray}
One can show that all off-diagonal terms w.r.t. the spatial directions
$(x, y, z)$  in this expression do not
contribute to the derivatives with respect to the wave vector. 

Omitting the terms giving no contribution to the second-order
derivatives with respect to $\vec{q}$, one obtains after some transformations the expression
\begin{eqnarray}
\Delta {\cal E}^{(2)} &=&  \frac{1}{2} \sum_{i,j} \bigg[ \Bigg({\cal
                          J}^{xx}_{ij}\cos\;(\vec{q}_1\cdot (\vec{R}_i -
                          \vec{R}_j)) + {\cal J}^{yy}_{ij} \Bigg) \cos\;(\vec{q}_2\cdot (\vec{R}_i -
                          \vec{R}_j))    \nonumber \\
                      && +    \Bigg({\cal J}^{xx}_{ij}
                         \cos\;(\vec{q}_1\cdot (\vec{R}_i -
                          \vec{R}_j)) - {\cal J}^{yy}_{ij} \Bigg) \cos\;(\vec{q}_2\cdot (\vec{R}_i +
                          \vec{R}_j))     + ...        \bigg] \;    
  \label{Eq_Free_Energy-J-D-3}
\end{eqnarray}
The second term does not contribute to the q-dependence of the energy (see
discussion above). 
Taking $q_1 = 0$ and evaluating the second-order derivative of the
energy with respect to $\vec{q}_2$ one obtains  
\begin{eqnarray}
  \frac{\partial^2}{\partial q_2^2}
  \Delta {\cal E}^{(2)} &=&  \frac{1}{2}\sum_{i,j} (\hat{q} \cdot
                            (\vec{R}_j - \vec{R}_i))^2 ({\cal
                            J}^{xx}_{ij} + {\cal J}^{yy}_{ij}) \;.     
  \label{Eq_Free_Energy-J-D-4}
\end{eqnarray}
Comparing this expression with the corresponding expression obtained
for the Heisenberg model leads to the expression for isotropic 
exchange parameter given by  Eq.\ (\ref{Eq_Js}), showing that both
modulations lead to the same result.

\subsection{Fourth-order interactions}

\subsubsection{Four-spin isotropic  interactions.  z-component of DMI-like
  exchnage interactions}

  Next, we consider four-spin terms. First, we will deal with the isotropic
  exchange interaction and DMI-like terms in the extended Heisenberg
  Hamiltonian 
 \begin{eqnarray}
   H^{(4)} &=&   - \sum_{i,j,k,l} \bigg[  J^{s}_{ijkl}(\hat{s}_i
               \cdot \hat{s}_j) (\hat{s}_k \cdot \hat{s}_l) +
               \vec{\cal D}_{ijkl} \cdot (\hat{s}_i \times \hat{s}_j)
               (\hat{s}_k \cdot \hat{s}_l) \bigg] \; .   
 \label{Eq_deriv_Heisenberg_3-spin}
 \end{eqnarray}

 For the sake of convenience we keep only the terms
 required to derive the expressions for the isotropic  and
 DMI-like (only z-component) exchange coupling parameters, following
 the same idea as used to derive the bilinear exchange parameters.
 The FM state with the
 magnetization along the $z$ axis is used as a reference state for a 
 spin-spiral characterized by the vector $\vec{q}$ leading to a
 $q$-dependent energy change. The terms associated with the 
 isotropic exchange interactions $J^{s}_{ijkl}$ are proportional to
 $q^4$. On the other hand, the terms proportional to $q^3$ are
 associated with the DMI-like interactions ${\cal D}^{z}_{ijkl}$,
 i.e. they depend on vector product and scalar product of two pairs of
 spin moments. We focus here on these 
 terms, for which we can find a one to one correspondence between the
 model and first-principles energy terms. Following the idea used to derive
the bilinear terms, for the sake of simplicity we consider now only the
terms giving raise to a q-dependence of the energy. 
 In this case, the Hamiltonian can be written as follows
 \begin{eqnarray}
   H^{(4)} &=&   - \sum_{i,j,k,l} \bigg[  J^{s}_{ijkl}(s^x_is^x_j + s^y_is^y_j)(s^x_ks^x_l + s^y_ks^y_l)\nonumber \\
     &&    + {\cal D}^{z}_{ijkl}(s^x_is^y_j - s^y_is^x_j)(s^x_ks^x_l +
        s^y_ks^y_l) + ... \bigg] \; ,  
 \label{Eq_deriv_Heisenberg_4site-3spin-1}
 \end{eqnarray}
with the summation over all sites in the lattice with $i \neq j$ and  $k \neq
l$.
A similar summation occurs also for the first-principles approach, 
leading to a one-to-one correspondence between various terms in 
the two approaches. In the following, however, we will focus at the
end on the three-spin and biquadratic interactions.

Using the spin modulation in the following form
\begin{eqnarray}
  \hat{s}_i &=&
 (\mbox{sin} \theta \,\mbox{cos}(\vec{q} \cdot \vec{R}_i),
                \mbox{sin}\theta\, \mbox{sin}(\vec{q} \, \cdot \vec{R}_i),
                \mbox{cos} \theta )
\label{eq:spin-spiral2}
\end {eqnarray}
the corresponding energy change within the Heisenberg model is given by
 \begin{eqnarray}
   \Delta E^{(4)} &=&   - \sum_{i,j,k,l} \bigg[ J^{s}_{ijkl} \; \mbox{sin}^4\theta\;
               \mbox{cos}\vec{q} \cdot (\vec{R}_i - \vec{R}_j)\; \mbox{cos}\vec{q} \cdot (\vec{R}_k - \vec{R}_l) \nonumber \\
           &&    + D^{z}_{ijkl} \;\mbox{sin}^4\theta \;\mbox{sin}\vec{q}
              \cdot (\vec{R}_i - \vec{R}_j)\; \mbox{cos}\vec{q} \cdot
              (\vec{R}_k - \vec{R}_l) + ... \bigg] \; .  
 \label{Eq_deriv_Heisenberg_3-spin-2}
 \end{eqnarray}

In order to derive the expressions for
the isotropic and DMI-like interaction parameters, we consider the energy derivatives  $\frac{\partial^2}{\partial q^2}
\frac{\partial^2}{\partial q^2} \Delta E^{(4)}$ and 
$\frac{\partial}{\partial q} \frac{\partial^2}{\partial q^2}
\Delta E^{(4)}$ in the limit of $q = 0$.
These derivatives are applied to the terms corresponding to different
pairs of spin moments. As a results, one obtains:
 \begin{eqnarray}
 \frac{\partial}{\partial q} \frac{\partial^2}{\partial q^2}\bigg|_{q \to 0} \Delta E^{(4)} &=&  
        \sum_{i,j,k,l}    {\cal D}^{z}_{ijkl} \;\mbox{sin}^4\theta \;(\vec{R}_i - \vec{R}_j)(\vec{R}_k - \vec{R}_l)^2  \; ,  
 \label{Eq_deriv_Heisenberg_3-spin-3}
 \end{eqnarray}
and the fourth-order derivative terms
%
 \begin{eqnarray}
 \frac{\partial^2}{\partial q^2} \frac{\partial^2}{\partial q^2}\bigg|_{q \to 0} \Delta E^{(4)} &=&  - \sum_{i,j,k,l}  J^{s}_{ijkl} \; \mbox{sin}^4\theta\;
               (\vec{R}_i - \vec{R}_j)^2 (\vec{R}_k - \vec{R}_l)^2 \; ,  
 \label{Eq_deriv_Heisenberg_3-spin-4}
 \end{eqnarray}
Note that we do not consider in the last expression the terms $\sim (\vec{R}_k -
\vec{R}_l)^4$ or  $\sim (\vec{R}_i - \vec{R}_j)^4$, which do not contribute
to the q-dependence of the energy which is associated with the second
pair of spin moments.



Next, we evaluate the first-principles energy change due to the same spin
spiral. For this purpose we use the fourth-order energy term with
respect to the perturbation.
\begin{eqnarray}
\Delta {\cal E}^{(4)} &=& -\frac{1}{\pi} \sum_{i,j,k,l}\mbox{Im}\,\mbox{Tr} \int^{E_F}
                          dE\, \Delta V G \Delta V G \Delta V G \Delta V G\; \nonumber\\
&=& -\frac{1}{\pi} \sum_{i,j,k,l}\mbox{Im}\,\mbox{Tr} \int^{E_F}
                          dE\,  \nonumber \\
                      && \times \langle Z_i| \Delta V | Z_i \rangle \tau_{ij}
                         \langle Z_j| \Delta V |Z_j \rangle \tau_{jk}
                         \langle Z_k| \Delta V |Z_k \rangle \tau_{kl}
                         \langle Z_l| \Delta V |Z_l \rangle \tau_{li} \;.
\label{Eq_Free_Energy-4spin}
\end{eqnarray}
Using the spin-spiral according to Eq.\ (\ref{eq:spin-spiral2}) and
keeping only terms leading to
derivatives in the form of Eqs. (\ref{Eq_deriv_Heisenberg_3-spin-3})
and (\ref{Eq_deriv_Heisenberg_3-spin-4}), we obtain
\begin{eqnarray}
\Delta {\cal E}^{(4)} &=&  -\frac{1}{\pi} \sum_{i,j,k,l}\mbox{Im}\,\mbox{Tr} \int^{E_F}
                          dE\, \mbox{sin}^4\theta \nonumber \\
                      && \times \bigg[ T^x_i \tau_{ij}  T^x_j \tau_{jk}  T^x_k
                         \tau_{kl} T^x_l \tau_{li} \;
                    \mbox{cos}(\vec{q} \cdot \vec{R}_i)\,  \mbox{cos}(\vec{q}
                         \cdot \vec{R}_j )\,  \mbox{cos}(\vec{q} \cdot
                         \vec{R}_k )\,  \mbox{cos}(\vec{q} \cdot \vec{R}_l)\nonumber \\
                      && +  T^x_i \tau_{ij}  T^x_j \tau_{jk}  T^y_k
                         \tau_{kl} T^y_l \tau_{li} \;
                    \mbox{cos}(\vec{q} \cdot \vec{R}_i)\,  \mbox{cos}(\vec{q}
                         \cdot \vec{R}_j) \,  \mbox{sin}(\vec{q} \cdot
                         \vec{R}_k ) \, \mbox{sin}\vec{q} \cdot \vec{R}_l)\nonumber\\
                      && + T^y_i \tau_{ij}  T^y_j \tau_{jk}  T^x_k
                         \tau_{kl} T^x_l \tau_{li} \;
                    \mbox{sin}(\vec{q} \cdot \vec{R}_i)\,  \mbox{sin}(\vec{q}
                         \cdot \vec{R}_j )\,  \mbox{cos}(\vec{q} \cdot
                         \vec{R}_k)\,   \mbox{cos}(\vec{q} \cdot \vec{R}_l)\nonumber\\
                      && +  T^y_i \tau_{ij}  T^y_j \tau_{jk}  T^y_k
                         \tau_{kl} T^y_l \tau_{li} \;
                    \mbox{sin}(\vec{q} \cdot \vec{R}_i)\,  \mbox{sin}(\vec{q}
                         \cdot \vec{R}_j )\,  \mbox{sin}(\vec{q} \cdot
                         \vec{R}_k )\,  \mbox{sin}(\vec{q} \cdot
                         \vec{R}_l) \nonumber\\
                      && + T^x_i \tau_{ij}  T^y_j \tau_{jk}  T^x_k
                         \tau_{kl} T^x_l \tau_{li} \;
                    \mbox{cos}(\vec{q} \cdot \vec{R}_i)\,  \mbox{sin}(\vec{q}
                         \cdot \vec{R}_j )\,  \mbox{cos}(\vec{q} \cdot
                         \vec{R}_k )\,  \mbox{cos}(\vec{q} \cdot \vec{R}_l)\nonumber \\
                      && +  T^y_i \tau_{ij}  T^x_j \tau_{jk}  T^y_k
                         \tau_{kl} T^y_l \tau_{li} \;
                    \mbox{sin}(\vec{q} \cdot \vec{R}_i)\,  \mbox{cos}(\vec{q}
                         \cdot \vec{R}_j)\,   \mbox{sin}(\vec{q} \cdot
                         \vec{R}_k )\,  \mbox{sin}\vec{q} \cdot \vec{R}_l)\nonumber\\
                      && + T^y_i \tau_{ij}  T^x_j \tau_{jk}  T^x_k
                         \tau_{kl} T^x_l \tau_{li} \;
                    \mbox{sin}(\vec{q} \cdot \vec{R}_i)\,  \mbox{cos}(\vec{q}
                         \cdot \vec{R}_j )\,  \mbox{cos}(\vec{q} \cdot
                         \vec{R}_k) \,  \mbox{cos}(\vec{q} \cdot \vec{R}_l)\nonumber\\
                      && +  T^x_i \tau_{ij}  T^y_j \tau_{jk}  T^y_k
                         \tau_{kl} T^y_l \tau_{li} \;
                    \mbox{cos}(\vec{q} \cdot \vec{R}_i)\,  \mbox{sin}(\vec{q}
                         \cdot \vec{R}_j )\,  \mbox{sin}(\vec{q} \cdot
                         \vec{R}_k )\,  \mbox{sin}(\vec{q} \cdot
                         \vec{R}_l) +  ...  \bigg]
\label{Eq_Free_Energy-4spin-2}
\end{eqnarray}
By using the definition
\begin{eqnarray}
{\cal J}_{ijkl}^{\alpha\beta\gamma\delta} &=&  \frac{1}{2\pi} \mbox{Im}\,\mbox{Tr} \int^{E_F}
                          dE\,  T^\alpha_i \tau_{ij}  T^\beta_j \tau_{jk}  T^\gamma_k
                         \tau_{kl} T^\delta_l \tau_{li} 
\label{Eq_j_ijkl}
\end{eqnarray}
with a factor $\frac{1}{2}$ to have the same form for the biquadratic term
in the Hamiltonian, as the bilinear term has, one gets after some 
transformations the expression
\begin{eqnarray}
\Delta {\cal E}^{(4)} &=& - \frac{1}{4} \sum_{i,j,k,l} \mbox{sin}^4\theta \nonumber \\
                      && \times \bigg[ ({\cal J}_{ijkl}^{xxxx} + {\cal
                         J}_{ijkl}^{xxyy} + {\cal J}_{ijkl}^{yyxx} + {\cal
                         J}_{ijkl}^{yyyy})   \; 
                    \mbox{cos}(\vec{q} \cdot (\vec{R}_i - \vec{R}_j))\,
                         \mbox{cos}(\vec{q} \cdot (\vec{R}_k - \vec{R}_l)) \nonumber\\
                      && + ({\cal J}_{ijkl}^{xxxx} + {\cal
                         J}_{ijkl}^{xxyy} -  {\cal J}_{ijkl}^{yyxx} - {\cal
                         J}_{ijkl}^{yyyy}) \;
                    \mbox{cos}(\vec{q} \cdot (\vec{R}_i + \vec{R}_j))\,
                         \mbox{cos}(\vec{q} \cdot (\vec{R}_k - \vec{R}_l)) \nonumber\\
                      && + ({\cal J}_{ijkl}^{xxxx} - {\cal
                         J}_{ijkl}^{xxyy} +  {\cal J}_{ijkl}^{yyxx} - {\cal
                         J}_{ijkl}^{yyyy}) \;
                    \mbox{cos}(\vec{q} \cdot (\vec{R}_i - \vec{R}_j))\,
                         \mbox{cos}(\vec{q} \cdot (\vec{R}_k + \vec{R}_l)) \nonumber\\
                      && + ({\cal J}_{ijkl}^{xxxx} - {\cal
                         J}_{ijkl}^{xxyy} -  {\cal J}_{ijkl}^{yyxx} + {\cal
                         J}_{ijkl}^{yyyy}) \;
                    \mbox{cos}(\vec{q} \cdot (\vec{R}_i + \vec{R}_j))\,
                         \mbox{cos}(\vec{q} \cdot (\vec{R}_k + \vec{R}_l)) \nonumber \\
                      && +   ({\cal J}_{ijkl}^{xyxx} + {\cal
                         J}_{ijkl}^{xyyy} - {\cal J}_{ijkl}^{yxxx} - {\cal
                         J}_{ijkl}^{yxyy})   ) \; 
                    \mbox{sin}(\vec{q} \cdot (\vec{R}_i - \vec{R}_j))\,
                         \mbox{cos}(\vec{q} \cdot (\vec{R}_k - \vec{R}_l)) \nonumber\\
                      && + ({\cal J}_{ijkl}^{xyxx} + {\cal
                         J}_{ijkl}^{xyyy} +  {\cal J}_{ijkl}^{yxxx} + {\cal
                         J}_{ijkl}^{yxyy}) \;
                    \mbox{sin}(\vec{q} \cdot (\vec{R}_i + \vec{R}_j))\,
                         \mbox{cos}(\vec{q} \cdot (\vec{R}_k - \vec{R}_l)) \nonumber\\
                      && + ({\cal J}_{ijkl}^{xyxx} - {\cal
                         J}_{ijkl}^{xyyy} +  {\cal J}_{ijkl}^{yxxx} - {\cal
                         J}_{ijkl}^{yxyy}) \;
                    \mbox{sin}(\vec{q} \cdot (\vec{R}_i - \vec{R}_j))\,
                         \mbox{cos}(\vec{q} \cdot (\vec{R}_k + \vec{R}_l)) \nonumber\\
                      && + ({\cal J}_{ijkl}^{xyxx} - {\cal
                         J}_{ijkl}^{xyyy} -  {\cal J}_{ijkl}^{yxxx} + {\cal
                         J}_{ijkl}^{yxyy}) \;
                    \mbox{sin}(\vec{q} \cdot (\vec{R}_i + \vec{R}_j))\,
                         \mbox{cos}(\vec{q} \cdot (\vec{R}_k + \vec{R}_l))
                         + ...
                         \bigg]
\label{Eq_Free_Energy-4spin-4}
\end{eqnarray}
Evaluating the derivatives $\frac{\partial}{\partial q}
\frac{\partial^2}{\partial q^2} \Delta {\cal E}^{(4)}$ and $\frac{\partial^2}{\partial q^2}
\frac{\partial^2}{\partial q^2} \Delta {\cal E}^{(4)}$ in the limit $q =
0$, and equating to corresponding terms in Eqs.\
(\ref{Eq_deriv_Heisenberg_3-spin-3}) and
(\ref{Eq_deriv_Heisenberg_3-spin-4}),
one obtains

\begin{eqnarray}
J_{ijkl}^{s} &=& \frac{1}{4} \bigg[ ({\cal J}_{ijkl}^{xxxx} + {\cal
                         J}_{ijkl}^{xxyy} + {\cal J}_{ijkl}^{yyxx} + {\cal
                         J}_{ijkl}^{yyyy})   )\bigg]
\label{Eq_SYM-spin-4}
\end{eqnarray}

\begin{eqnarray}
D_{ijkl}^{z} &=& \frac{1}{4} \bigg[{\cal J}_{ijkl}^{xyxx} + {\cal
                         J}_{ijkl}^{xyyy} -  {\cal J}_{ijkl}^{yxxx} - {\cal
                         J}_{ijkl}^{yxyy}) \;\bigg]
\label{Eq_DMI-spin-4}
\end{eqnarray}

 \subsubsection{Biquadratic and 3-spin  DMI-like interactions: x- and y- components}

To obtain the $x$- and $y$-components of the 4-spin DMI-like interactions
we will follow the scheme used to derive the corresponding components of the
bilinear DMI. 
The corresponding term in the Heisenberg Hamiltonian has the form
\begin{eqnarray}
  {H}^{a(4)} &=&  - \sum_{i,j,k,l}  \vec{{\cal D}}_{ijkl} \cdot (\hat{s}_i \times
      \hat{s}_j) (\hat{s}_k \cdot \hat{s}_l) \;.  
\label{Eq_Heisenberg_H4_DMIXY2}
\end{eqnarray}
Here we use the spin modulation characterized by two $\vec{q}$-vectors,
which allows the simultaneous tilting of spin moments towards the $x$- and $y$-axes:
\begin{eqnarray}
  \hat{s}_i &=&
 (\mbox{sin}(\vec{q}_1 \cdot \vec{R}_i) \;\mbox{cos}(\vec{q}_2 \cdot \vec{R}_i)\,,\;
                \mbox{sin}(\vec{q}_2 \cdot \vec{R}_i) \,,\; \mbox{cos}(\vec{q}_1 \cdot
 \vec{R}_i)\mbox{cos}(\vec{q}_2 \cdot \vec{R}_i) ) \;,
\label{spiral2-2}
\end {eqnarray}
The corresponding contribution to the energy in model approach
is now:
\begin{eqnarray}
  \Delta {E}^{a(4)} &=&  -  \sum_{i,j,k,l}  \bigg[
                 {\cal D}^x_{ijkl}
                 [\mbox{sin}(\vec{q}_2 \cdot \vec{R}_i)\, \mbox{cos}(\vec{q}_1 \cdot
 \vec{R}_j)\mbox{cos}(\vec{q}_2 \cdot \vec{R}_j) - \mbox{sin}(\vec{q}_2 \cdot \vec{R}_j)\, \mbox{cos}(\vec{q}_1 \cdot
 \vec{R}_i)\,\mbox{cos}(\vec{q}_2 \cdot \vec{R}_i) ] \nonumber  \\
             &&      + {\cal D}^y_{ijkl} \; \mbox{sin}(\vec{q}_1 \cdot
(\vec{R}_j - \vec{R}_i))\; \mbox{cos}(\vec{q}_2 \cdot \vec{R}_i)\;\mbox{cos}(\vec{q}_2 \cdot
                 \vec{R}_j)     \nonumber  \\
             &&   
                 + {\cal D}^z_{ijkl} [\mbox{sin}(\vec{q}_1 \cdot \vec{R}_i)\,\mbox{cos}(\vec{q}_2 \cdot \vec{R}_i)\,
                \mbox{sin}(\vec{q}_2 \cdot \vec{R}_j) -
                \mbox{sin}(\vec{q}_1 \cdot \vec{R}_j)\,
                \mbox{cos}(\vec{q}_2 \cdot \vec{R}_j)\,
                \mbox{sin}(\vec{q}_2 \cdot \vec{R}_i)  ]
                \bigg] \nonumber \\
             && \times  [\mbox{cos}(\vec{q}_1 \cdot (\vec{R}_k -\vec{R}_l))\,
                 \mbox{cos}(\vec{q}_2 \cdot \vec{R}_k)\,
                 \mbox{cos}(\vec{q}_2 \cdot \vec{R}_l) + \mbox{sin}(\vec{q}_2 \cdot \vec{R}_k)\,
                 \mbox{sin}(\vec{q}_2 \cdot \vec{R}_l) ]\,.    
\label{Eq_Heisenberg_H4_DMIXY4-1}
\end{eqnarray}
The 4-spin DMI-like terms involve a cross-product of the spin moments on sites 
$i$ and $j$ and a scalar product of the spin moments on sites $k$ and $l$.
Therefore, to derive the expression for the exchange parameters, one has
to consider the first-order derivative with respect to the wave vector of the
part associated with the atoms $i$ and $j$ and the second order derivative of
the part associated with the atoms $k$ and $l$. Taking first $q_1 = 0$,
one obtains 
\begin{eqnarray}
\frac{\partial}{\partial q_2} \frac{\partial^2}{\partial q_2^2} \bigg|_{q_2 = 0}  \Delta {E}^{a(4)}
  &=&
       \sum_{i,j,k,l} {\cal D}^x_{ijkl}\,
   ({\hat{q}_2}\cdot (\vec{R}_j -\vec{R}_i))\,  ({\hat{q}_2}\cdot (\vec{R}_k -\vec{R}_l))^2 \,.
\label{Eq_Heisenberg_H4_Deriv-1}
\end{eqnarray}

Calculating alternatively first derivatives
 $\frac{\partial}{\partial q_1}|_{q_1 = 0} \Delta {E}^{a(4)}$ and then $ \frac{\partial^2}{\partial
   q_2^2}|_{q_2 = 0}\Delta {E}^{a(4)}$, one obtains
\begin{eqnarray}
\frac{\partial}{\partial q_1}\bigg|_{q_1 = 0} \frac{\partial^2}{\partial q_2^2}\bigg|_{q_2 = 0} \Delta {E}^{a(4)}
  &=&
       \sum_{i,j,k,l} {\cal D}^y_{ijkl}\,
    ({\hat{q}_1}\cdot (\vec{R}_j -\vec{R}_i)) \, ({\hat{q}_2}\cdot (\vec{R}_k -\vec{R}_l))^2 
\label{Eq_Heisenberg_H4_Deriv-2}
\end{eqnarray}

In the case $l=j$ we obtain the three-spin interactions, while
in the case  $l=j, k=i$ we will get the DMI-like biquadratic
interactions.

Let us focus on the three-spin DMI-like interaction  $D^{x(y)}_{ijkj}$.
In the first-principles approach, let's consider the perturbation as
follows, 
\begin{eqnarray}
 \Delta V  &=& \sum_{m} \delta v_m\;,
\label{spiral3}
\end {eqnarray}
with $\delta v_m =  \beta \big( \vec{\sigma}\cdot\hat{s}_m
  -  \sigma_z\big) B_{xc}(\vec{r})$ and $m$ running over all lattice
  sites, implying a spatial spin modulation in the form as given by Eq.\ (\ref{spiral2-2}). 
We use the third-order term of the energy expansion
\begin{eqnarray}
\Delta {\cal E}^{(3)} &=& -\frac{1}{\pi} \sum_{i,j,k,l}\mbox{Im}\,\mbox{Tr} \int^{E_F}
                          dE\, (E - E_F) G \Delta V G \Delta V G \Delta
                          V G\;\\
&=& -\frac{1}{\pi} \sum_{i,j,k,l}\mbox{Im}\,\mbox{Tr} \int^{E_F}
                          dE\,  (E - E_F) \nonumber \\
                      && \times \langle Z_i| Z_i \rangle \tau_{ij}
                         \langle Z_j| \Delta V |Z_j \rangle \tau_{jk}
                         \langle Z_k| \Delta V |Z_k \rangle \tau_{kj}
                         \langle Z_j| \Delta V |Z_j \rangle \tau_{ji} \;
\label{Eq_Free_Energy-DX1}
\end{eqnarray}
with the summation over all indexes $i \neq j$ and $k \neq j$.

Assuming that the coupling of the spin moments $i$ and $j$ has the form of
a cross-product and that of the spin moments $k$ and $j$ has the form of
a scalar product, this expression can be written in the symmetrized form (in
analogy to the DMI) taking into account that the perturbation $\Delta
v_m$ can be applied either to site $i$ or to site $j$.
Using the trace invariance with respect to circular permutations, this
leads to the expression 
\begin{eqnarray}
\Delta {\cal E}^{(3)} &=&
 -\frac{1}{\pi}\frac{1}{4} \sum_{i,j,k,j}\mbox{Im}\,\mbox{Tr} \int^{E_F}
                          dE\,  (E - E_F) \nonumber \\
                      && \times \bigg[
                        \tau_{ji} \langle Z_i| Z_i \rangle \tau_{ij}
                         \langle Z_j| \delta v_{j} |Z_j\rangle  \bigg( \tau_{jk}  
                         \langle Z_k| \delta v_{k} |Z_k \rangle \tau_{kj}
                         \langle Z_j| \delta v_{j} |Z_j \rangle \bigg)
                           \nonumber \\  
                      && +  \tau_{ji} \langle Z_i| \delta v_{i} |Z_i\rangle  \tau_{ij}
                            \langle Z_j| Z_j \rangle \bigg( \tau_{jk}  
                         \langle Z_k| \delta v_{k} |Z_k \rangle \tau_{kj}
                         \langle Z_j| \delta v_{j} |Z_j \rangle \bigg)
                           \nonumber \\  
                      && + \langle Z_j| Z_j\rangle
                         \tau_{ji} \langle Z_i|\delta v_{j}| Z_i \rangle
                         \tau_{ij}  \bigg( \langle Z_j|\delta v_{j}|Z_j )
                         \tau_{jk}
                         \langle Z_k|\delta v_{k}|Z_k \rangle \tau_{kj}
                         ) \bigg) 
                           \nonumber \\  
                      && + \langle Z_j|\delta v_{j}| Z_j \rangle
                         \tau_{ji} \langle Z_i| Z_i\rangle 
                         \tau_{ij}  \bigg( \langle Z_j|\delta v_{j}|Z_j )
                         \tau_{jk}
                         \langle Z_k|\delta v_{k}|Z_k \rangle \tau_{kj}
                         ) \bigg)  \bigg] \;,
\label{Eq_Free_Energy-DX1}
\end{eqnarray}
where the  terms in the parentheses are interpreted as those
corresponding to the scalar product of the spin moments $k$ and $j$.
Using the spin modulation according to Eq.\ (\ref{spiral2-2}), this part has to be
reduced in the analogy to the bilinear case leading to Eq. (\ref{Eq_Free_Energy-J-D-4}).

Taking the derivatives for this energy term with respect to $\vec{q}_1$ and
$\vec{q}_2$, as in the case of Eqs. (\ref{Eq_Heisenberg_H4_Deriv-1}) and
(\ref{Eq_Heisenberg_H4_Deriv-2}), one obtains the expressions for
three-spin DMI-like interaction parameter, as given in Eqs.\
(\ref{Eq:D-4-4_XYZ}, \ref{Eq:D-4-2_XYZ})

\subsection{Three-spin chiral interactions}

To derive the tree-spin chiral interactions, we follow the idea used
to derive the $x$- and $y$- components of the DMI-like interactions discussed above.
In this case also one has to use the spin modulation given by Eq.\
(\ref{spiral2-2}) and 
characterized by two wave vectors to allow spin moment tiltings towards the
$x$ and $y$ axes simultaneously.  
The term associated with the three-spin chiral exchange interaction in the
extended Heisenberg Hamiltonian is given by
\begin{eqnarray}
  H^{(3)} &=&  - \sum_{i \neq j\neq k}
                   J_{ijk} \hat{s}_i\cdot (\hat{s}_j \times \hat{s}_k) \; .
\label{Eq_Heisenberg_3-spin-TCI-1}
\end{eqnarray}
The three-spin energy term in both approaches has to have the same
properties with respect to permutation. The Heisenberg term
Eq.\ (\ref{Eq_Heisenberg_3-spin-TCI-1}) can be written in the form
\begin{eqnarray}
  H^{(3)} &=&  - \frac{1}{3}\sum_{i \neq j\neq k}
              J_{ijk} \Big[\hat{s}_i\cdot (\hat{s}_j \times \hat{s}_k) +
              \hat{s}_k\cdot (\hat{s}_j \times \hat{s}_j)  +
              \hat{s}_j\cdot (\hat{s}_k \times \hat{s}_i) \Big]              \; .
\label{Eq_Heisenberg_3-spin-TCI-2}
\end{eqnarray}

On first principles level, we use also the second-order term of the 
energy expansion given by the expression 
\begin{eqnarray}
\Delta {\cal E}^{(2)} &=& -\frac{1}{\pi} \mbox{Im}\,\mbox{Tr} \int^{E_F}
                    dE (E - E_F)\, G_0 \Delta V G_0 \Delta V G_0 \;.
\label{Eq_Free_Energy-TCI-1}
\end{eqnarray}

To cover different forms of the triple scalar product in the Eq.\
(\ref{Eq_Heisenberg_3-spin-TCI-2}), the first-principles energy Eq.\
(\ref{Eq_Free_Energy-TCI-1}) (in analogy to the case of 4-spin DMI-like
expression) should be written as follows
\begin{eqnarray}
\Delta {\cal E}^{(3)} &=&
 -\frac{1}{\pi}\frac{1}{3} \sum_{i \neq j\neq k}\mbox{Im}\,\mbox{Tr} \int^{E_F}
                          dE\,  (E - E_F) \nonumber \\
                      && \times \bigg[                         
                         \langle Z_i| Z_i \rangle \tau_{ij}
                         \langle Z_j| \delta v_{j} |Z_j\rangle \tau_{jk}  
                         \langle Z_k| \delta v_{k} |Z_k \rangle \tau_{ki}
                           \nonumber \\  
                      &&                        
                        + \langle Z_i|\delta v_{i} |Z_i \rangle \tau_{ij}
                         \langle Z_j|Z_j\rangle \tau_{jk} 
                         \langle Z_k| \delta v_{k} |Z_k \rangle \tau_{ki}
                           \nonumber \\  
                      &&                        
                        + \langle Z_i|\delta v_{i} | Z_i \rangle \tau_{ij}
                         \langle Z_j| \delta v_{j} |Z_j\rangle \tau_{jk}  
                         \langle Z_k| Z_k \rangle \tau_{ki} \bigg] \;.
\label{Eq_Free_Energy-TCI-2}
\end{eqnarray}

Using the spin modulation according to Eq.\ (\ref{spiral2-2}) the model and
first-principles energy expressions have to be reduced to a form 
having a corresponding q-dependence of the terms giving non-vanishing second-order
derivatives with respect to $\vec{q}_1$ and $\vec{q}_2$ in the limit $q_1 \to 0$,
$q_2 \to 0$, which are proportional to $\big( \hat{z} \cdot [(\vec{R}_i
- \vec{R}_j) \times(\vec{R}_k - \vec{R}_j) ] \big) $. Equating these
expressions give access to the three-spin chiral interactions as given
by Eq.\ (\ref{Eq:J_XYZ}).

\end{widetext}


%

\end{document}